\documentclass[twocolumn]{aastex631}
\usepackage{multirow}
\usepackage{array}
\usepackage{amsmath}    
\usepackage{hyperref}

\shorttitle{Periodicity Search for Repeating FRBs}
\shortauthors{Chen Du et al.}
\graphicspath{{./}}

\begin{document}

\title{A Thorough Search for Short Timescale Periodicity in Four Active Repeating Fast Radio Bursts}

\correspondingauthor{Yong-Feng Huang}
\email{hyf@nju.edu.cn}

\author{Chen Du}
\affiliation{School of Astronomy and Space Science, Nanjing University, Nanjing 210023, China}

\author{Yong-Feng Huang}
\affiliation{School of Astronomy and Space Science, Nanjing University, Nanjing 210023, China}
\affiliation{Key Laboratory of Modern Astronomy and Astrophysics (Nanjing University), Ministry of Education, China}

\author{Zhi-Bin Zhang}
\affiliation{School of Physics and Physical Engineering, Qufu Normal University, Qufu 273165, China}

\author{Alexander Rodin}
\affiliation{Pushchino Radio Astronomy Observatory, Astro Space Center, P.N.Lebedev Physical Institute, Russian Academy of Sciences,\\
Moscow region 142290, Russia}

\author{Viktoriya Fedorova}
\affiliation{Pushchino Radio Astronomy Observatory, Astro Space Center, P.N.Lebedev Physical Institute, Russian Academy of Sciences,\\
Moscow region 142290, Russia}

\author{Abdusattar Kurban}
\affiliation{Xinjiang Astronomical Observatory, Chinese Academy of Sciences, Urumqi 830011, China}
\affiliation{Key Laboratory of Radio Astronomy, Chinese Academy of Sciences, Urumqi 830011, China}
\affiliation{Xinjiang Key Laboratory of Radio Astrophysics, Urumqi 830011, China}

\author{Di Li}
\affiliation{National Astronomical Observatories, Chinese Academy of Sciences, Beijing 100101, China}
\affiliation{Key Laboratory of Radio Astronomy and Technology, Chinese Academy of Sciences, Beijing 100101, China}
\affiliation{Zhejiang Lab, Hangzhou 311121, China}



\begin{abstract}

Fast Radio Bursts (FRBs) are bright radio transients with
millisecond durations which typically occur at extragalactic
distances. The association of FRB 20200428 with the Galactic
magnetar SGR J1935+2154 strongly indicates that they could
originate from neutron stars, which naturally leads to the
expectation that periodicity connected with the spinning of
magnetars should exist in the activities of repeating FRBs.
However, previous studies have failed to find any signatures
supporting such a conjecture. Here we perform a thorough search
for short timescale periodicity in the four most active repeating
sources, i.e. FRBs 20121102A, 20200120E, 20201124A, and 20220912A.
Three different methods are employed, including the phase folding
algorithm, the H-test and the Lomb-Scargle periodogram. For the
three most active repeaters from which more than 1000 bursts have
been detected, i.e. FRBs 20121102A, 20201124A, and 20220912A, more
in-depth period searches are conducted by considering various
burst properties such as the pulse width, peak flux, fluence, and
the brightness temperature. No clear periodicity is found in a period range of
0.001--1000 s in all the efforts. Implications of such a
null result on the theoretical models of FRBs are discussed.

\end{abstract}




\section{Introduction}
\label{sect:intro}

Fast radio bursts (FRBs) are mysterious radio transients with an extremely high
brightness temperature, and with a short duration of a few
milliseconds \citep{thornton2013population}. Since the first discover of FRBs
in 2007 \citep{lorimer2007Bright}, a new field is opened in astronomy, which concentrates
on the study of such extremely rapid transients. Exciting progresses have been acquired
in the past decade. Over 700 FRB sources are
discovered \citep{petroff2016FRBCATa,chime/frb2021First,xu2023Blinkverse}, along with a
diverse burst phenomenology \citep{pleunis2021Fast,zhang2023FRBs,hu2023Comprehensive}. Several important databases
have been established for FRBs, such as FRBCAT \footnote{\url{https://www.frbcat.org/}},
Blinkverse \footnote{\url{https://blinkverse.alkaidos.cn/}}, and the
CHIME/FRB Catalog \footnote{\url{https://www.chime-frb.ca/}}, which have greatly
facilitated the study of FRBs by integrating a large amount of observational data.
Due to the high dispersion measures of FRBs, they should occur at cosmological
distances \citep{chime/frb2021First}, connecting with some kinds of violent activities
and/or even catastrophic events \citep{moroianu2023assessmenta}. However, the
triggering mechanisms and radiation processes behind FRBs are still under
debate \citep{platts2019livinga,zhang2020physical,hu2023Comprehensive, zhang2023physics}.

Notably, although most FRBs seem to be one-off events, a small fraction
of them do produce repeating bursts. The discovery of the first repeating
source, FRB 20121102A, was a milestone in the field \citep{spitler2016repeatinga}.
It provides a direct hint for classifying FRBs, suggesting that at least some
FRBs are non-catastrophic events. Models in which FRBs originate from compact
objects such as relatively young neutron stars are favored. The connection
between FRBs and magnetars is further confirmed by the discovery of FRB 20200428,
which is found to be associated with an X-ray burst from SGR J1935+2154, a
magnetar in the Milky Way \citep{li2021HXMT,mereghetti2020INTEGRAL}.

The waiting time of active repeating FRBs shows a clear bimodal
log-normal distribution \citep{li2021bimodal,xu2022fast}. The main component peaks at around tens of seconds, reflecting
the active level of the source, while the secondary component peaks at
a few tens of milliseconds, possibly reflecting the intrinsic radiation
characteristics of the central engine. Recently, it was argued
that the secondary waiting time peak might be due to some burst
substructures being misclassified as separate
bursts \citep{jahns2022FRB,zhou2022FAST,wang2024Memory}. Observations on these
active repeaters raise two interesting questions. The first is
whether all FRBs are actually repeating sources, answering which
requires long-term continuous observations and thus it is not yet
conclusive \citep{katz2022sources}. The second is whether repeating FRBs show any
periodicity, which is a natural expectation if FRBs originate from
neutron stars that are generally rotating at an extremely stable
period.


Interestingly, recent observations on FRB 20180916B show that it has a periodic
activity. The period is $\sim 16.35$ days, with an active window of $\sim 5$
days \citep{chime/frb2020Periodic}. A tentative period of $\sim 160$ days has
also been suggested for FRB 20121102 \citep{rajwade2020Possible}. These long periods
are unlikely due to the rotation of neutron stars \citep{beniamini2020Periodicity,xu2021Periodic},
but could be explained as a signature of the orbital motion of binary
systems \citep{zhang2020What,ioka2020binarya,li2021Periodic,sridhar2021Periodic,geng2021Repeating,kurban2022Periodic},
or the precession of neutron stars \citep{yang2020Orbitinduced,tong2020Periodicity,levin2020Precessing,zanazzi2020Periodic,chen2021Reconciling}, or even the precession of accretion discs \citep{katz2022Precessiona,katz2024Orbital},
although a decisive conclusion still could not be drawn yet \citep{katz2021Testing}. It should also be noted that the
observations of FRBs are subjected to the daily-sampling problem, i.e., each source is
generally monitored by a particular radio telescope only in a relatively fixed time span
every day due to the rotation of the Earth. Such an observational strategy might lead to
some false features in the periodicity, which should be carefully examined.


Possible short-time periodic features were also reported in some repeating FRBs recently.
For example, a series of nine pulses were observed in a total duration of $\sim 3$ s from
FRB 20191221A. As a result, it seems to have a short period of $\sim 200$ ms at a confidence
level of 6.5 $\sigma$ \citep{chime/frb2022Subsecond}. However, inspection of the lightcurve of FRB 20191221A suggests that the
substructures are not obviously periodic, making the periodicity highly doubtful.
FRB 20201020A was reported to be consisted of five pulses with a total duration of $\sim 2.14$ ms,
which even shows a sub-millisecond periodic behavior \citep{pastor-marazuela2022fasta}, .
It is worth mentioning that these phenomena could only be regarded as quasi-periodic
behaviors that are observed in a very short time window. These temporal structures could
also be multiple components of a single burst. The possibility that they are
due to some oscillations in the magnetosphere of magnetars could not be excluded, and their
connection with the spin period cannot be firmly established. On the other hand, some
attempts to search for short time periodicity (from milliseconds to seconds) in the most
active repeating sources such as FRBs 20201124A and 20121102A, have failed to present any
positive results \citep{li2021bimodal,xu2022fast,niu2022FAST}. The
lack of any signatures connecting to the spinning of magnetars has troubled theorists
deeply, which thus needs further investigation with more data and samples.

In this study, we will analyze the dedispersed arrival times of four active repeating FRBs,
namely FRBs 20121102A, 20200120E, 20201124A and 20220912A, paying special
attention on the possible existence of short-time periodicity (0.001--1000 s). Various period searching methods are engaged for the purpose,
including the phase folding algorithm, the H-test and the Lomb-Scargle
periodogram. Observational data obtained by various telescopes such as FAST, Arecibo, uGMRT, Effelsberg and GBT are collected as far as possible. Different weights connected to the pulse width, peak flux, fluence, and brightness temperature
are also considered during the period analysis.

The structure of this paper is organized as follows. In Section 2, the data samples
of the four FRBs are briefly introduced. The three period searching methods used in
this study are described in Section 3. Numerical results based on the single-day period searching are presented in Section 4. More elaborated period searches are conducted in Section 5 for the three most active repeaters. Finally, Section 6 presents our conclusions and discussion.

\section{Samples}
\label{sect:Sam}

\begin{figure*}[h!]
   \centering
   \includegraphics[width=1\textwidth]{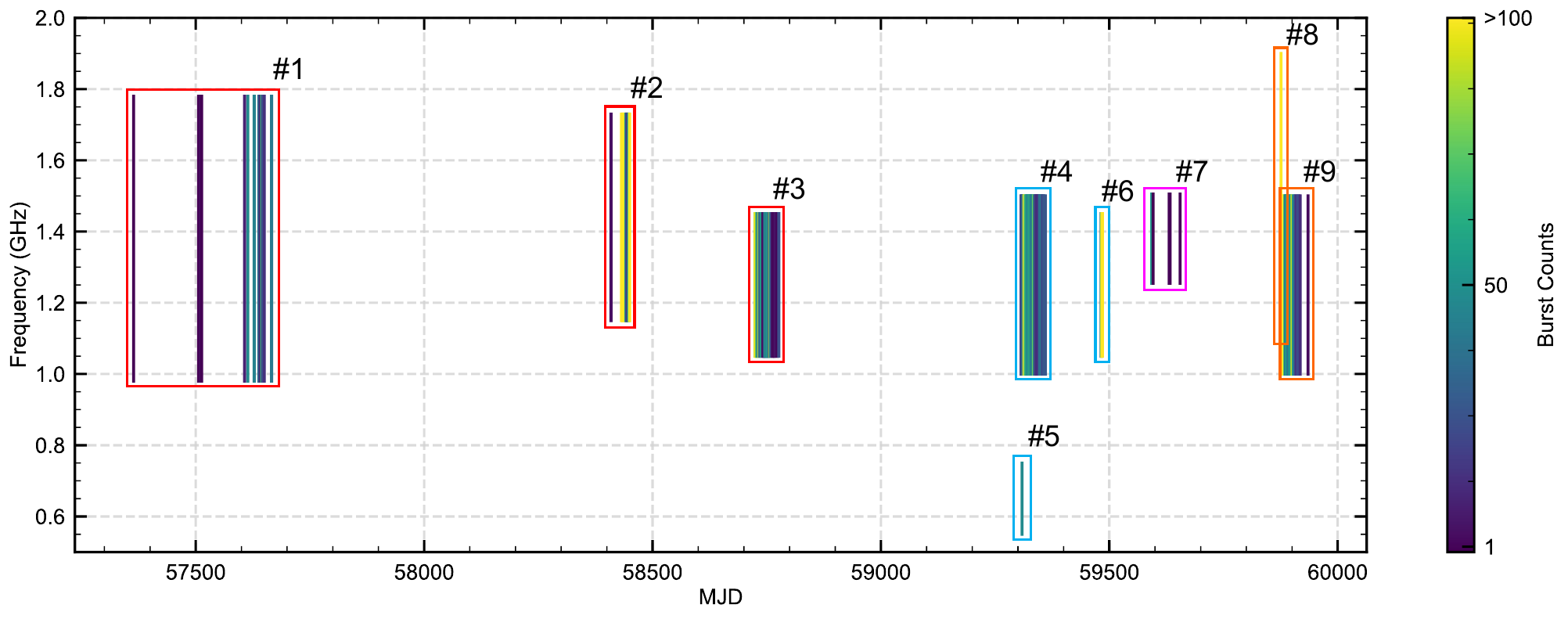}
   \caption{\footnotesize Overview of the observations of all
   the nine datasets from the four repeaters.
   The vertical axis represents the observation
   frequency range of different telescopes, and the horizontal axis
   represents the MJD time. The number of bursts observed on each day
   is indicated by the color of the data bars.
   Each dataset is enclosed in a box.
   Datasets \#1, \#2 and \#3 (red boxes) are bursts from FRB 20121102A:
   \#1 shows Arecibo observations that includes a total of 478 bursts, \#2 shows Arecibo observations with a total of 849 bursts, and \#3 corresponds to FAST observations with a total of 1652 bursts. Datasets \#4, \#5 and \#6 (cyan boxes) are bursts from FRB 20201124A:
   \#4 corresponds to FAST observations with a total of 1863 bursts,
   \#5 shows uGMRT observations with a total of 48 bursts,
   and \#6 shows FAST observations with a total of 881 bursts.
   Dataset \#7 (magenta box) includes 60 bursts from FRB 20200120E, observed by Effelsberg.
   Datasets \#8 and \#9 (orange boxes) are bursts from FRB 20220912A:
   \#8 corresponds to GBT observations with a total of 128 bursts,
   and \#9 shows FAST observations that includes a total of 1075 bursts.}
   \label{Fig1}
\end{figure*}

To search for periodicity in a particular FRB source, enough
bursts need to be detected. Among all the repeating FRBs, four
repeaters are very active and are suitable for short timescale
period analysis. We have collected the key parameters of all the
FRBs from these four repeaters. Below, we present a detailed
description of the data samples. Table ~\ref{Tab1} summarizes the
basic information of these datasets. Figure ~\ref{Fig1} shows an
overview of the observations for all selected datasets.

\subsection{FRB 20121102A}

FRB 20121102A is the first repeating source discovered
\citep{spitler2016repeatinga}. Extensive follow-up observations
show that it locates near a star-forming region in a dwarf galaxy
at a redshift of $z = 0.193$
\citep{chatterjee2017directa,marcote2017Repeating}. It is
interestingly found to be associated with a persistent compact
radio source \citep{tendulkar2017hosta}. More strikingly, it seems
to have long-term periodic activities. Analysis of burst data over
5 years shows a tentative period of $\sim$ 160 days and a duty
cycle of $\sim$ 55 percent \citep{rajwade2020Possible}. Recently,
an extremely active phase was observed by FAST, which reported the
detection of 1652 individual bursts in 59.5 hours spanning 47 days
since August 2019, with a peak burst rate of 122 hr$^{-1}$
\citep{li2021bimodal}. These bursts are detected in a frequency
range of 1.05--1.45 GHz. Arecibo also detected a burst storm
from FRB 20121102A in September 2016. A total number of 478 bursts
were recorded in 59 hours, with a peak burst rate of 49.3
hr$^{-1}$, in the frequency range of 0.98--1.78 GHz
\citep{hewitt2022Arecibo}. In another active episode around
November 2018, Arecibo detected 849 bursts from FRB 20121102A
during 10 observation epoches, with a peak burst rate of > 200
hr$^{-1}$, in the frequency range of 1.15--1.73 GHz
\citep{jahns2022FRB}. With a total number of $> 2000$ bursts being
observed, FRB 20121102A is a unique repeater for which the
possible existence of short-time periodicity can be examined in
great detail. In short, for FRB 20121102A, three datasets are
available for our periodicity analysis, which are listed as
Datasets \#1, \#2 and \#3 in Table ~\ref{Tab1} (also shown in
Figure ~\ref{Fig1}).

\subsection{FRB 20200120E}

A total number of 60 bursts were detected from FRB 20200120E by
Effelsberg over a time span of $\sim$ 62 days
\citep{nimmo2023burst}. Especially, a burst storm of 53 events
occurring in a short epoch of 40 minutes (equivalent to a burst
rate of $\sim$ 80 hr$^{-1}$) are detected from this source on Jan
14, 2022. The observation frequency range is 1.255--1.505 GHz for
Effelsberg. FRB 20200120E is not far from us. Its distance is only
3.6 Mpc. It is associated with a globular cluster in a nearby
galaxy \citep{kirsten2022repeating}. On average, the pulse widths
of the bursts from FRB 20200120E are much narrower than that of
others (by about one order of magnitude). The burst luminosity is
also significantly lower. The spectra of its bursts are generally
well described by a steep power-law function. Comparing with other
repeaters, both the DM and the variation of DM of FRB 20200120E
are quite low, which means a relatively clean local magneto-ionic
environment \citep{nimmo2023burst}. To conclude, for FRB
20200120E, only one dataset is available for our periodicity
analysis, which is listed as Dataset \#7 in Table ~\ref{Tab1}
(also shown in Figure ~\ref{Fig1}).

\subsection{FRB 20201124A}

FRB 20201124A is an extremely active repeater, with nearly $\sim
2000$ bursts being detected. The observations were conducted
mainly by FAST and uGMRT. FAST detected 1863 bursts from it over a
time span of only $\sim 53$ days beginning in April 2021
\citep{xu2022fast}. Note that all these bursts occurred in a total
time period of 88 hours, with a peak burst rate of 46 hr$^{-1}$.
The observation frequency range is 1.0--1.5 GHz. uGMRT also
detected 48 bursts from it on a single day (April 5, 2021), with a
burst rate of 16 hr$^{-1}$, covering a frequency range of 0.55--0.75 GHz \citep{marthi2021Burst}. A few months later, during
another extremely active episode in September 2021, FAST observed
881 bursts within just 4 days in the frequency range of 1.05--1.45 GHz, with a peak burst rate reaching an astonishing 542
hr$^{-1}$ \citep{zhang2022FASTa}. The RM of FRB 20201124A varies
significantly on a short timescale. A high degree of circular
polarization is observed in a good fraction of bursts. FRB
20201124A is associated with a barred spiral galaxy at a redshift
of $z = 0.098$, located in a region of low stellar density
\citep{xu2022fast}. So, for FRB 20201124A, three datasets are
available for our periodicity analysis, which are listed as
Datasets \#4, \#5 and \#6 in Table ~\ref{Tab1} (also shown in
Figure ~\ref{Fig1}).

\subsection{FRB 20220912A}

FRB 20220912A is a recently discovered extremely active repeating
source, first reported by CHIME \citep{2022ATel15679....1M}. Its
burst activities have been observed by many telescopes
\citep{zhang2023FAST,feng2023extreme,fedorova2023Observations,hewitt2023Densea}.
During a time span of 55 days around October 2022, FAST detected a
total of 1076 bursts in 17 observations totaling 8.67 hours, with
the peak burst rate reaching 390 hr$^{-1}$, in the frequency range
of 1.0--1.5 GHz \citep{zhang2023FAST}. GBT also observed a total
of 128 bursts within a 1.4-hour observation (equivalent to a burst
rate of $\sim$ 90 hr$^{-1}$) on October 24, 2022, in the frequency
range of 1.1--1.9 GHz \citep{feng2023extreme}. The host galaxy
of FRB 20220912A has been located at a redshift of $z = 0.0771$
\citep{ravi2023Deep}. Unlike other active repeating sources such
as FRB 20121102A, this source has an absolute RM value close to
zero, suggesting that it appears to be in a clean environment
\citep{feng2023extreme}. Most bursts from FRB
20220912A in the L Band have nearly 100\% linear polarization, and
a significant fraction of bursts exhibit circular polarization
\citep{zhang2023FAST,feng2023extreme}. Additionally, observations of this source have revealed microshots with widths < 32 ns \citep{hewitt2023Densea,katz2024Radiation}. As a result, for FRB
20220912A, two datasets are available for our periodicity
analysis, which are listed as Datasets \#8 and \#9 in Table
~\ref{Tab1} (also shown in Figure ~\ref{Fig1}).

\section{Methods}
\label{sect:Met}

In this study, three different methods are used for periodicity
analysis of the above four active repeaters: the phase folding
algorithm, the H-test, and the Lomb-Scargle periodogram. Among
them, the phase folding is a time-domain method, while the H-test
and Lomb-Scargle periodogram are frequency-domain methods. Here we
describe the three methods briefly. The period range and period
step used in our analysis are introduced in the final part of this
section.

\subsection{Phase folding algorithm}

The phase folding algorithm is a traditional but useful and
practical method to analyze the time of arrivals of FRBs. Many
other analysis methods such as those based on Fourier transforms
require additional intensity data. However, the phase folding
algorithm only requires a sequence of timing data as the inputs.
Considering the diversity of the FRB data sources and the
incompleteness and disunity of the data sample itself, the phase
folding algorithm is even more credible to some extent.

To begin the analysis, we first assume a trial period ($p$) for
the timing data. The time of arrivals of the FRBs is then folded
with respect to this assumptive period. Usually, the first data of
the sequence ($t_0$) is taken as the zero point. Then the phase of
the $i$-th burst that occurs at $t_i$ can be calculated as:
\begin{equation}
\phi_i=\frac{\big[(t_i-t_0) \bmod p\big]}{p},
\end{equation}
where $\bmod$ means a remainder operation. If $p$ is really the
intrinsic period of the timing sequence, then a large number of
bursts will concentrate at some particular phase. On the other
hand, if $p$ is not the period, then the bursts will randomly
distribute at all phases so that no prominent structures will be
seen in the phase space. The classic Pearson's $\chi^2$ test can
be used to assess whether $p$ is a potential period or not.

For a particular trial period $p$, when the phases of all the bursts have been
determined, we group the bursts into $n$ bins in the phase space. The $\chi^2$ value is then calculated as:
\begin{equation}
\chi ^2=\sum_{j=1}^{n}\frac{(O_j-E_j)^2}{E_j},
\end{equation}
where $O_j$ is the observed count in the $j$-th bin, and $E_j$ is the expected count in each bin for a uniform phase distribution. Here the uniform phase distribution means all the bursts distribute uniformly in the whole phase space. Usually, a large $\chi^2$ value indicates that $p$ could potentially be the period of the burst activity, while a small $\chi^2$ value means that $p$ is unlikely the intrinsic period. Varying $p$ in all the possible period ranges, we would be able to find the true period
if it really exists.

The phase folding method is relatively insensitive to the randomly
occurring gaps in the observational data as long as the exposure
is reasonably uniform across the phase space. Note that the event
rate of FRBs is connected to the detection rate of bursts, i.e.
the number of bursts observed by a particular telescope during a
unit time period. This factor can be considered in the phase
folding procedure by implementing a more precise calculation of
$\chi^2$, in which $E_j$ should be substituted by $T_j\cdot \sum
O_j / \sum T_j$, where $T_j$ is the exposure time for the $j$-th
bin \citep{chime/frb2020Periodic}. For this purpose, we need to
know the telescope's observation time interval associated with
each burst. However, in our data sample, such information is not
always available for all the bursts. So, we simply calculate the
$\chi ^2$ in its original form in this study.

\subsection{H-test}

The H-test method is based on the $Z_m^2$-test, which computes the
sum of the Fourier power spectra of the first $m$ harmonics, and
is used to search for periodicity in a time series. The
$Z_m^2$-test is defined as \citep{1983A&A...128..245B}
\begin{equation}
Z_m^2 = \frac{2}{N} \sum_{k=1}^m \left( \left( \sum_{i=1}^N \cos(k \phi_i) \right)^2 + \left( \sum_{i=1}^N \sin(k \phi_i) \right)^2 \right),
\end{equation}
where $\phi_i$ is the phase of the $i$-th data point, and $N$ is
the total number of data points. When $m = 1$, it reduces to the
well-known Rayleigh test. Unlike the $\chi^2$ test, the
$Z_m^2$-test is bin-free. However, an additional smoothing
parameter $m$ is involved. The H-test offers a straightforward
method for determining $m$, which is to find the $m$ value that
maximizes $Z_m^2$ \citep{1989A&A...221..180D},
\begin{equation}
H = \max_{1 \leq m \leq m_{\max}} \left( Z_m^2 - 4m + 4 \right),
\end{equation}
where $m_{\max}$ is the maximum number of harmonics considered,
typically set as $m_{\max} = 20$. The significance of the H-test
is determined by comparing the observed $H$ value with the
distribution of $H$ obtained from a large number of simulations
under the null hypothesis (i.e., no periodic signal). The
probability distribution of H statistic can be obtained as
\citep{2010A&A...517L...9D}
\begin{equation}
Prob(> H) \approx \exp\left( -0.4 H \right).
\end{equation}
The H-test is more sensitive to periodic signals than the $\chi^2$ test, especially when the signal-to-noise ratio is low.

\subsection{Lomb-Scargle periodogram}

Based on the classic Schuster periodogram \citep{schuster1898investigation}, a more effective periodogram
was developed by \citet{lomb1976Leastsquares} and \citet{scargle1982studies},
which calculate the power as:
\begin{eqnarray}
\begin{split}
P_{\rm{LS}}(\omega) = \frac{1}{2}&\Bigg\{\dfrac{\big[\sum_i g_i \cos \omega (t_i-\tau)\big]^2}{\sum_i \cos^2\omega (t_i-\tau)}\\
&+ \dfrac{\big[\sum_i g_i \sin\omega (t_i-\tau)\big]^2}{\sum_i \sin^2\omega (t_i-\tau)}\Bigg\},
\end{split}
\end{eqnarray}
where $\omega$ is the angular frequency, $g_i$ is the observed signal value of the $i$-th data point, $t_i$ is the time of the $i$-th data point, and $\tau$ is specified for each $\omega$ to ensure the time-shift invariance:
\begin{equation}
\tau = \frac{1}{2\omega}\tan^{-1}\Bigg(
\frac{\sum_i \sin2\omega t_i}{\sum_i \cos2\omega t_i}\Bigg),
\end{equation}
where $i$ ranges from 1 to $N$ in the summation. $N$ is the total number of data points.

A pronounced peak in the Lomb-Scargle periodogram at a certain frequency usually indicates the presence of a potential periodicity at that frequency. In our study, we use the \texttt{LombScargle} \footnote{\url{https://docs.astropy.org/en/stable/timeseries/lombscargle.html}} class provided by \texttt{astropy} to compute the periodogram. False Alarm Probability (FAP) are used to estimate the significance of the peaks in the periodogram. FAP represents the probability that a peak in the power spectrum reaches or exceeds the current level due to random noise under the null hypothesis that no true period signal is present, as given by \citet{vanderplas2018Understanding}
\begin{equation}
\text{FAP}(Z) \approx 1 - \left[ 1 - \exp{(-Z)} \right] ^{(N-1)/2} ,
\end{equation}
where $Z$ represents the normalized periodogram power value. We
use the \texttt{false\_alarm\_level} code of the
\texttt{LombScargle} class to compute the power threshold for a
given FAP, which employs extreme value statistics to compute an
upper bound of the false alarm probability for the alias-free case
\citep{2008MNRAS.385.1279B}.

Compared to the traditional Schuster periodogram, the Lomb-Scargle
periodogram is more effective in analyzing time-series data that
have non-uniformly distributed sampling intervals. In reality,
especially in FRB observations, data collection may be affected by
complicated factors such as observation conditions and instrument
limitations, resulting in unevenly spaced intervals. The
Lomb-Scargle method is more suitable in these cases.

\subsection{Strategy for period searches}

In the process of periodicity analysis, it is crucial to set the
target period range and the time step appropriately
\citep{2013ApJ...779L..11P}. In our study, we aim to search for
short timescale periodicities in the four active repeating FRBs.
To ensure search accuracy with affordable computational costs, it
is necessary to limit the time span of the dataset being used.
To ensure accuracy, we will only conduct single-day searching. Hundreds of bursts can be
detected from active repeaters within a single day, making the
periodicity searches within a single day feasible. The advantage
of performing such a single-day period search is that it can
effectively eliminate possible systematic timing errors on
different days. For example, if 100 bursts were detected by FAST
in a two-hour monitoring, then the time intervals between every
two successive FRBs will be recorded with an extremely high
accuracy. The relative arrival times of these 100 bursts will not
be subject to any serious systematic errors when they are
calibrated to the barycenter of the solar system, thus they can
effectively reserve their intrinsic period information. Another
advantage of single-day period searches is the avoidance of errors
caused by variations in the source's DM.


To further reduce computation costs, we divide the period
range into three segments: 0.001--0.1 s, 0.1--10 s and 10--1000 s. For the 0.001--0.1 s period range, we use a time
resolution of $10^{-6}$ s, which reaches the highest precision of
the recorded ToAs in our collected datasets. For the 0.1--10 s
period range, we use a time resolution of $10^{-4}$ s. For the 10--1000 s range, we use a time resolution of $10^{-2}$ s.

We have tested the effectiveness of our period search
methods and strategies using simulation data. For this purpose, we
generated various simulated datasets that include the arrival
times of mock FRBs. In the datasets, different numbers of FRBs are
generated, in different observation time spans, with different
pre-assumed intrinsic periodicity. Observational truncation
effects during the data acquisition are also considered, and
noises (i.e., random offsets) are injected into the arrival times.
We then applied our search methods and strategies to the simulated
datasets. In all the cases, the pre-assumed periodicity can be
correctly recovered. Especially, in the 0.001--0.1 s segment,
our analysis is effective as long as the burst rate exceeds $\sim$
100 hr$^{-1}$ and the arrival time fluctuation is less than 100
$\mu$s.

It is also worth mentioning that the waiting time of repeating
FRBs usually exhibits a log-normal bimodal distribution, which is
an important temporal characteristic of repeating FRBs. Typically,
the left peak of the waiting time distribution is on the order of
tens of milliseconds. Such closely spaced bursts can interfere
with the search for period larger than 0.1 s. Therefore, in our
searches for periods larger than 0.1 s, we exclude the bursts with
a waiting time less than 0.1 s. For example, if Burst 1 and Burst
2 have a waiting time of 0.05 s, we retain Burst 1 and discard
Burst 2. When we search for periods less than 0.1 s, we do
not exclude any bursts. However, it should be noted that the
intrinsic pulse widthes of FRBs are typically $\sim 10$ ms, and
many FRBs have multi-peak structures, which lead to substantial
uncertainties in determining the exact arrival time. As a result,
searching for periods shorter than 0.1 s is quite difficult.

\section{Periodicity results for the four FRBs}
\label{sect:analysis1}

In this section, we present our results of period searching for
the four active repeating FRBs, namely FRBs 20121102A, 20200120E,
20201124A, and 20220912A. All the three methods described above
are applied to each repeater. In our study, we concentrate
on short-time periodicity, with the period ranging in 0.001--1000 s. As mentioned above, to improve the computing efficiency,
the whole period range is divided into three segments, i.e. 0.001--0.1 s, 0.1--10 s and 10--1000 s, and the searching time
step is taken as $10^{-6}$ s, $10^{-4}$ s and $10^{-2}$ s,
respectively. Such a strategy can meet the requirement of the
expected accuracy and avoid unnecessary computational burden. To
avoid interference from closely spaced bursts, we excluded bursts
with waiting times less than 0.1 s in searches for periods larger
than 0.1 s. As previously mentioned, to ensure the reliability of
the results, we need to limit the time span of the bursts used in
our searches. To balance the accuracy and efficiency, we perform
our analysis on single day bursts, which typically arrive within a
time span of less than 2 hours.

\subsection{FRB 20121102A}

The ToAs of the bursts from FRB 20121102A, obtained by Arecibo (Datasets \#1 and \#2) and FAST
(Dataset \#3) and , are used to
search for potential short-time periodicity. Here, the ToA
corresponds to the peak of each burst, calibrated to the
barycenter of the solar system. The ToAs of Arecibo datasets \#1 and \#2
are barycentre corrected and de-dispersed to infinite frequency,
and the timing precision is $\sim 10^{-9}$d (86.4 $\mathrm{\mu
s}$) and $\sim 10^{-10}$ d (8.64 $\mathrm{\mu s}$), respectively. For the FAST Dataset \#3, the ToAs
are de-dispersed to 1.5 GHz, and the timing precision (estimated
from the last digit of the recorded ToAs) is $\sim 10^{-9}$ d
(86.4 $\mathrm{\mu s}$).

For the Arecibo Dataset \#1, we selected the bursts from
the two days with the highest number of bursts detected (56 bursts
on MJD 57644, of which 48 bursts have a waiting time > 0.1 s; and
76 bursts on MJD 57645, of which 69 bursts have a waiting time >
0.1 s) to perform period searches. For the Arecibo Dataset \#2,
bursts from the four days with the highest number of bursts
detected are selected (203 bursts on MJD 58432, of which 179
bursts have a waiting time > 0.1 s; 180 bursts on MJD 58435, of
which 160 bursts have a waiting time > 0.1 s; 227 bursts on MJD
58439, of which 218 bursts have a waiting time > 0.1 s; and 101
bursts on MJD 58450, of which 96 bursts have a waiting time > 0.1
s). For the FAST \#3 dataset, bursts from the four days with the
highest number of bursts detected are used (121 bursts on MJD
58726, of which 99 bursts have a waiting time > 0.1 s; 110 bursts
on MJD 58727, of which 86 bursts have a waiting time
> 0.1 s; 106 bursts on MJD 58733, of which 94 bursts have a waiting
time > 0.1 s; and 117 bursts on MJD 58757, of which 78 bursts have
a waiting time > 0.1 s).

\begin{figure*}
   \centering
   \includegraphics[width=0.32\textwidth]{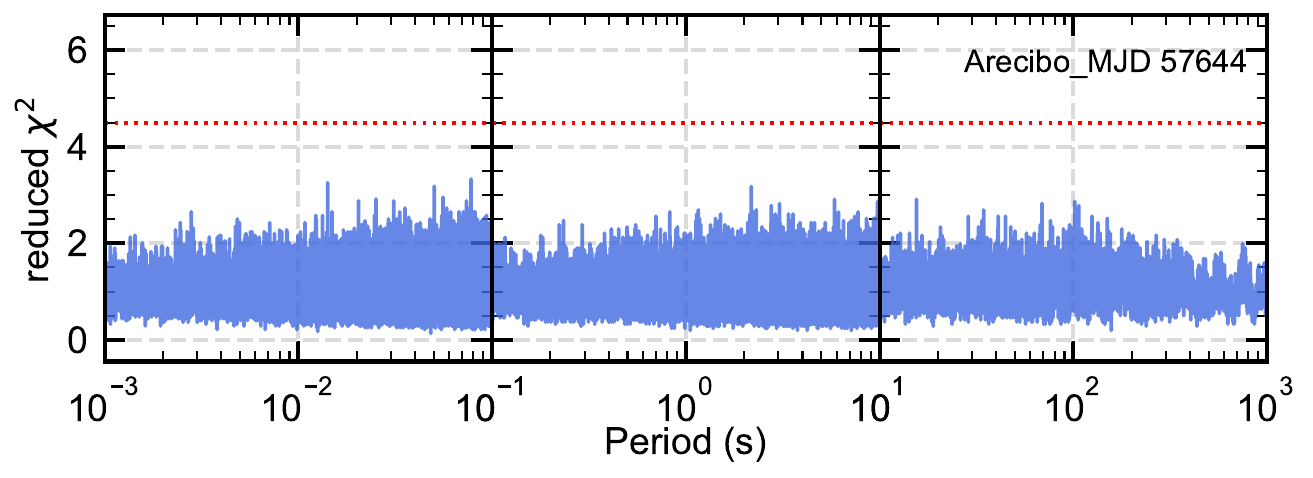}
   \includegraphics[width=0.32\textwidth]{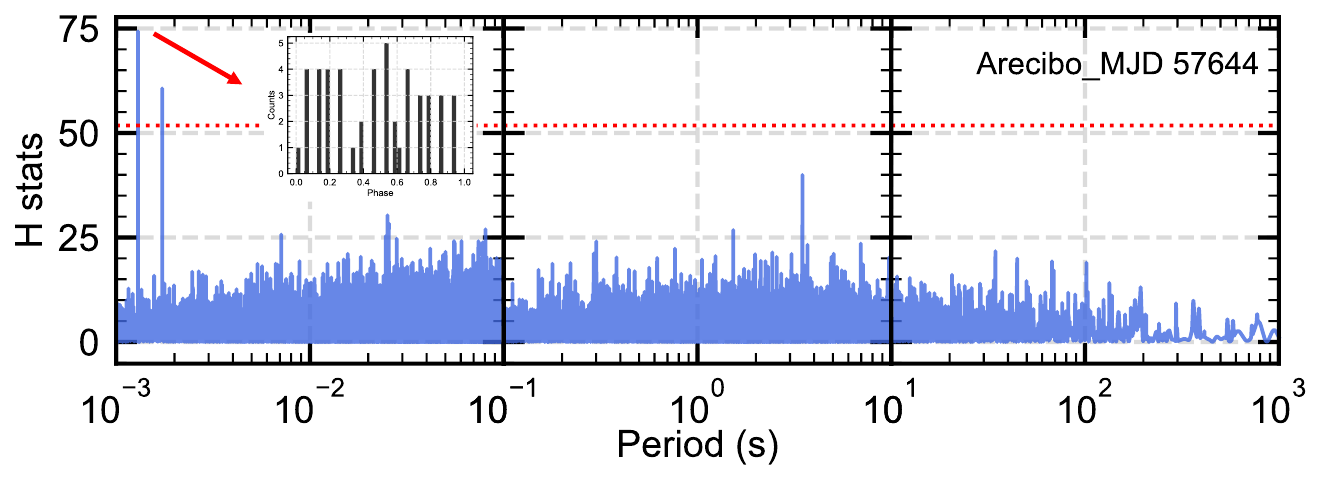}
   \includegraphics[width=0.32\textwidth]{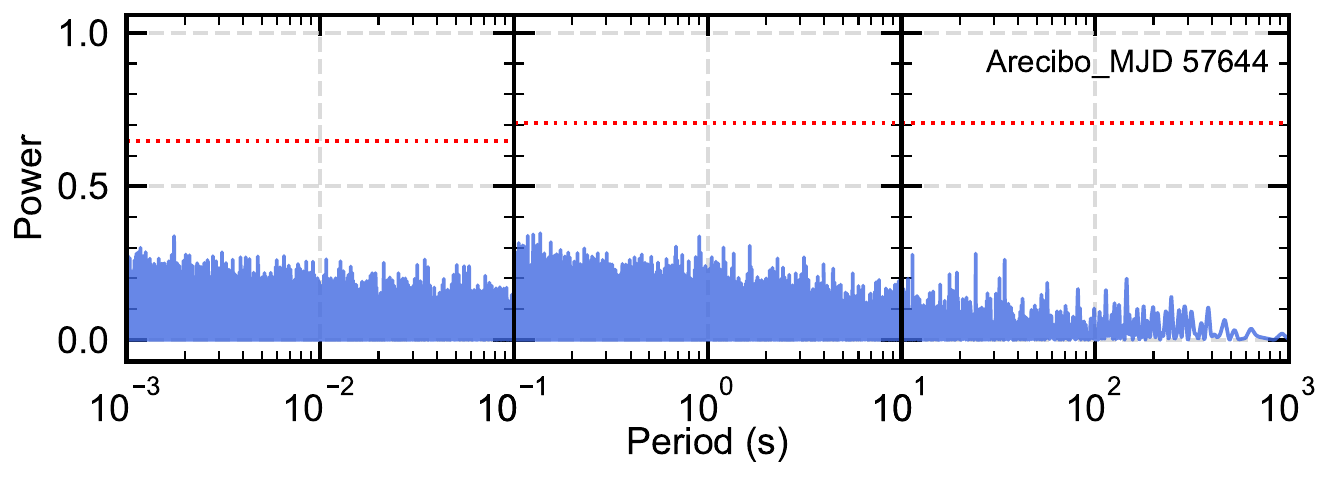}
   \includegraphics[width=0.32\textwidth]{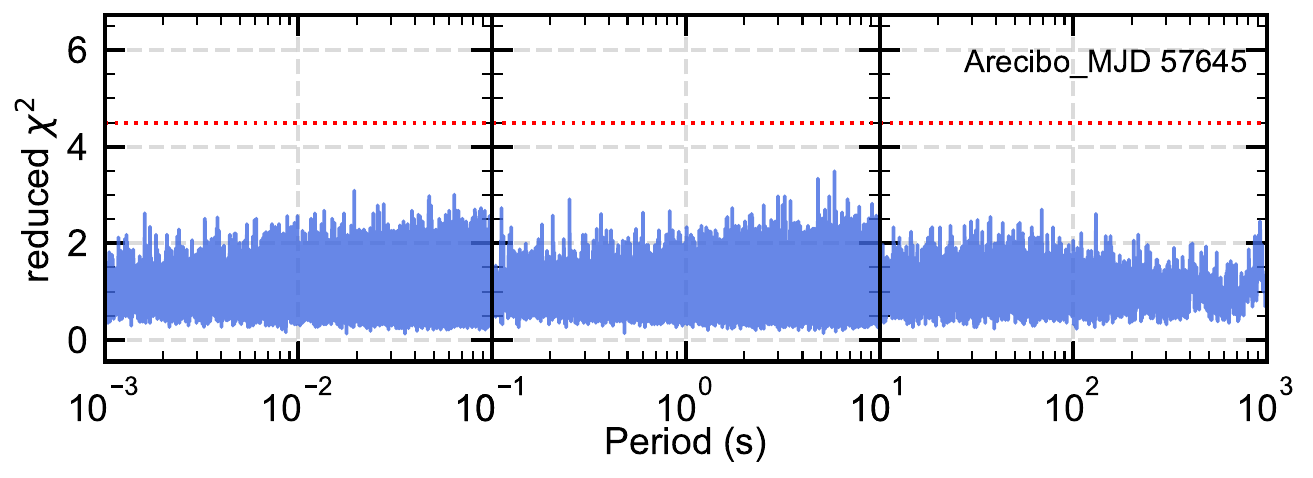}
   \includegraphics[width=0.32\textwidth]{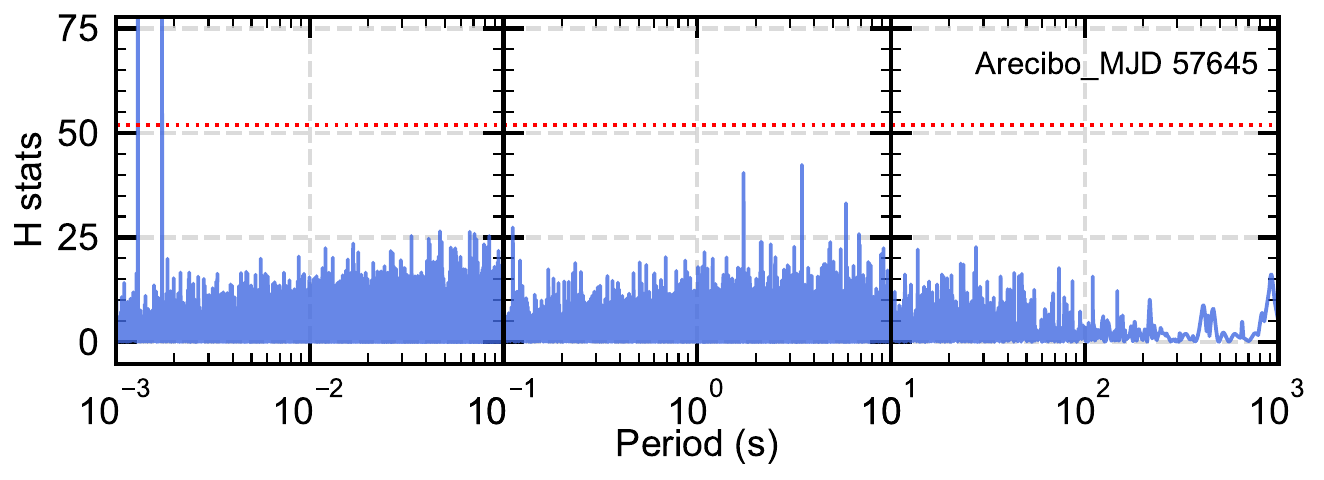}
   \includegraphics[width=0.32\textwidth]{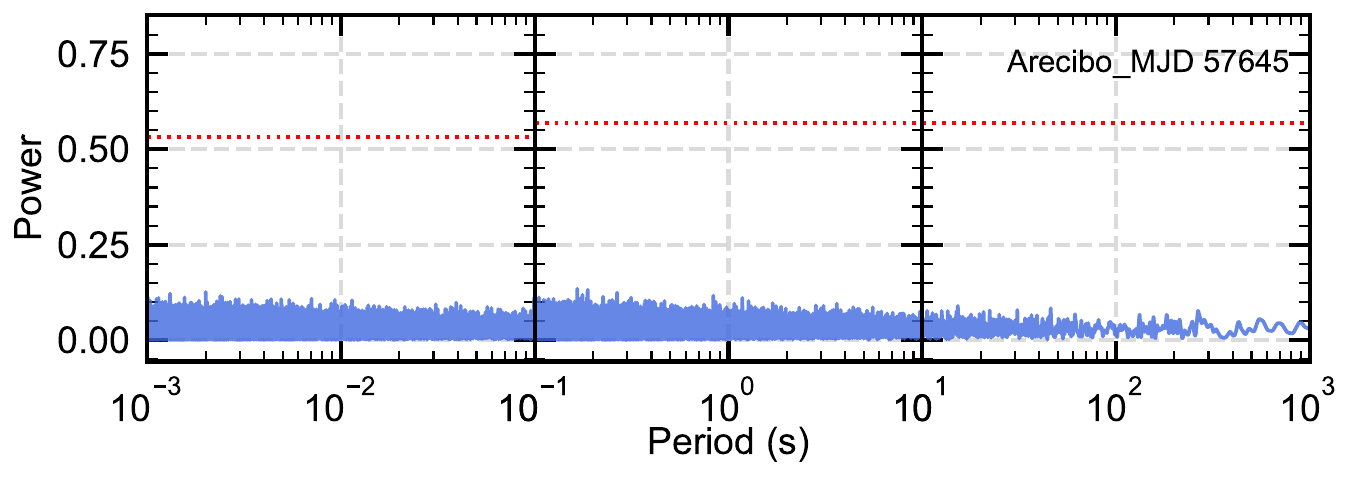}
   \caption{\footnotesize The period search results based on the
   Arecibo Dataset \#1 of FRB 20121102A. The three columns from left to right
   correspond to the three search methods, i.e. the phase folding algorithm,
   the H-test, and the Lomb-Scargle periodogram, respectively. The two rows
   from top to bottom correspond to the two selected days (MJD 57644 and MJD 57645).
   In the cases of phase folding and the H-test methods, a horizontal dotted line is plotted to
   mark the p-value of $10^{-9}$. In the cases of Lomb-Scargle periodogram,
   a horizontal dotted line indicating a FAP level of $10^{-9}$ is plotted.
   No clear evidence of periodicity is found in these plots. Note that in the two middle panels,
   the peak structures in the 1 ms--2 ms
   range are fake signals due to the limited timing accuracy. The inset in the upper-middle panel
   shows the folded histogram at one of the peaks (0.001296 s = 1.5 $\times 10^{-8}$ d).}
   \label{Fig2}
\end{figure*}

\begin{figure*}
   \centering
   \includegraphics[width=0.32\textwidth]{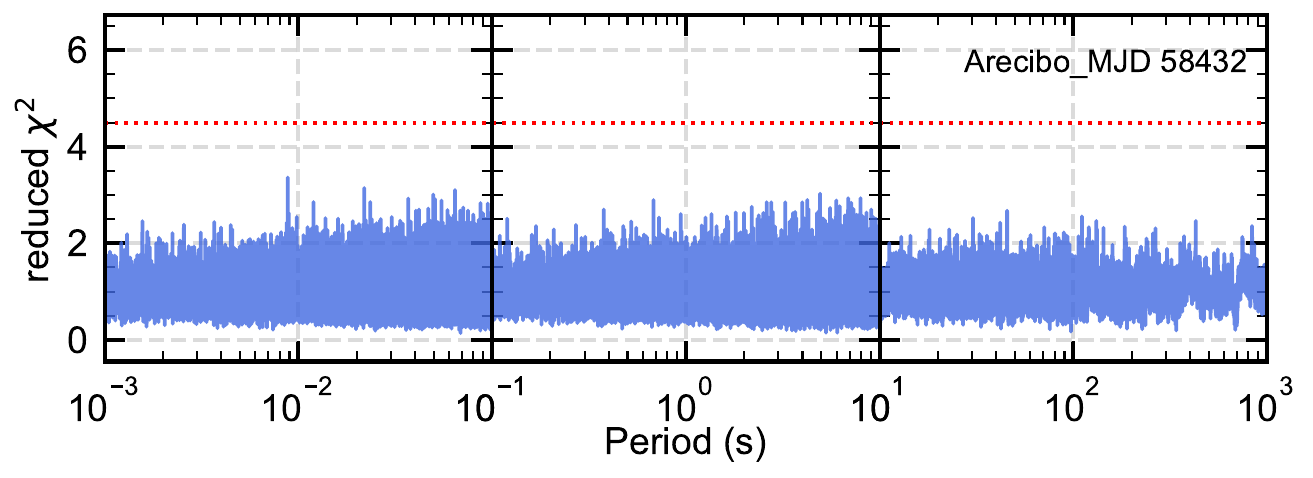}
   \includegraphics[width=0.32\textwidth]{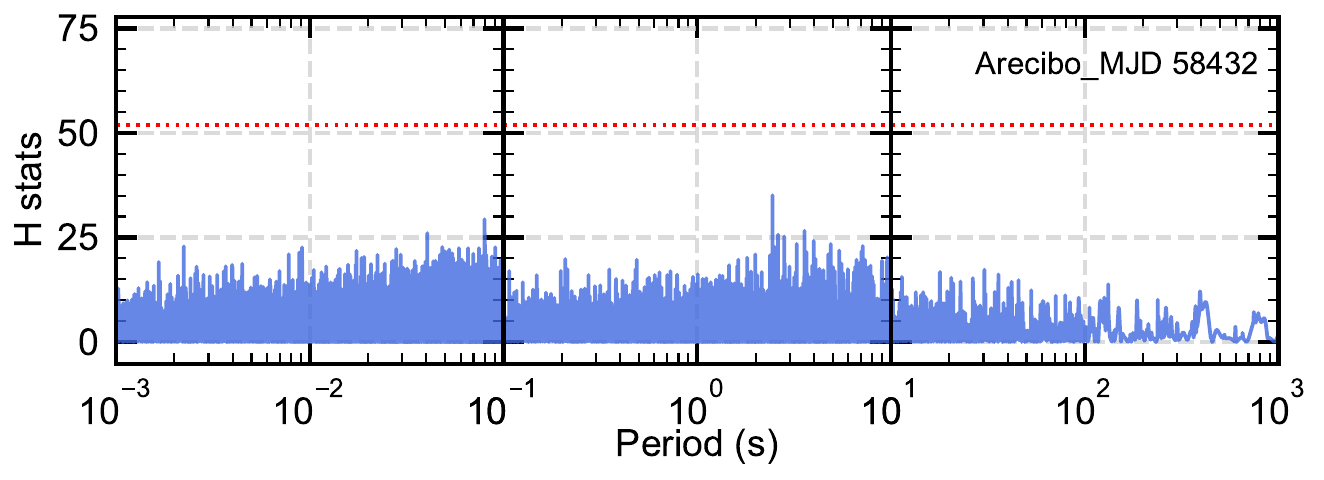}
   \includegraphics[width=0.32\textwidth]{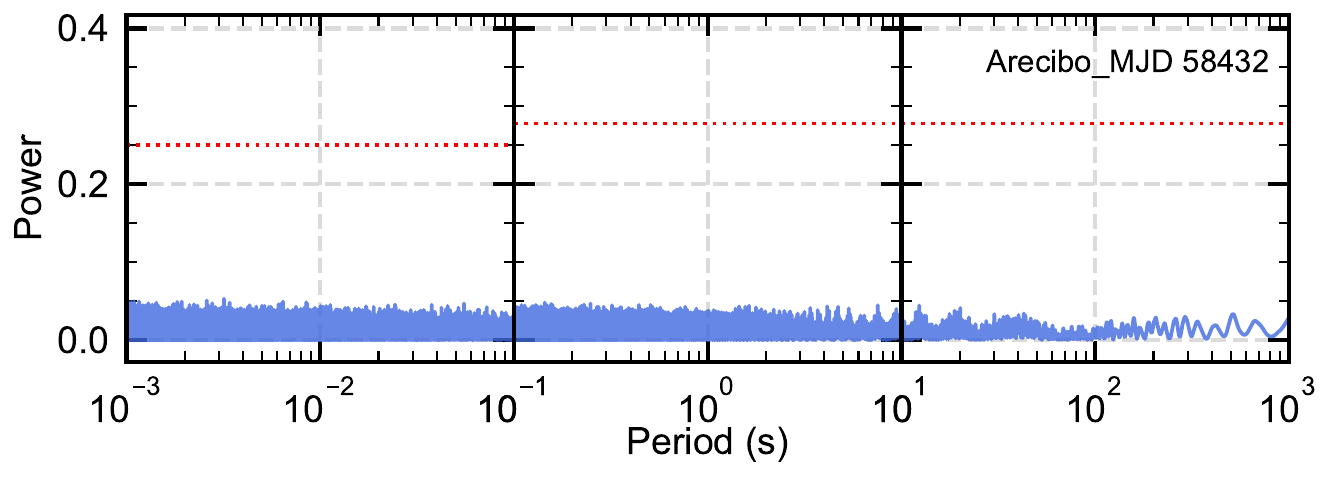}
   \includegraphics[width=0.32\textwidth]{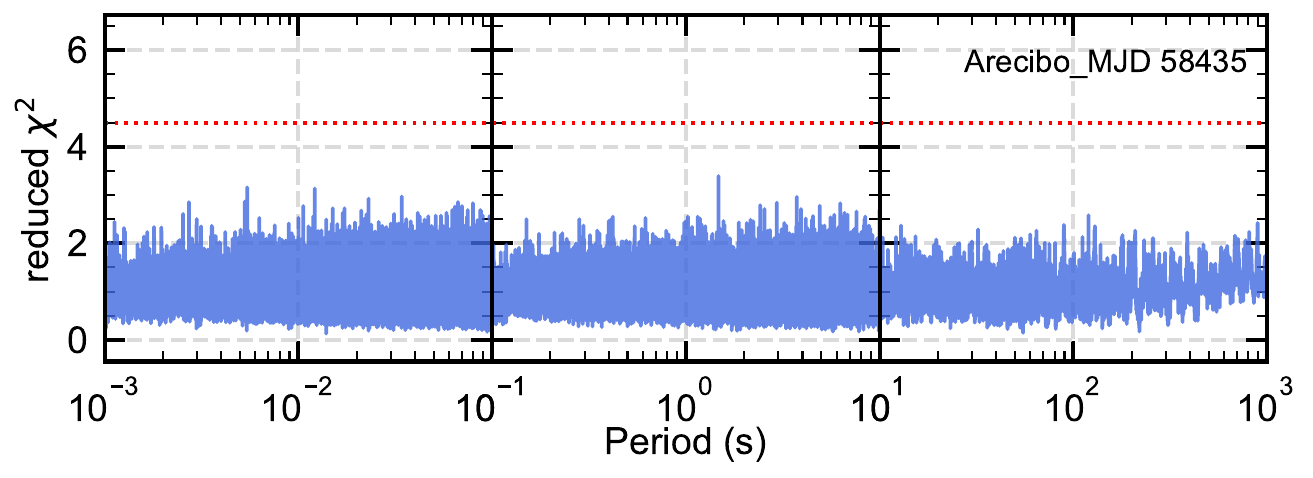}
   \includegraphics[width=0.32\textwidth]{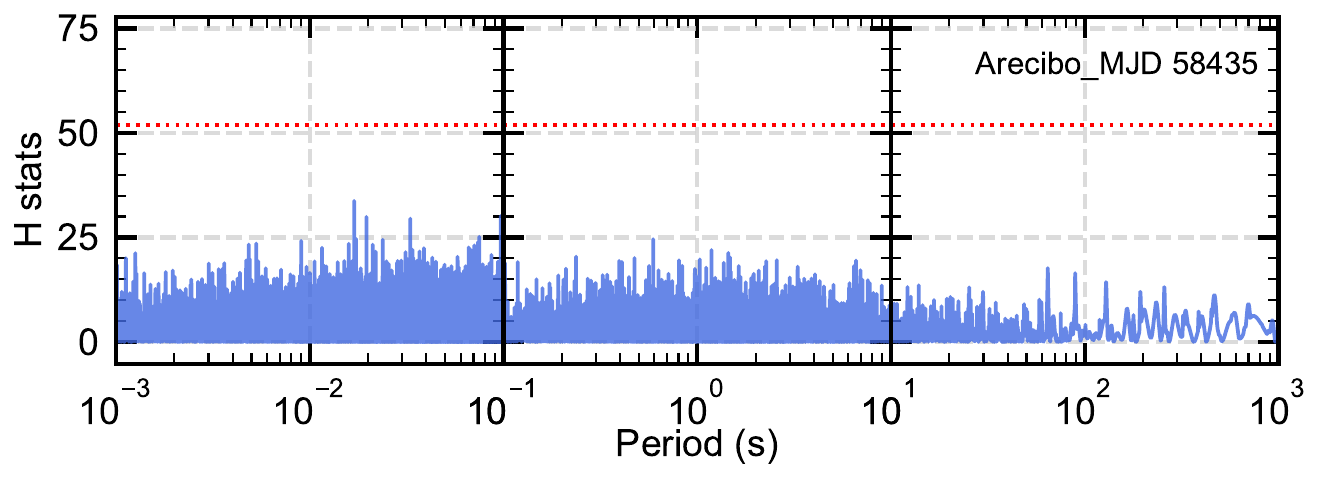}
   \includegraphics[width=0.32\textwidth]{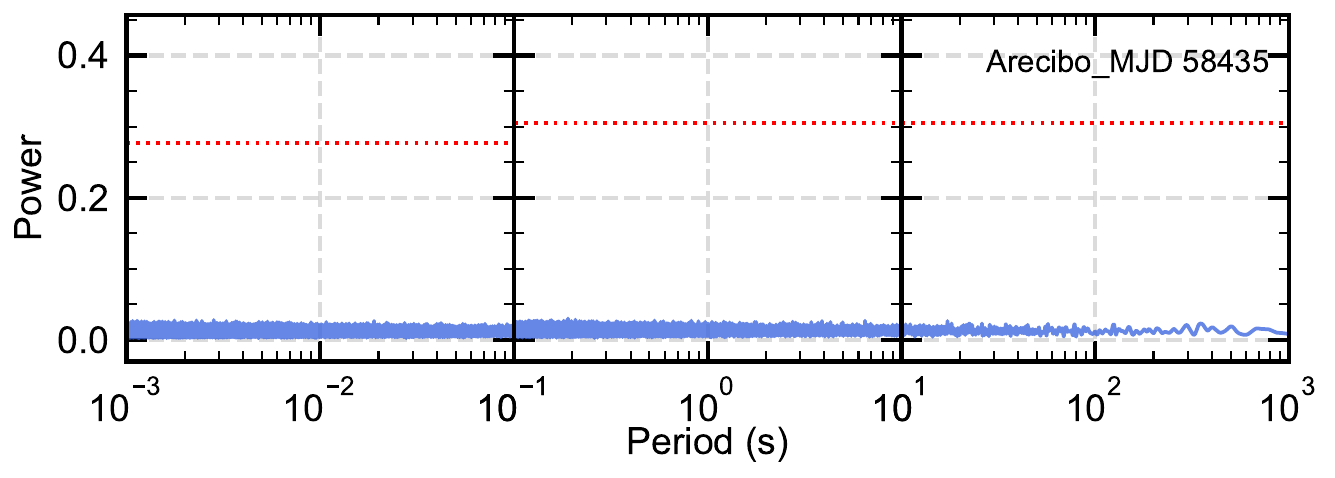}
   \caption{\footnotesize The period search results based on the
   Arecibo Dataset \#2 of FRB 20121102A. The three columns from left to right
   correspond to the three search methods, i.e. the phase folding algorithm,
   the H-test, and the Lomb-Scargle periodogram, respectively. The two rows
   from top to bottom correspond to the two selected days (MJD 58432 and MJD 58435).
   In the cases of phase folding and the H-test methods, a horizontal dotted line is plotted to
   mark the p-value of $10^{-9}$. In the cases of Lomb-Scargle periodogram,
   a horizontal dotted line indicating a FAP level of $10^{-9}$ is plotted.
   No clear evidence of periodicity is found in these plots.}
   \label{Fig3}
\end{figure*}

\begin{figure*}
   \centering
   \includegraphics[width=0.32\textwidth]{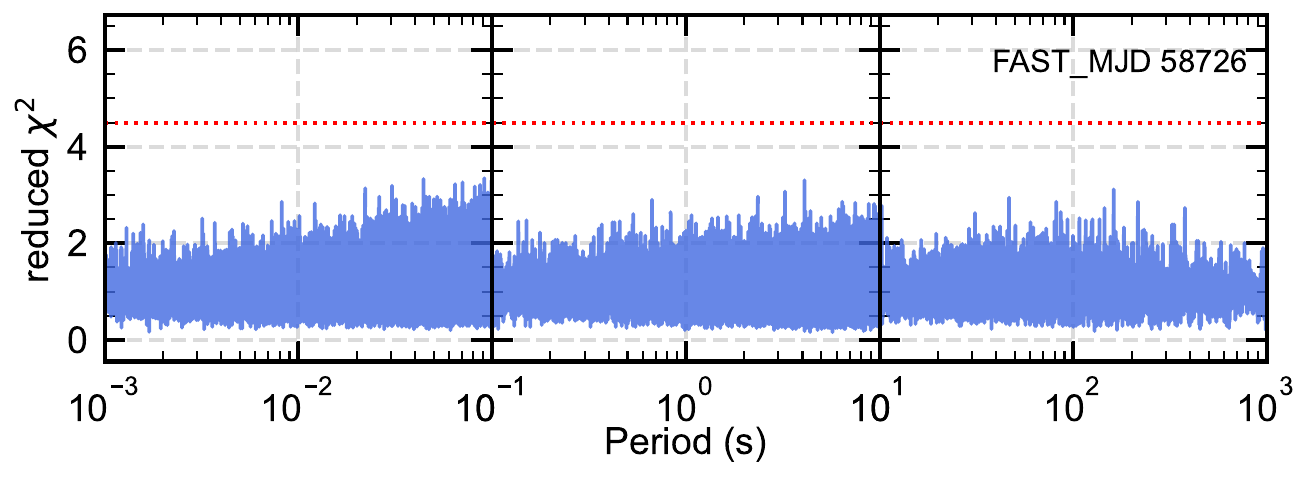}
   \includegraphics[width=0.32\textwidth]{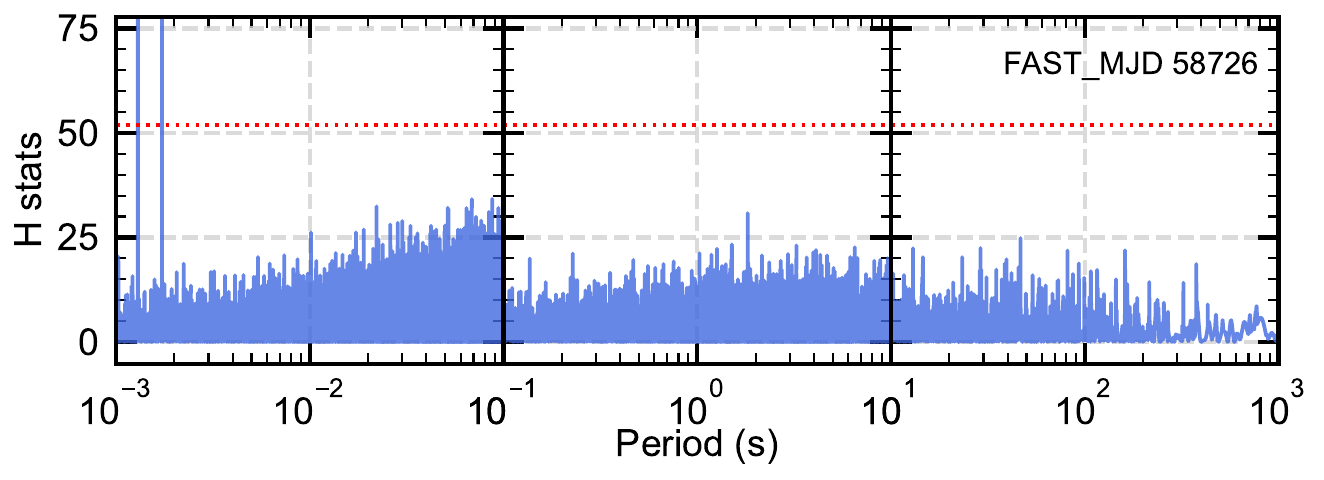}
   \includegraphics[width=0.32\textwidth]{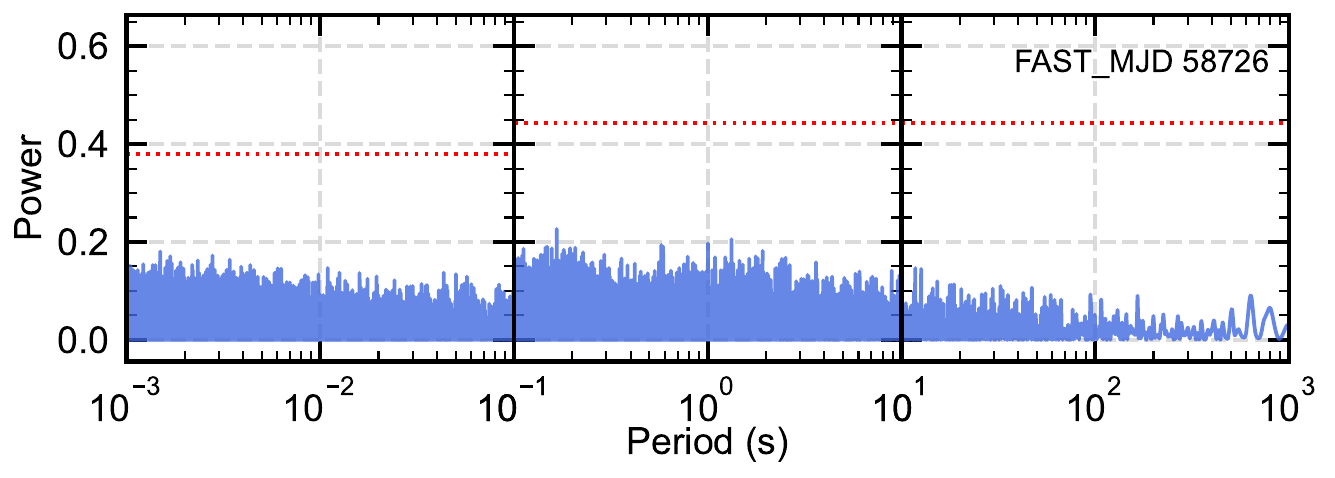}
   \includegraphics[width=0.32\textwidth]{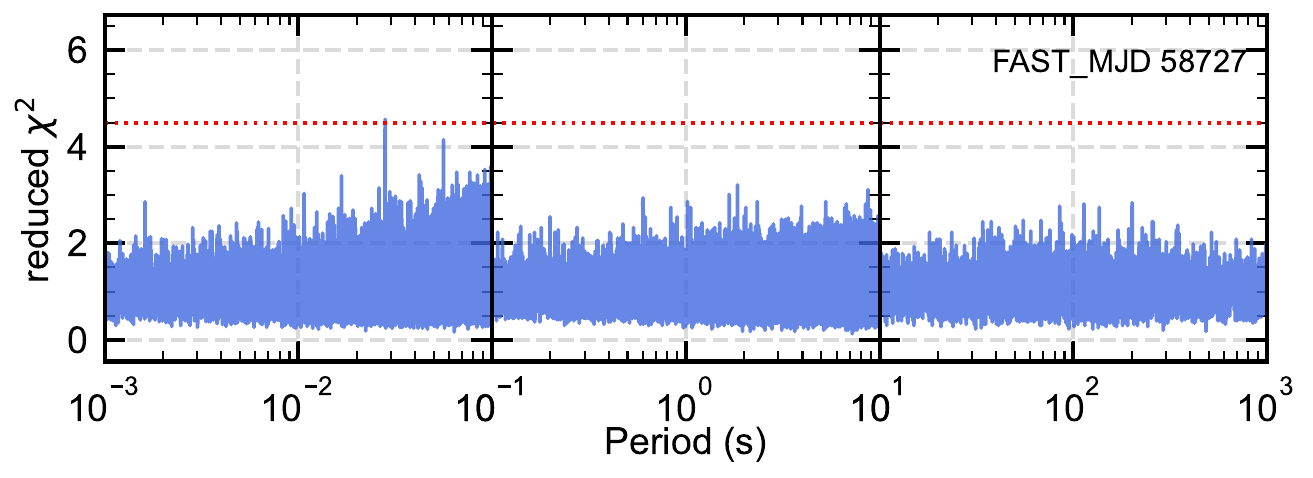}
   \includegraphics[width=0.32\textwidth]{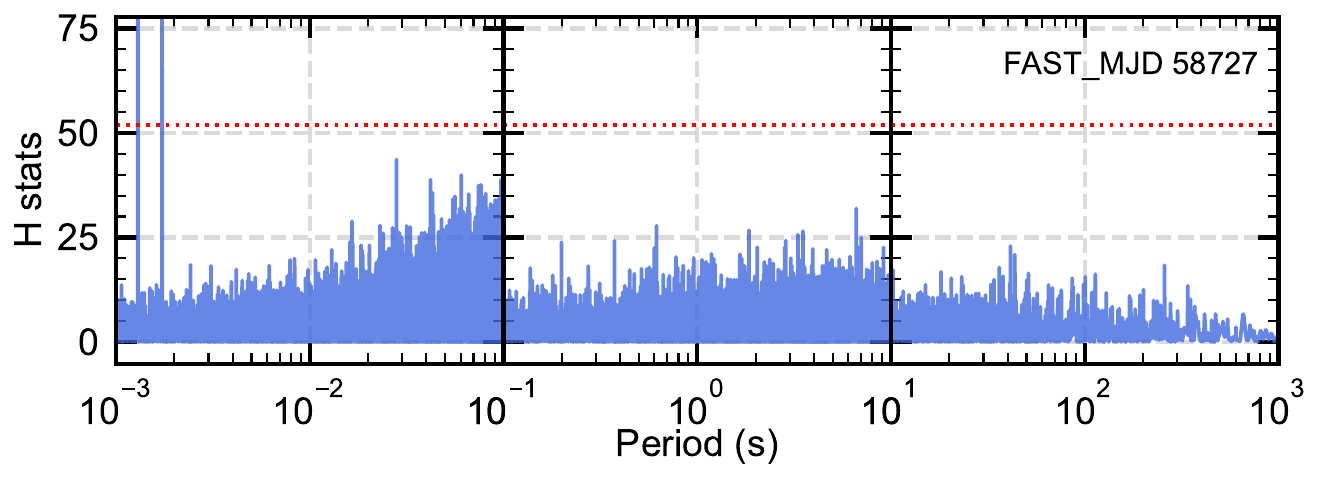}
   \includegraphics[width=0.32\textwidth]{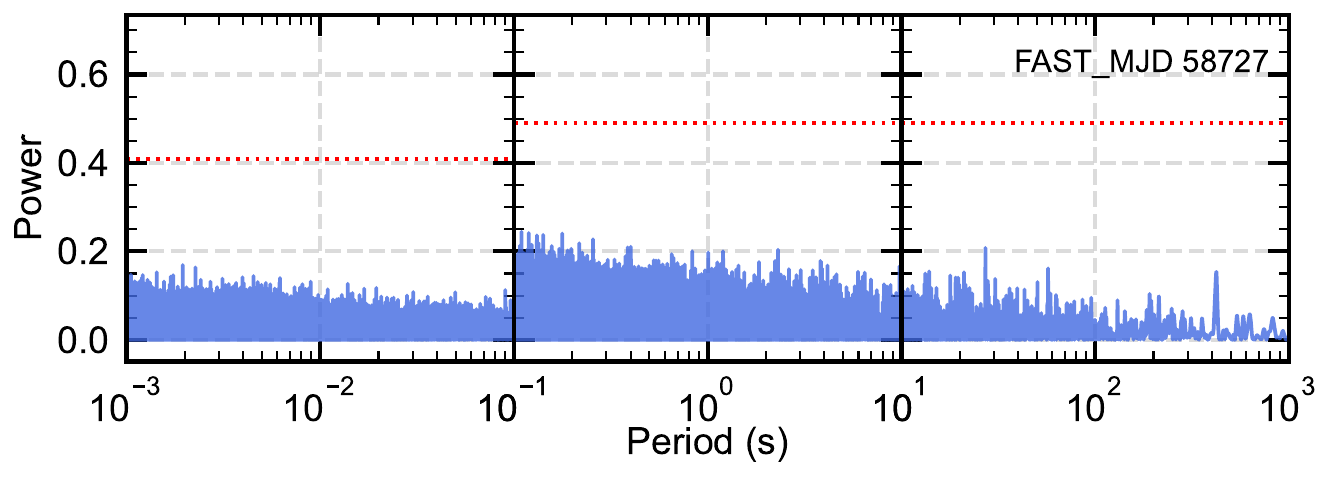}
   \includegraphics[width=0.32\textwidth]{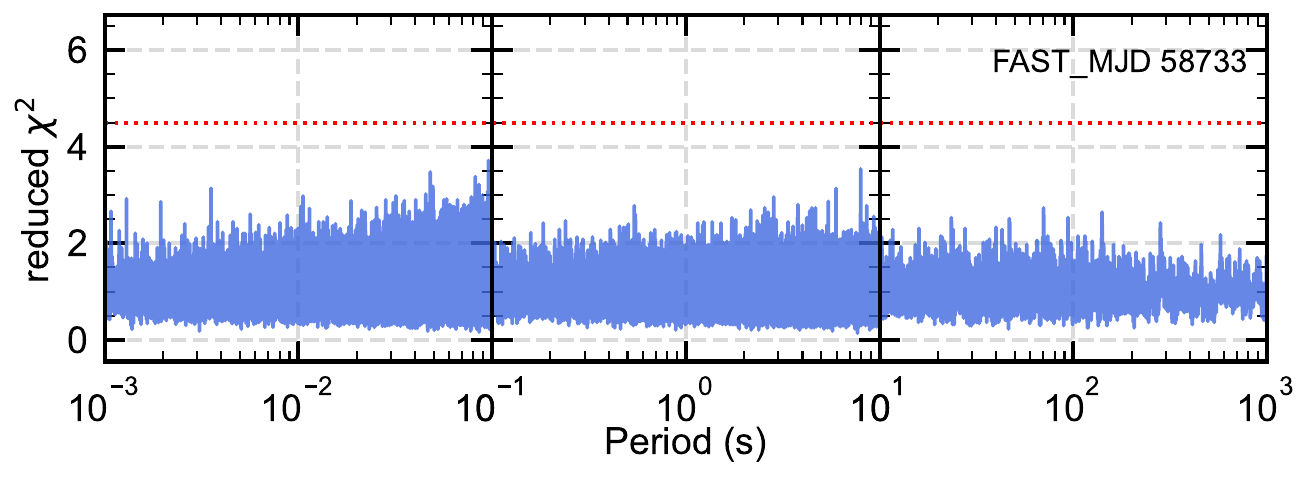}
   \includegraphics[width=0.32\textwidth]{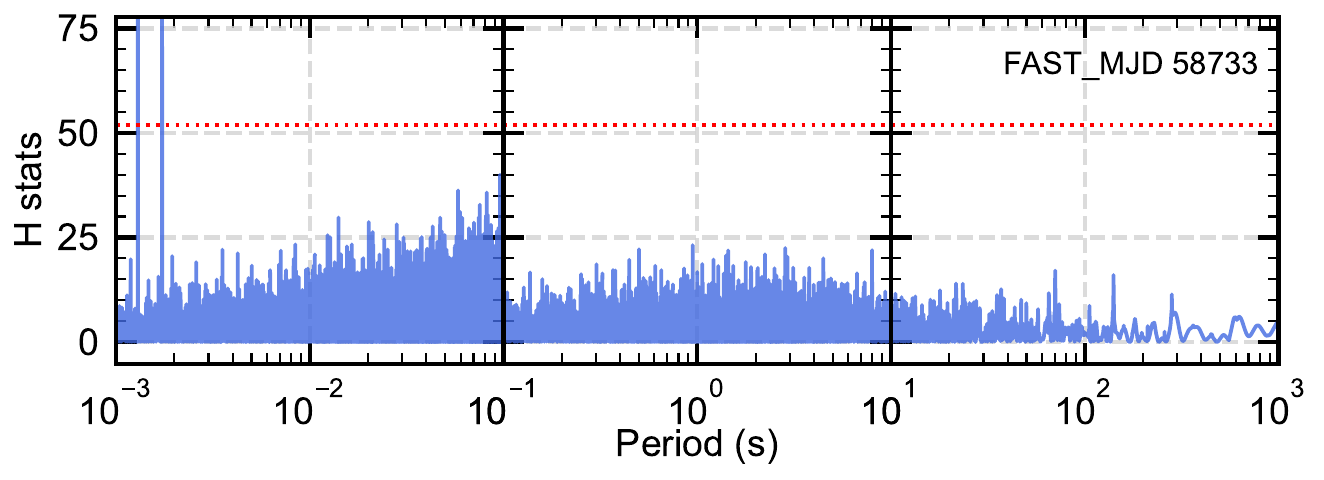}
   \includegraphics[width=0.32\textwidth]{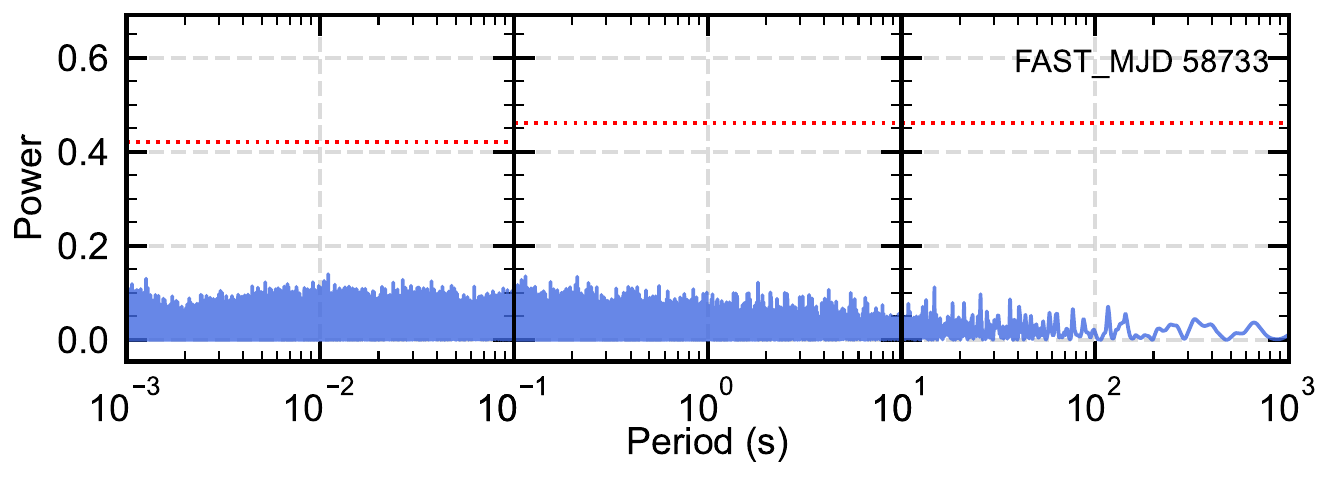}
   \includegraphics[width=0.32\textwidth]{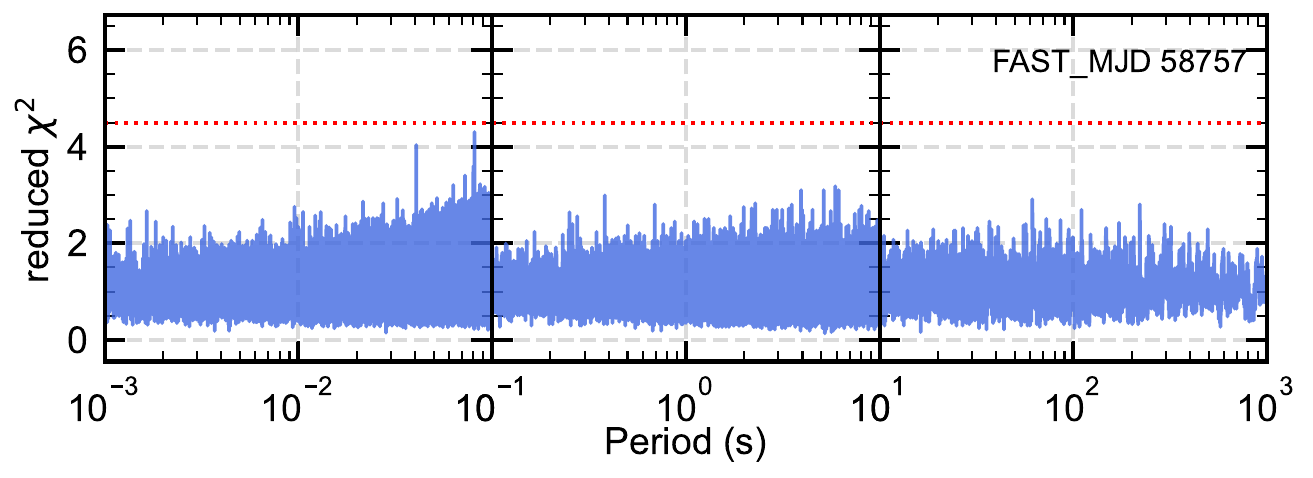}
   \includegraphics[width=0.32\textwidth]{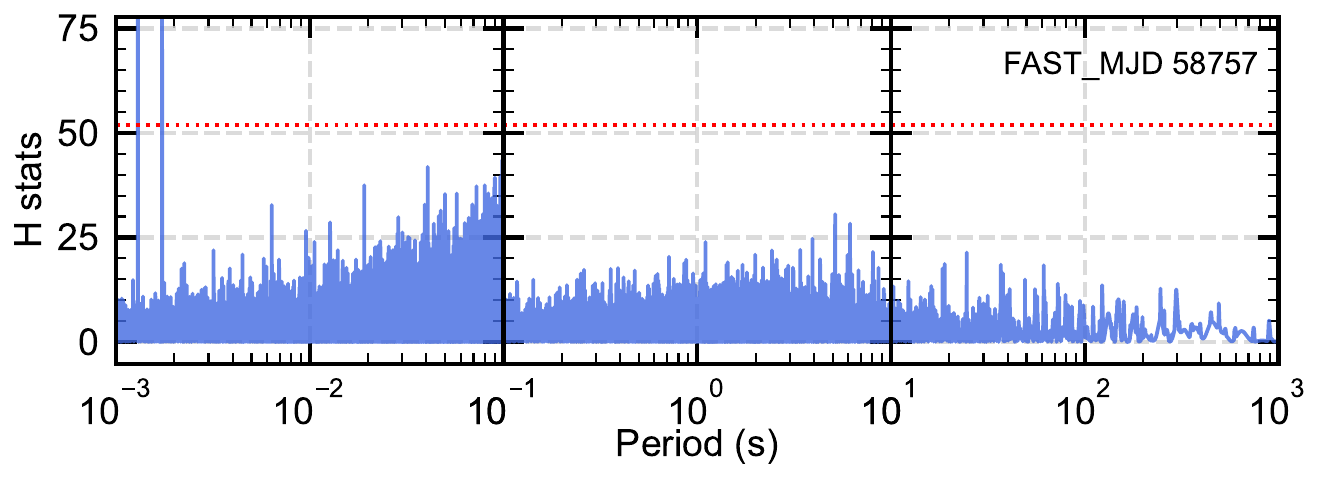}
   \includegraphics[width=0.32\textwidth]{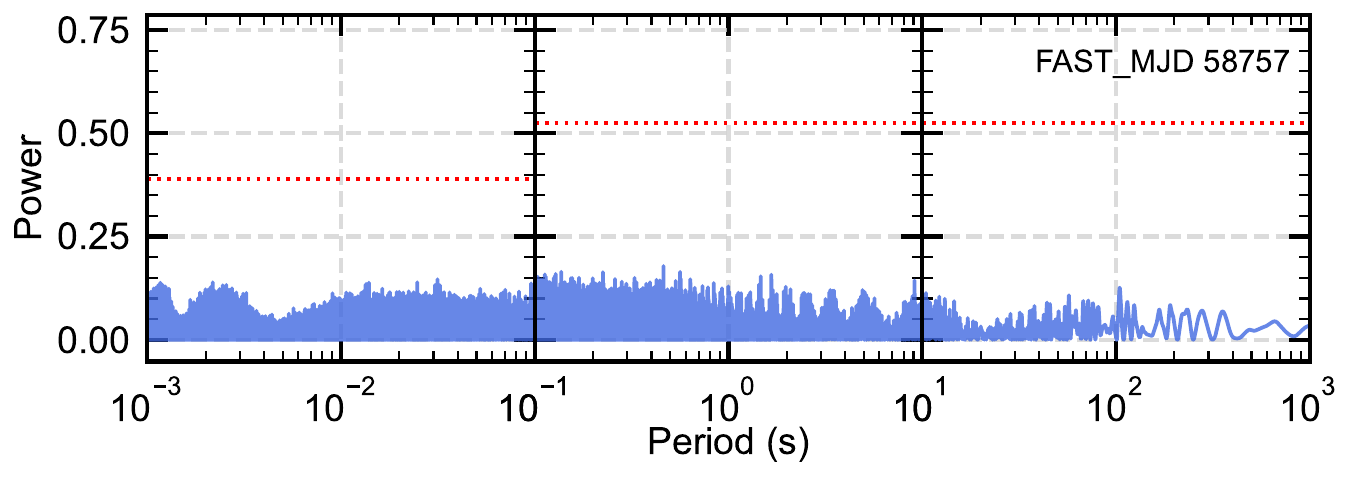}
   \caption{\footnotesize The period search results based on the
   FAST Dataset \#3 of FRB 20121102A. The three columns from left to right
   correspond to the three search methods, i.e. the phase folding algorithm,
   the H-test, and the Lomb-Scargle periodogram, respectively. The four
   rows from top to bottom correspond to the four selected days
   (MJD 58726, MJD 58727, MJD 58733 and MJD 58757). In the cases of phase
   folding and the H-test methods, a horizontal dotted line is plotted to
   mark the p-value of $10^{-9}$. In the cases of Lomb-Scargle periodogram,
   a horizontal dotted line indicating a FAP level of $10^{-9}$ is plotted.
   No clear evidence of periodicity is found in these plots. Note that in the middle panels, the peak structures in the 1 ms--2 ms
   range are fake signals due to the limited timing accuracy.}
   \label{Fig4}
\end{figure*}

We have used all the three methods mentioned above to analyze the
periodicity. In the phase folding algorithm, the number of bins is
taken as $n = 20$. In the H-test method, the maximum number of
harmonics is set as $m_{\max} = 20$. In the method of the
Lomb-Scargle periodogram, we use the peak flux of each burst as a
necessary input for the intensity. These specific parameters are
consistently applied in all searches. The results from Datasets
\#1, \#2 and \#3 are shown in Figures ~\ref{Fig2}, ~\ref{Fig3},
and ~\ref{Fig4}, respectively. For the phase folding method and
the H-test method, a horizontal dotted line is plotted to mark the
p-value of $10^{-9}$ (corresponding to a gaussian significance of
6.11$\sigma$). For the Lomb-Scargle periodogram, a horizontal
dotted line showing a FAP level of $10^{-9}$ is also plotted. Note that in some plots in the 1 ms--2 ms range, several
extremely high peaks could be observed. The inset in Figure
~\ref{Fig2} shows the folded histogram at one of the peaks. Upon
further examination, we notice that they are all harmonics of
0.864 ms ($10^{-8}$ d). They should be fake signals caused by
limited timing accuracy of observations. For other datasets with a
higher ToA recording precision, these fake signals disappear. In
each plot, we have also examined by hand the phase-folded
histograms at the positions that have a relatively high peak in
the $\chi^2$/H-statistic/Lomb-Scargle periodograms, but found that
no marked features appear in the corresponding histograms.
Additionally, a peak occurring on one day usually does not appear
on other days, which further excludes its possibility of being a
true period. From Figures ~\ref{Fig2}, ~\ref{Fig3}, and
~\ref{Fig4}, we conclude that no credible periodicity exceeding
the confidence level is found in the whole range of 0.001--1000
s for FRB 20121102A.

\subsection{FRB 20200120E}

The ToAs of the bursts from FRB 20200120E, obtained by Effelsberg
(Dataset \#7), are used to search for potential short-time
periodicity. Here, the ToA corresponds to the peak of each burst,
calibrated to the barycenter of the solar system. The ToAs are
de-dispersed to infinite frequency, and the timing precision
(estimated from the last digit of the recorded ToAs) is $\sim
10^{-8}$ d (864 $\mathrm{\mu s}$).

We selected the bursts from the day with the highest
number of bursts detected (53 bursts on MJD 59594, of which all
bursts have a waiting time > 0.1 s) to perform period searches.

\begin{figure*}
   \centering
   \includegraphics[width=0.32\textwidth]{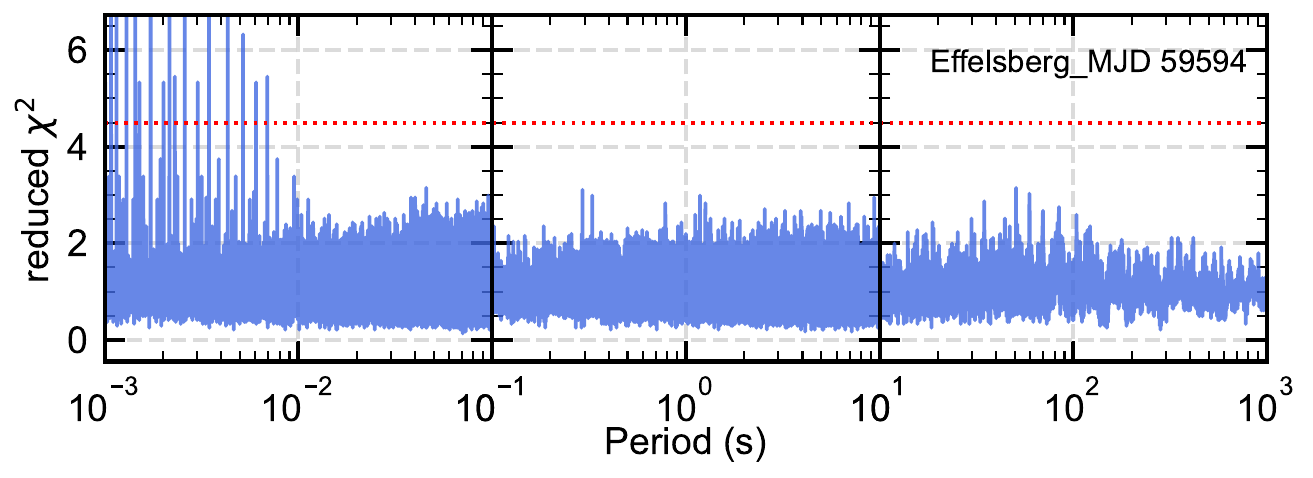}
   \includegraphics[width=0.32\textwidth]{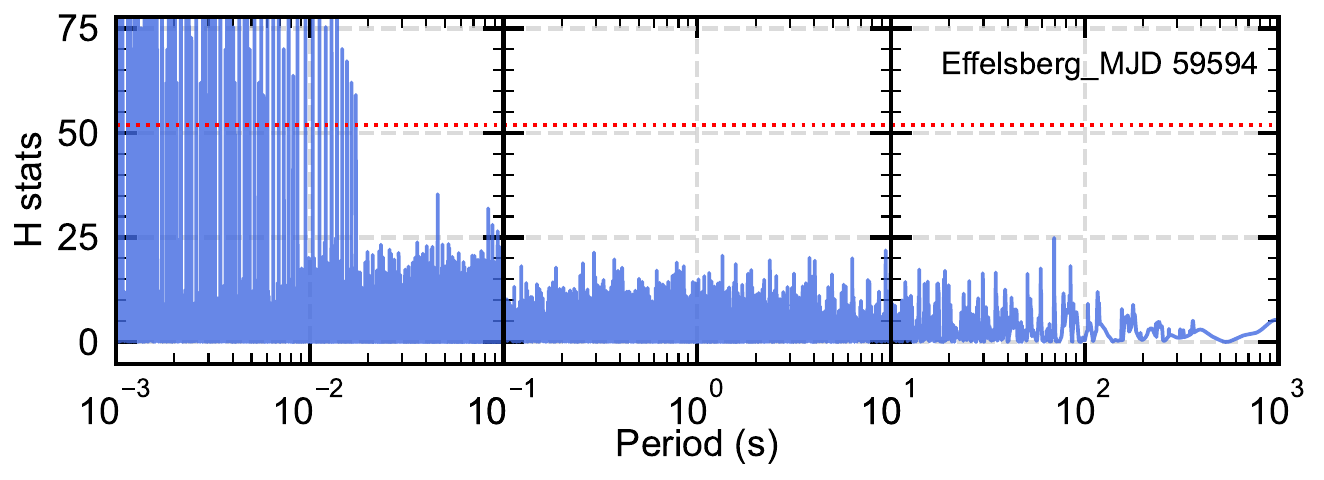}
   \includegraphics[width=0.32\textwidth]{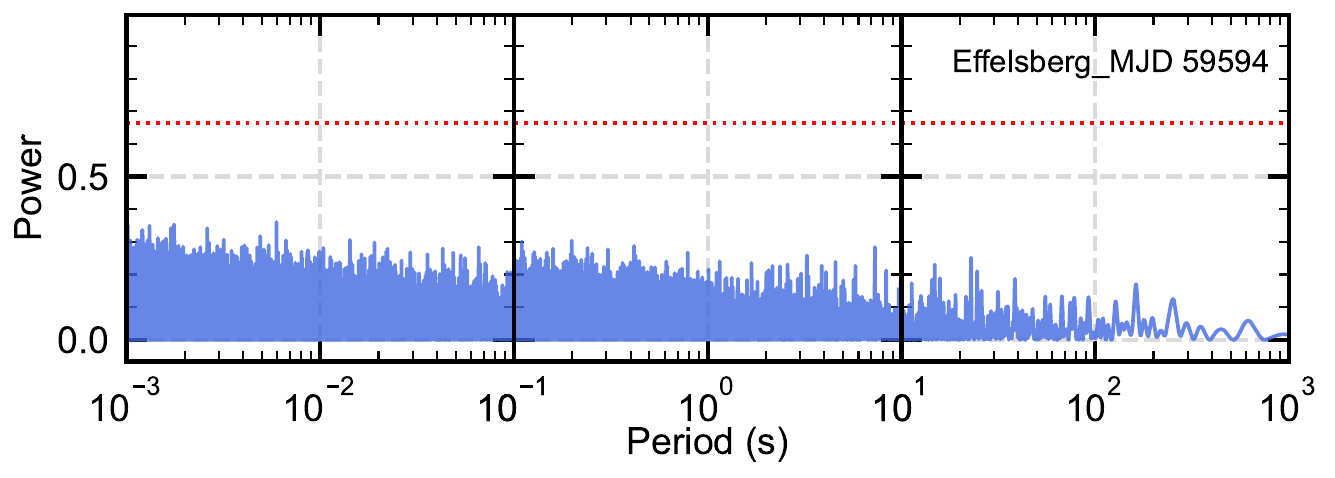}
   \caption{\footnotesize The period search results based on the
   Effelsberg Dataset \#7 of FRB 20200120E. Bursts on MJD 59594 are analyzed.
   The three columns from left to right correspond to the three search methods, i.e.
   the phase folding algorithm, the H-test, and the Lomb-Scargle periodogram, respectively.
   In the cases of phase folding and the H-test methods, a horizontal dotted line is
   plotted to mark the p-value of $10^{-9}$. In the cases of Lomb-Scargle periodogram,
   a horizontal dotted line indicating a FAP level of $10^{-9}$ is plotted.
   No clear evidence of periodicity is found in these plots. Note that in the left and middle panels,
   the peak structures in the 1 ms--20 ms
   range are fake signals due to the limited timing accuracy. }
   \label{Fig5}
\end{figure*}

We have used all the three methods mentioned above to analyze the
periodicity. In the phase folding algorithm, the number of bins is
set as $n = 20$. In the H-test method, the maximum number of
harmonics is taken as $m_{\max}=20$. In the method of the
Lomb-Scargle periodogram, the peak flux of each burst is used as a
necessary input for the intensity. The results from Dataset \#7
are shown in Figure ~\ref{Fig5}. For the phase folding method and
the H-test method, a horizontal dotted line is plotted to mark the
p-value of $10^{-9}$ (corresponding to a gaussian significance of
6.11$\sigma$). For the Lomb-Scargle periodogram, a horizontal
dotted line showing a FAP level of $10^{-9}$ is also plotted. Note that in the 1--20 ms range, some extremely high
peaks could be observed. Upon further examination, again we find
that they are all harmonics of 0.864 ms ($10^{-8}$ d), and are
thus fake signals due to limited timing accuracy of observations.
In each plot, we have also examined by hand the phase-folded
histograms at the positions that have a relatively high peak in
the $\chi^2$/H-statistic/Lomb-Scargle periodograms, but found that
no marked features appear in the corresponding histograms. From
Figure ~\ref{Fig5}, we conclude that no periodicity exceeding the
confidence level is found in the whole range of 0.001--1000 s
for FRB 20200120E.

\subsection{FRB 20201124A}

The ToAs of the bursts from FRB 20201124A, obtained by FAST
(Datasets \#4 and \#6) and uGMRT (Dataset \#5), are used to
search for potential short-time periodicity. Here, the ToAs were calibrated to the
barycenter of the solar system. For the FAST Datasets \#4 and \#6,
the ToAs are de-dispersed to 1.5 GHz, and the timing precision
(estimated from the last digit of the recorded ToAs) is $\sim
10^{-8}$d (864 $\mathrm{\mu s}$) and $\sim 10^{-9}$ d (86.4
$\mathrm{\mu s}$), respectively. The ToAs of uGMRT Dataset \#5 are
barycentre corrected and de-dispersed to 0.55 GHz, and the
precision is $\sim 10^{-8}$d (864 $\mathrm{\mu s}$).

For the FAST Dataset \#4, we selected the bursts from the
four days with the highest number of bursts detected to perform
period searches (82 bursts on MJD 59313, of which 67 bursts have a
waiting time > 0.1 s; 102 bursts on MJD 59314, of which 88 bursts
have a waiting time > 0.1 s; 108 bursts on MJD 59315, of which 99
bursts have a waiting time
> 0.1 s; and 77 bursts on MJD 59334, of which 72 bursts have a
waiting time > 0.1 s). For the FAST Dataset \#6, we selected the
bursts from the two days with the highest number of bursts
detected (232 bursts on MJD 59485, of which 188 bursts have a
waiting time > 0.1 s; and 542 bursts on MJD 59486, of which 409
bursts have a waiting time > 0.1 s). For the uGMRT Dataset \#5, 48
bursts from the only observation day MJD 59310 are utilized. All
the bursts have a waiting time > 0.1 s.

\begin{figure*}
   \centering
   \includegraphics[width=0.32\textwidth]{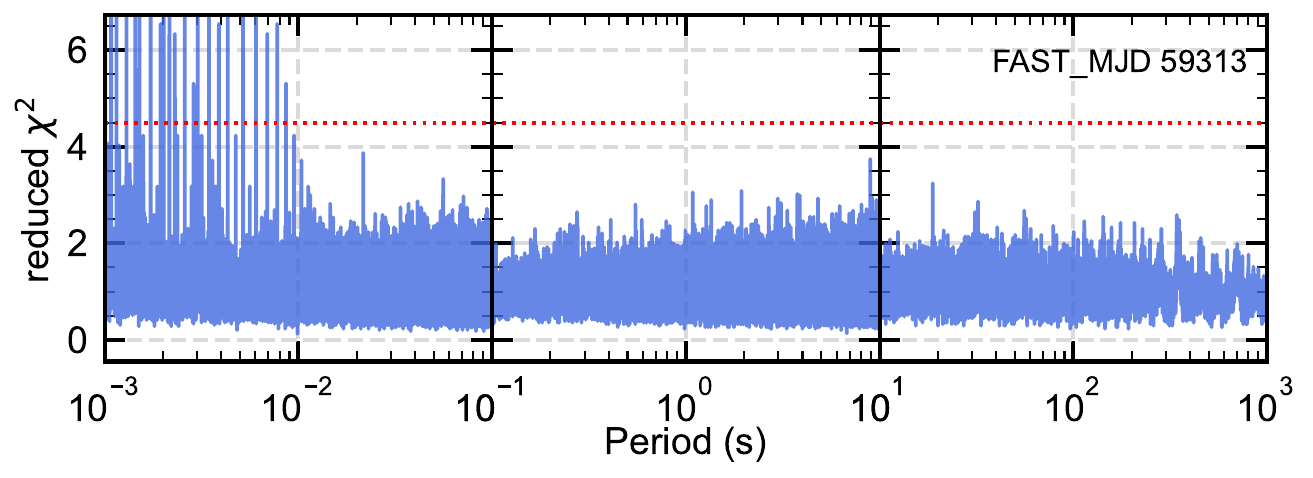}
   \includegraphics[width=0.32\textwidth]{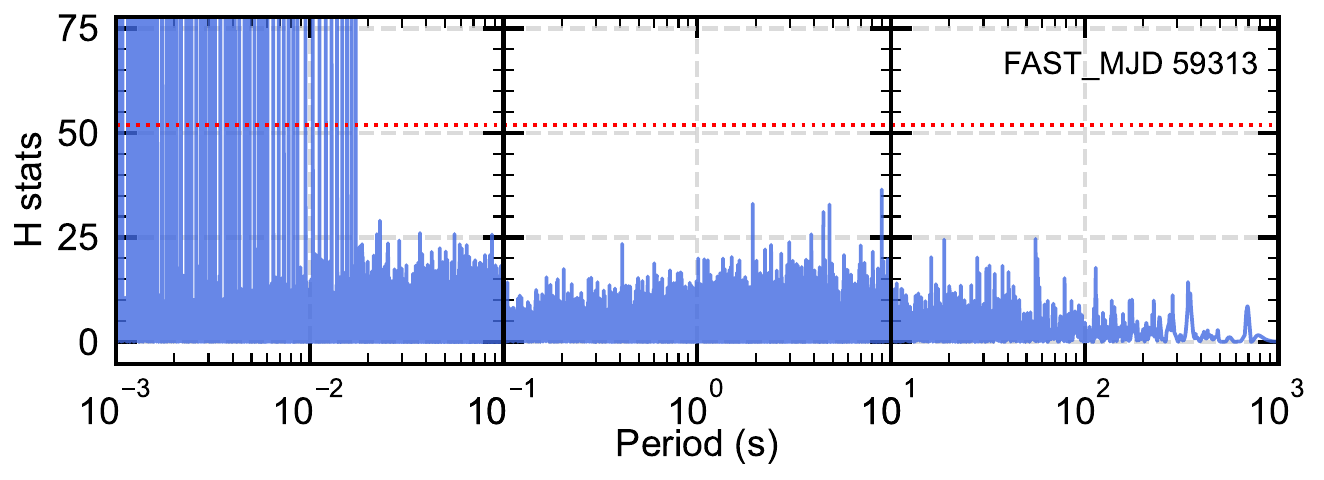}
   \includegraphics[width=0.32\textwidth]{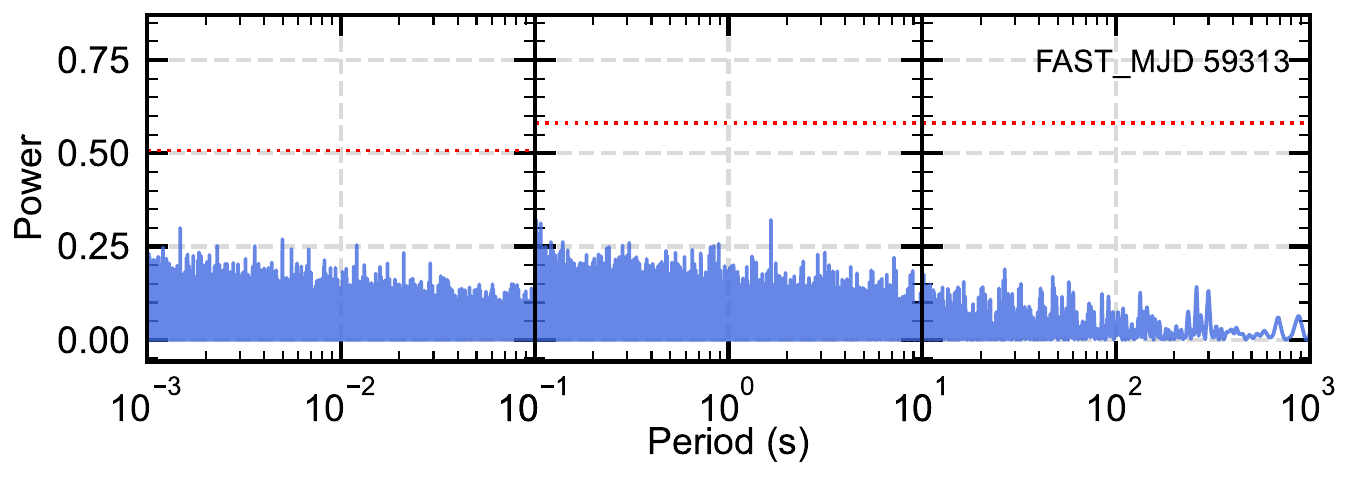}
   \includegraphics[width=0.32\textwidth]{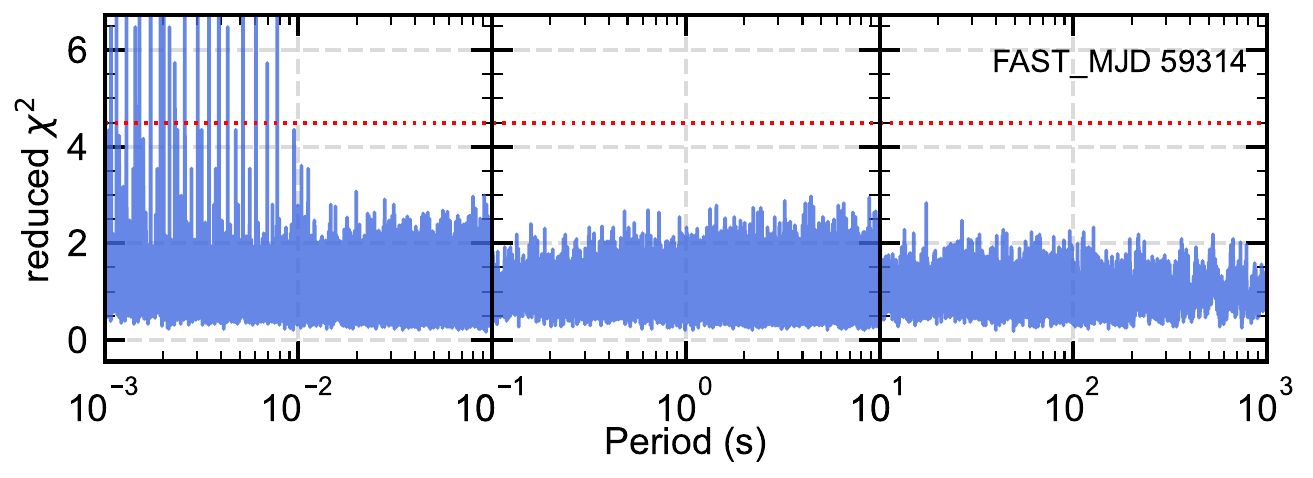}
   \includegraphics[width=0.32\textwidth]{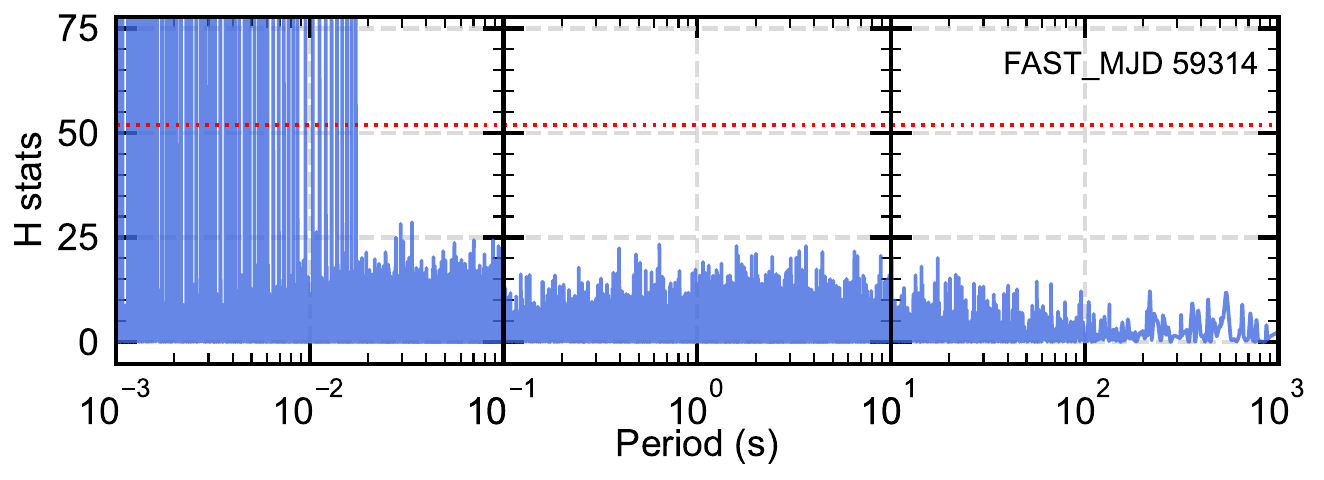}
   \includegraphics[width=0.32\textwidth]{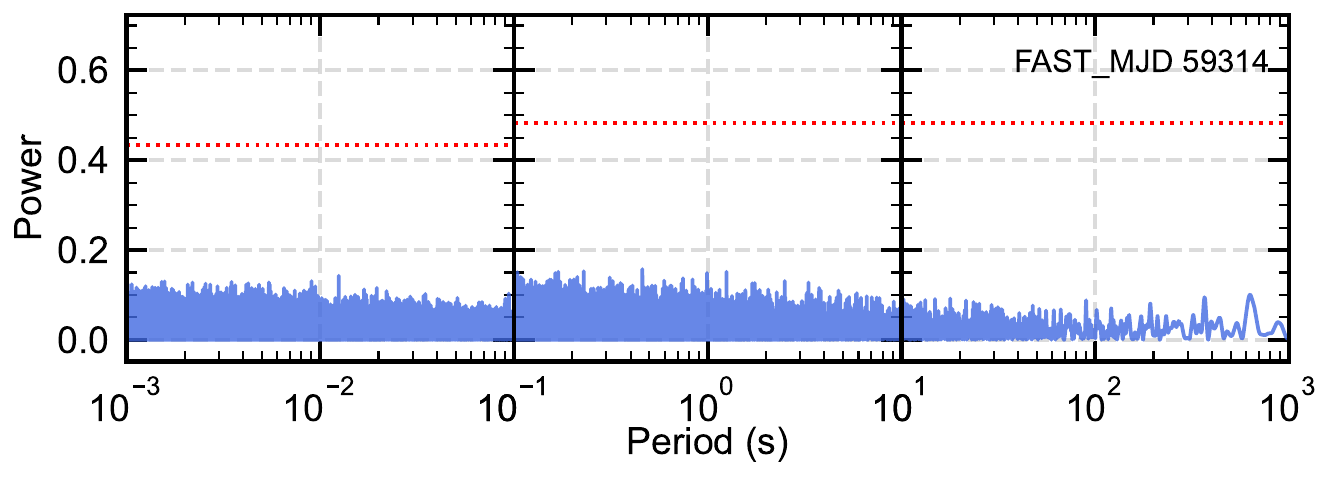}
   \includegraphics[width=0.32\textwidth]{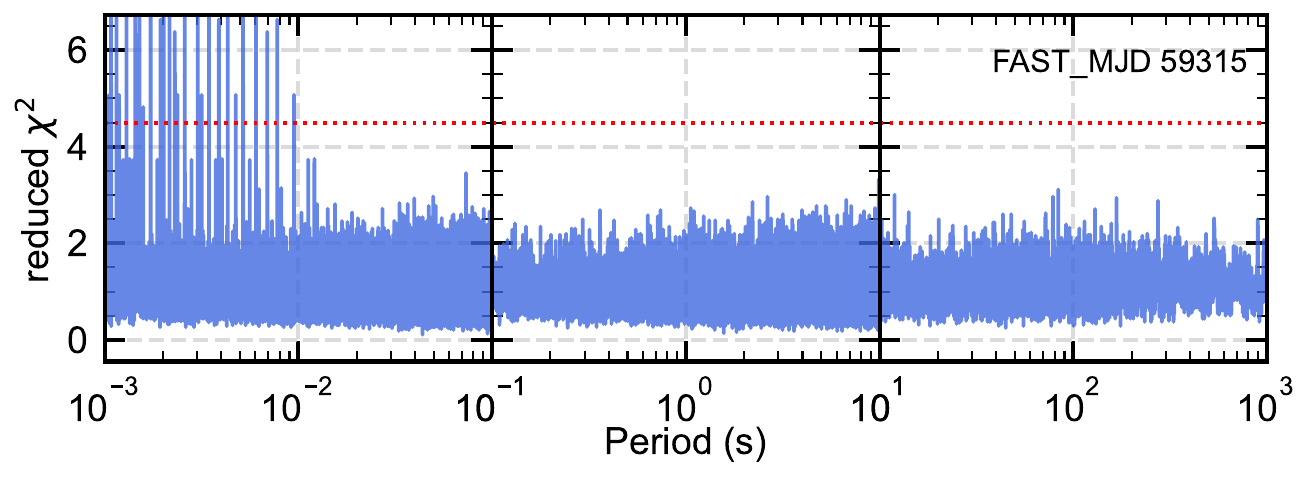}
   \includegraphics[width=0.32\textwidth]{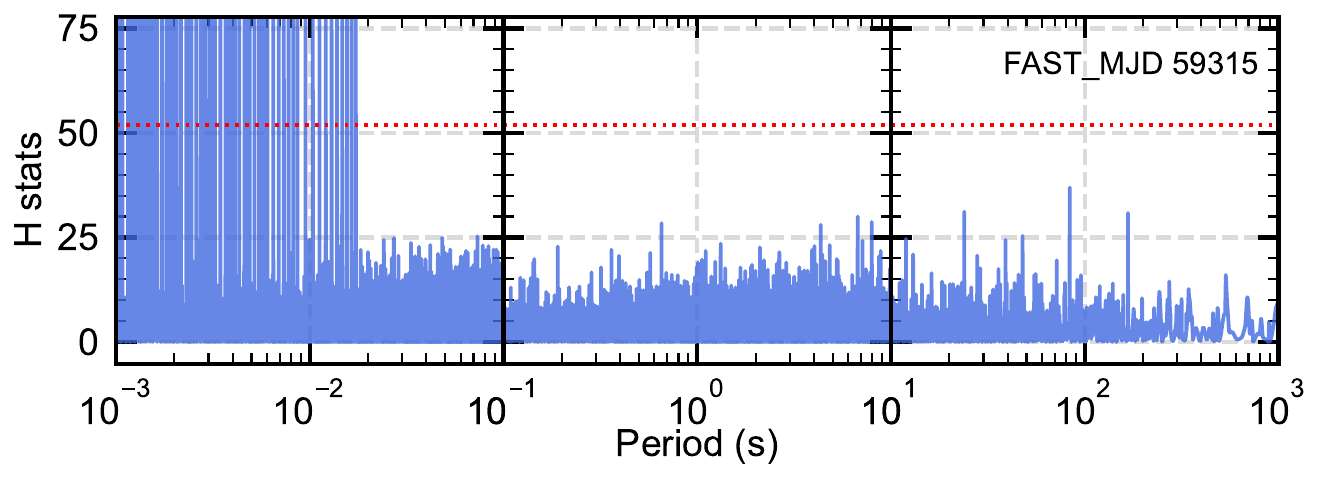}
   \includegraphics[width=0.32\textwidth]{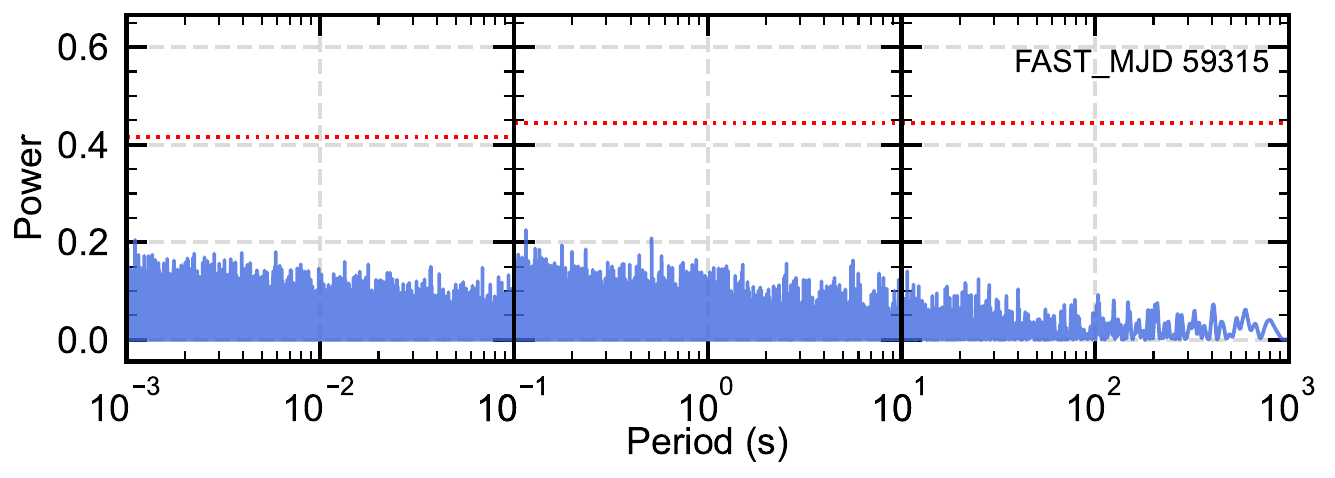}
   \includegraphics[width=0.32\textwidth]{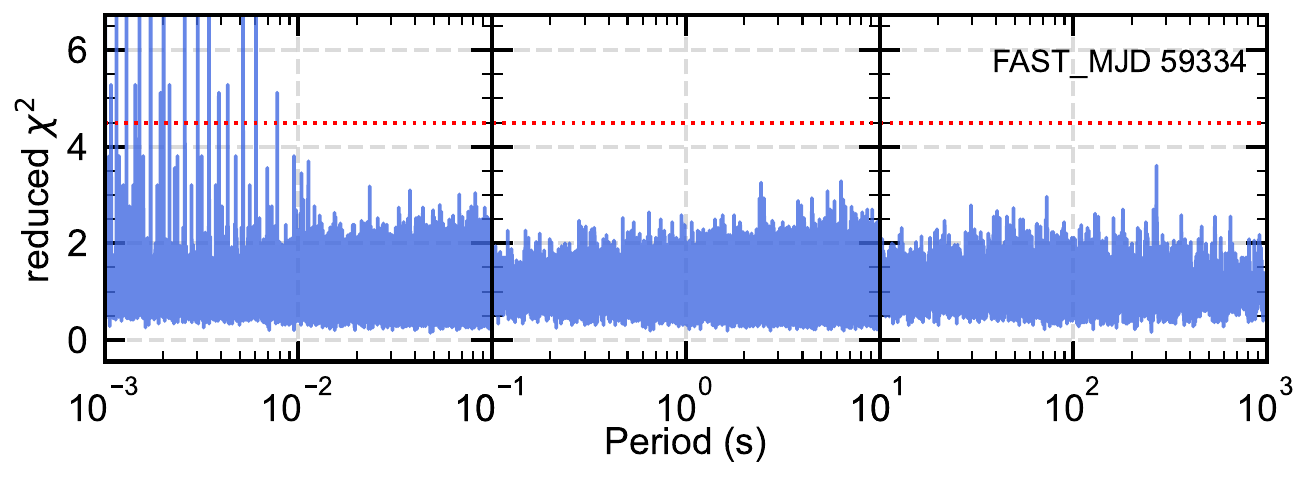}
   \includegraphics[width=0.32\textwidth]{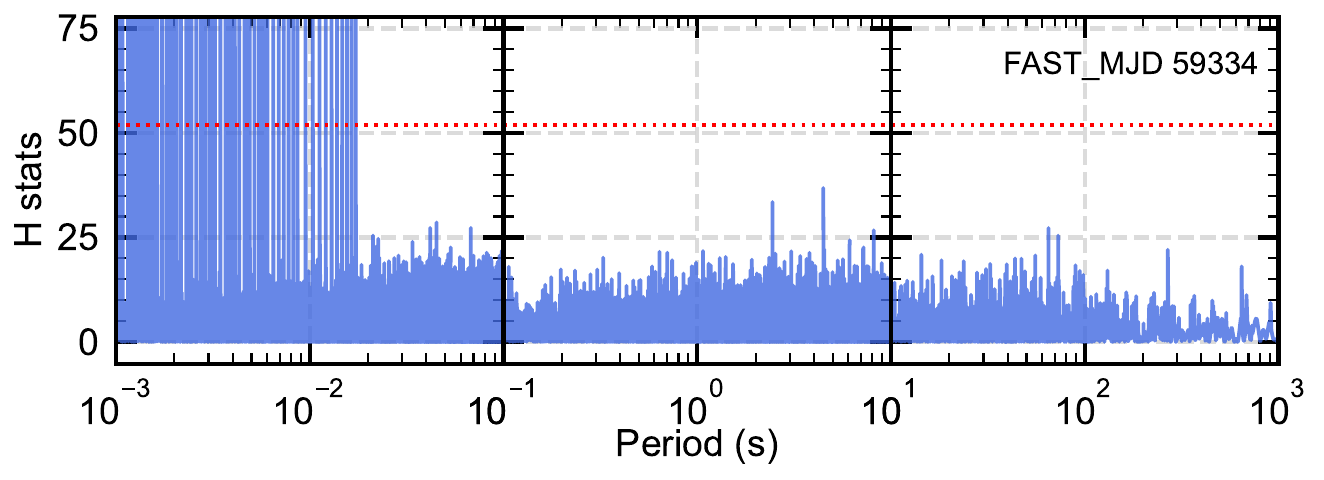}
   \includegraphics[width=0.32\textwidth]{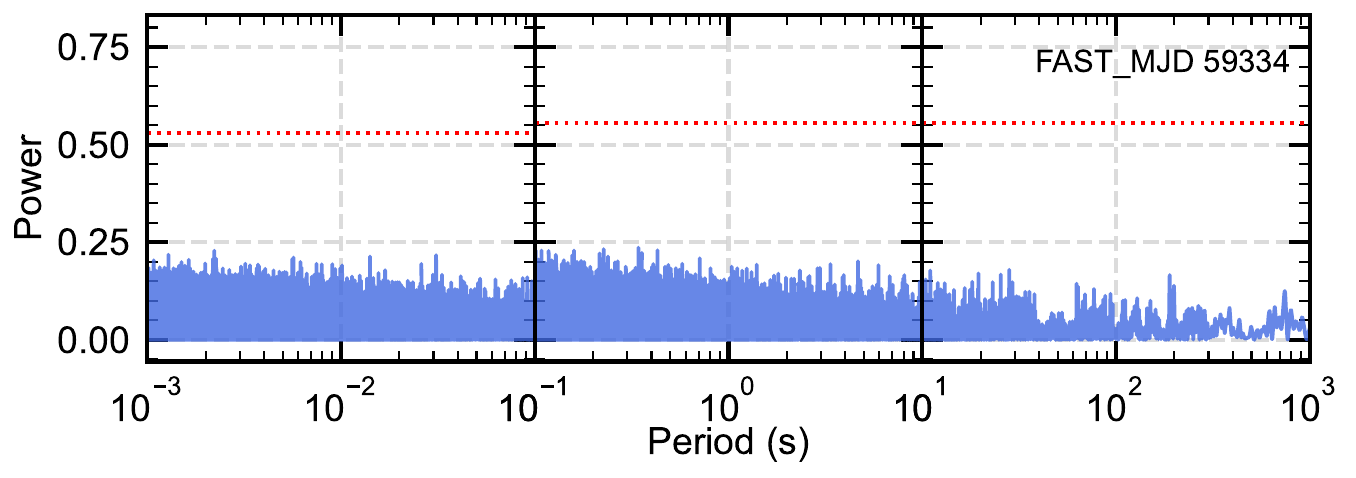}
   \caption{\footnotesize The period search results based on the FAST Dataset \#4
   of FRB 20201124A. The three columns from left to right correspond to the three search methods, i.e.
   the phase folding algorithm, the H-test, and the Lomb-Scargle periodogram, respectively.
   The four rows from top to bottom correspond to the four selected days (MJD 59313, MJD 59314,
   MJD 59315 and MJD 59334). In the cases of phase folding and the H-test methods,
   a horizontal dotted line is plotted to mark the p-value of $10^{-9}$. In the cases of
   Lomb-Scargle periodogram, a horizontal dotted line indicating a FAP level of $10^{-9}$
   is plotted. No clear evidence of periodicity is found in these plots. Note that in the left and middle panels,
   the peak structures in the 1 ms--20 ms
   range are fake signals due to the limited timing accuracy.  }
   \label{Fig6}
\end{figure*}

\begin{figure*}
   \centering
   \includegraphics[width=0.32\textwidth]{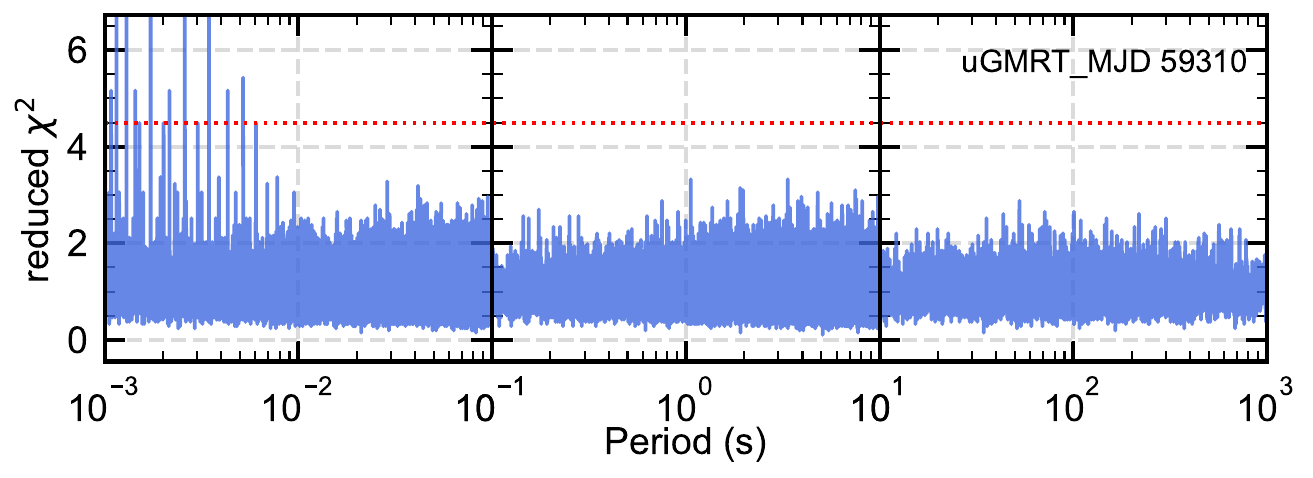}
   \includegraphics[width=0.32\textwidth]{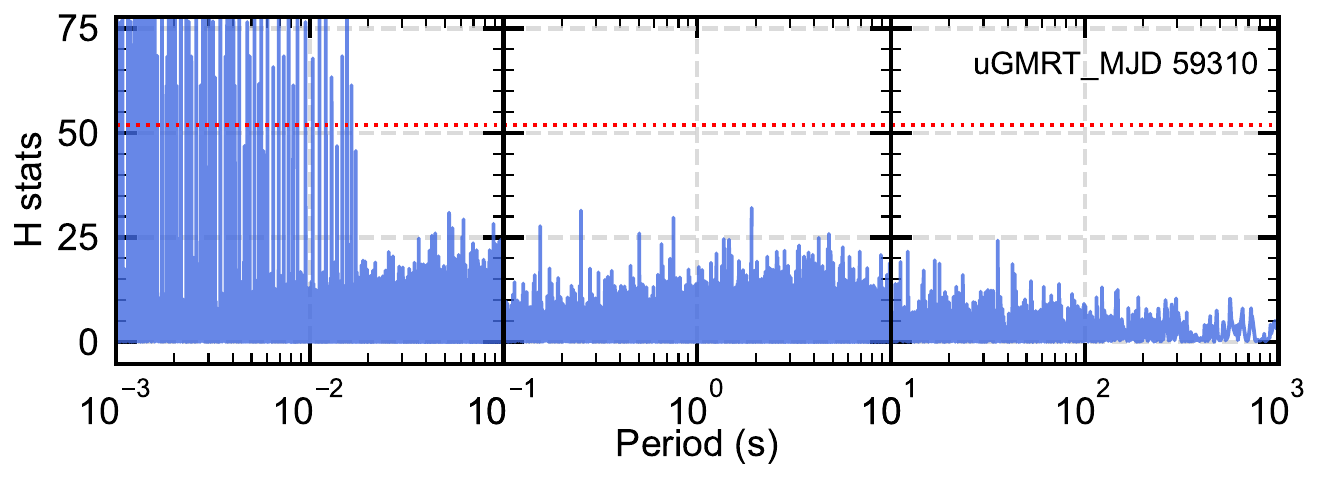}
   \includegraphics[width=0.32\textwidth]{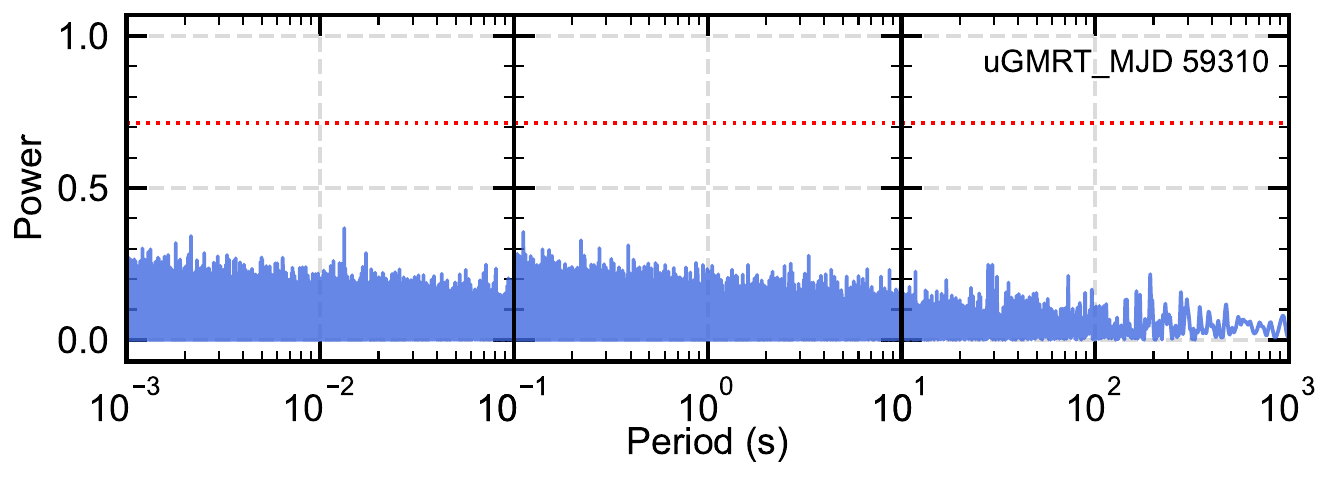}
   \caption{\footnotesize The period search results based on the uGMRT Dataset \#5
   of FRB 20201124A. Bursts on MJD 59310 are analyzed. The three columns from left to right
   correspond to the three search methods, i.e. the phase folding algorithm, the H-test, and
   the Lomb-Scargle periodogram, respectively. In the cases of phase folding and the H-test methods,
   a horizontal dotted line is plotted to mark the p-value of $10^{-9}$. In the cases of
   Lomb-Scargle periodogram, a horizontal dotted line indicating a FAP level of $10^{-9}$
   is plotted. No clear evidence of periodicity is found in these plots. Note that in the left and middle panels,
   the peak structures in the 1 ms--20 ms
   range are fake signals due to the limited timing accuracy. }
   \label{Fig7}
\end{figure*}

\begin{figure*}
   \centering
   \includegraphics[width=0.32\textwidth]{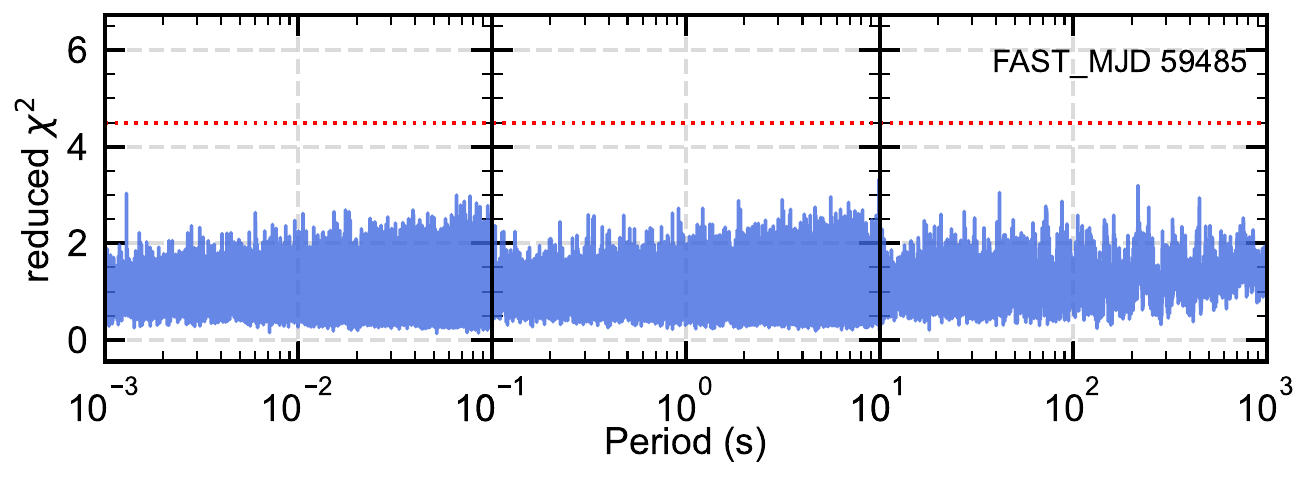}
   \includegraphics[width=0.32\textwidth]{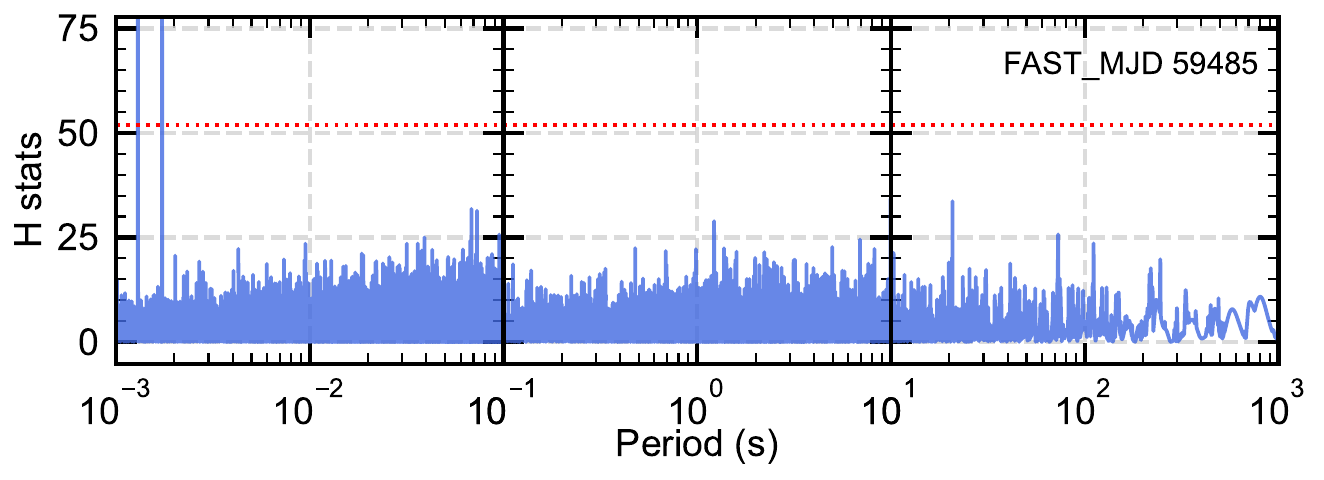}
   \includegraphics[width=0.32\textwidth]{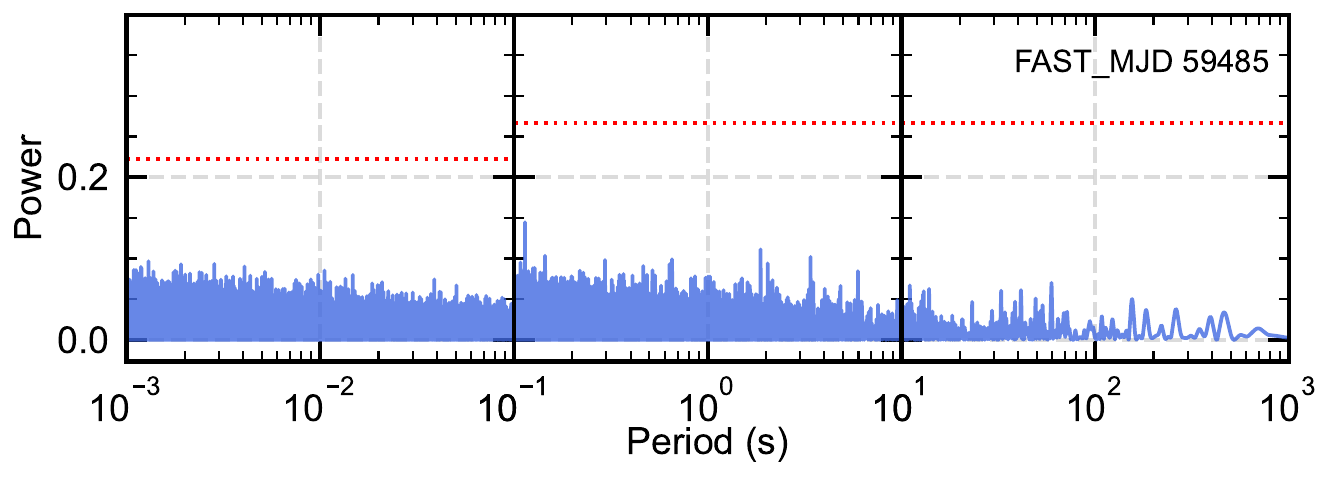}
   \includegraphics[width=0.32\textwidth]{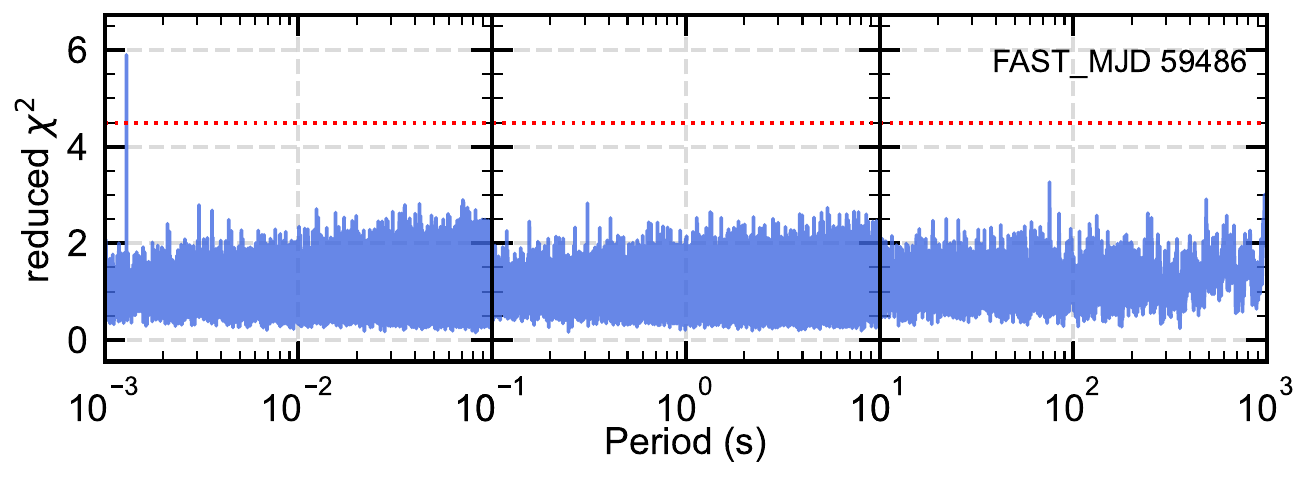}
   \includegraphics[width=0.32\textwidth]{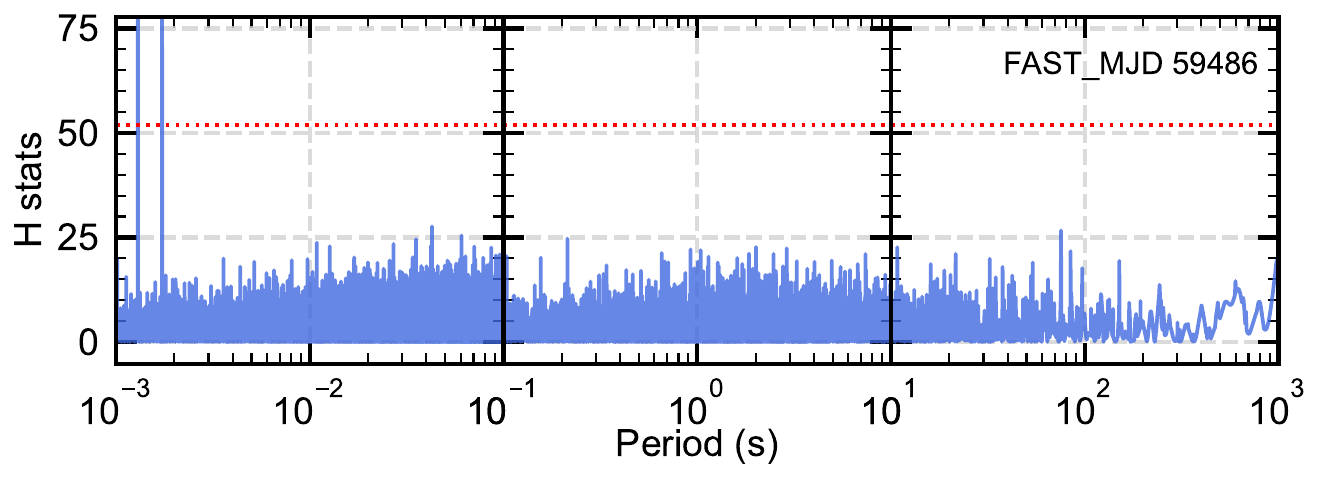}
   \includegraphics[width=0.32\textwidth]{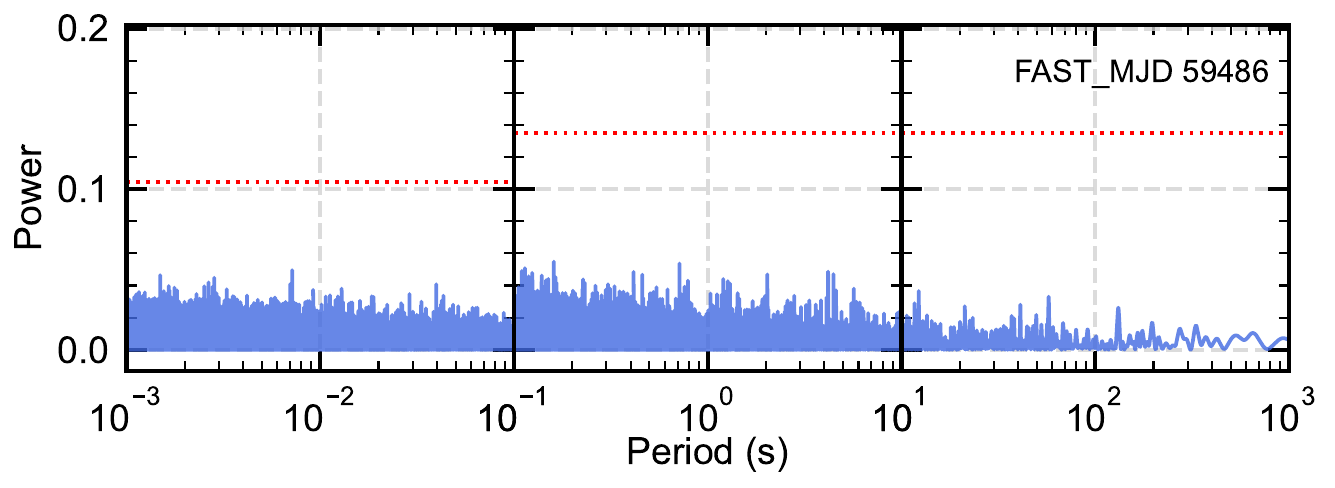}
   \caption{\footnotesize The period search results based on the FAST Dataset \#6
   of FRB 20201124A. The three columns from left to right correspond to the three search methods
   i.e. the phase folding algorithm, the H-test, and the Lomb-Scargle periodogram, respectively.
   The two rows from top to bottom correspond to the two selected days (MJD 59485 and MJD 59486).
   In the cases of phase folding and the H-test methods, a horizontal dotted line is plotted to
   mark the p-value of $10^{-9}$. In the cases of Lomb-Scargle periodogram, a horizontal
   dotted line indicating a FAP level of $10^{-9}$ is plotted.
   No clear evidence of periodicity is found in these plots. Note that in the left and middle panels,
   the peak structures in the 1 ms--2 ms
   range are fake signals due to the limited timing accuracy. }
   \label{Fig8}
\end{figure*}

We have applied all the three methods mentioned above to analyze
the periodicity. In the phase folding algorithm, the number of
bins is set as $n = 20$. In the H-test method, the maximum number
of harmonics is taken as $m_{\max}=20$. In the method of the
Lomb-Scargle periodogram, the peak flux of each burst is used as a
necessary input for the intensity. The results of Datasets \#4,
\#5 and \#6 are shown in Figures ~\ref{Fig6}, \ref{Fig7} and
\ref{Fig8}, respectively. For the phase folding method and the
H-test method, a horizontal dotted line is plotted to mark the
p-value of $10^{-9}$ (corresponding to a gaussian significance of
6.11$\sigma$). For the Lomb-Scargle periodogram, a horizontal
dotted line showing a FAP level of $10^{-9}$ is also plotted.
Note that in the 1--20 ms range, some extremely high
peaks could be observed. Again they are all harmonics of 0.864 ms
($10^{-8}$ d), and are fake signals due to limited timing accuracy
of observations. In each plot, we have also examined by hand the
phase-folded histograms at the positions that correspond to a
relatively high peak in the $\chi^2$/H-statistic/Lomb-Scargle
periodograms, but found that no marked features appear in the
histograms. Additionally, a peak occurring on one day usually does
not appear on other days, which further excludes its possibility
of being a true period. From Figures ~\ref{Fig6}, \ref{Fig7}, and
\ref{Fig8}, we conclude that no periodicity exceeding the
confidence level is found in the whole range of 0.001--1000 s
for FRB 20201124A.

\subsection{FRB 20220912A}

The ToAs of the bursts from FRB 20220912A, obtained by GBT
(Dataset \#8) and FAST (Dataset \#9), are used to search for
potential short-time periodicity. Here, the ToA corresponds to the
peak of each burst, calibrated to the barycenter of the solar
system. For the GBT Dataset \#8, the ToAs are de-dispersed to
infinite frequency, and the timing precision (estimated from the
last digit of the recorded ToAs) is $\sim 10^{-8}$ d (864
$\mathrm{\mu s}$). The ToAs of FAST Dataset \#9 are de-dispersed
to 1.5 GHz, and the precision is $\sim 10^{-9}$ d (86.4
$\mathrm{\mu s}$).

For the GBT Dataset \#8, we selected the 128 bursts from
the only observation day MJD 59876 to perform period searches, of
which 117 bursts have a waiting time > 0.1 s. For the FAST Dataset
\#9, the bursts from the three days with the highest number of
bursts detected are utilized (195 bursts on MJD 59880, of which
155 bursts have a waiting time > 0.1 s; 277 bursts on MJD 59883,
of which 247 have a waiting time > 0.1 s; and 91 bursts on MJD
59901, of which 75 bursts have a waiting time > 0.1 s).

\begin{figure*}
   \centering
   \includegraphics[width=0.32\textwidth]{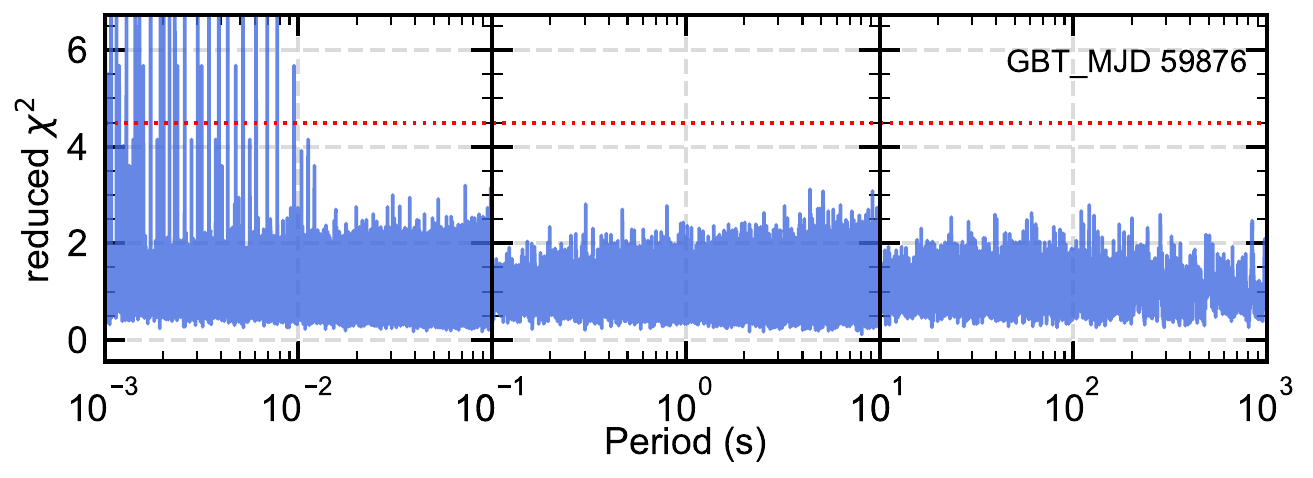}
   \includegraphics[width=0.32\textwidth]{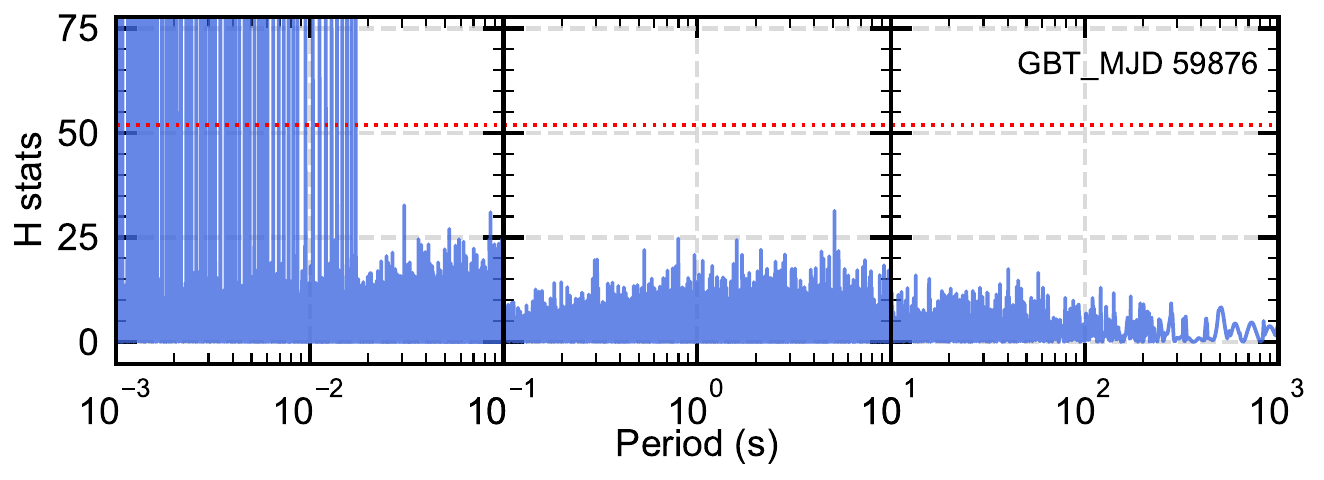}
   \includegraphics[width=0.32\textwidth]{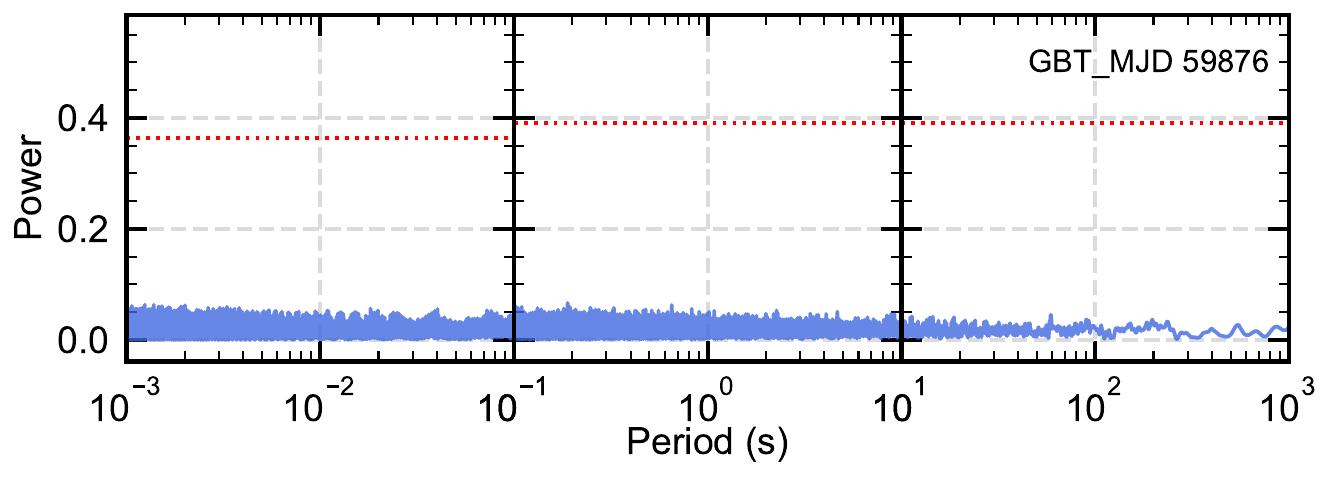}
   \caption{\footnotesize The period search results based on the GBT Dataset \#8
   of FRB 20220912A. Bursts on MJD 59876 are analyzed. The three columns from left to right
   correspond to the three search methods, i.e. the phase folding algorithm, the H-test,
   and the Lomb-Scargle periodogram, respectively. In the cases of phase folding and the H-test methods,
   a horizontal dotted line is plotted to mark the p-value of $10^{-9}$. In the cases of
   Lomb-Scargle periodogram, a horizontal dotted line indicating a FAP level of $10^{-9}$ is
   plotted. No clear evidence of periodicity is found in these plots. Note that in the left and middle panels,
   the peak structures in the 1 ms--20 ms
   range are fake signals due to the limited timing accuracy. }
   \label{Fig9}
\end{figure*}

\begin{figure*}
   \centering
   \includegraphics[width=0.32\textwidth]{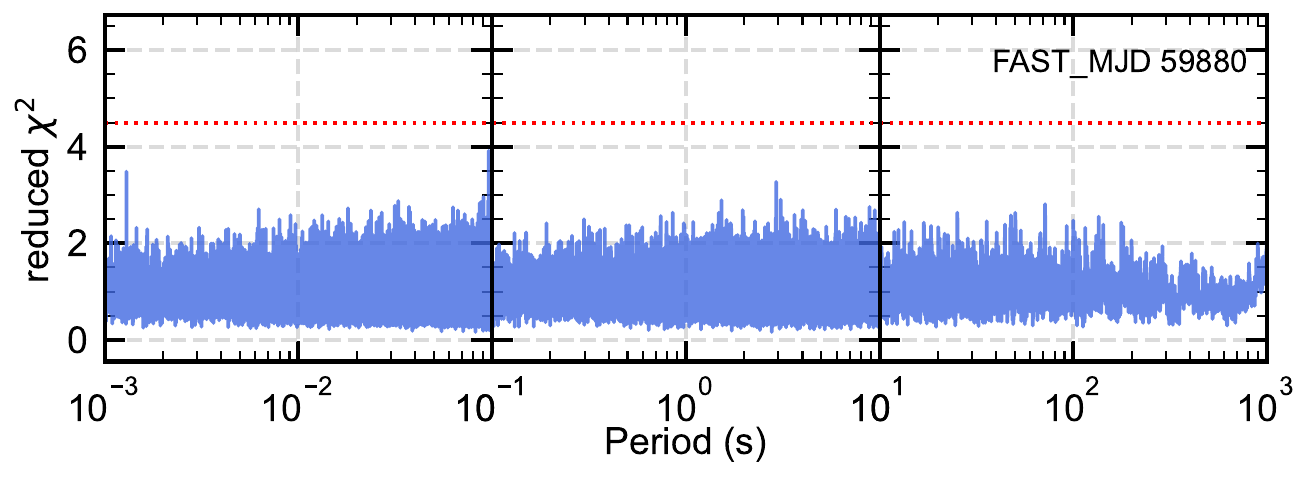}
   \includegraphics[width=0.32\textwidth]{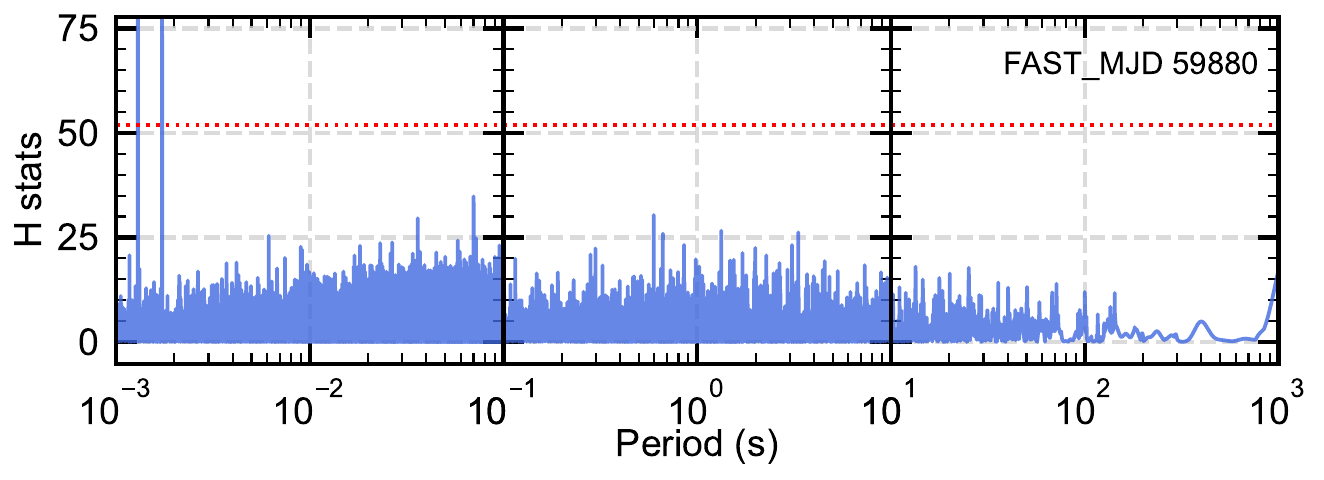}
   \includegraphics[width=0.32\textwidth]{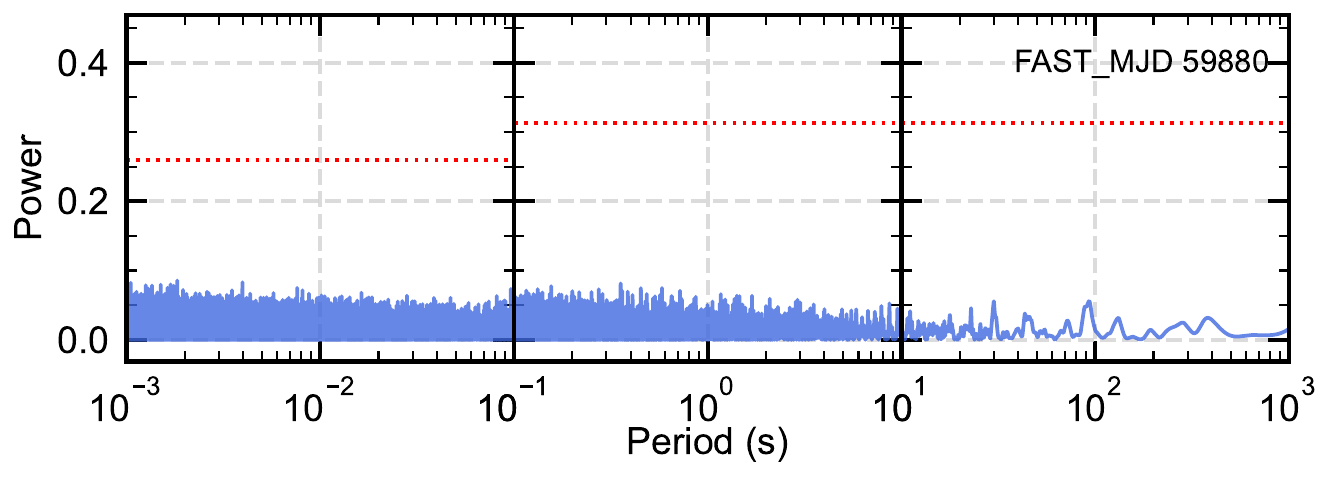}
   \includegraphics[width=0.32\textwidth]{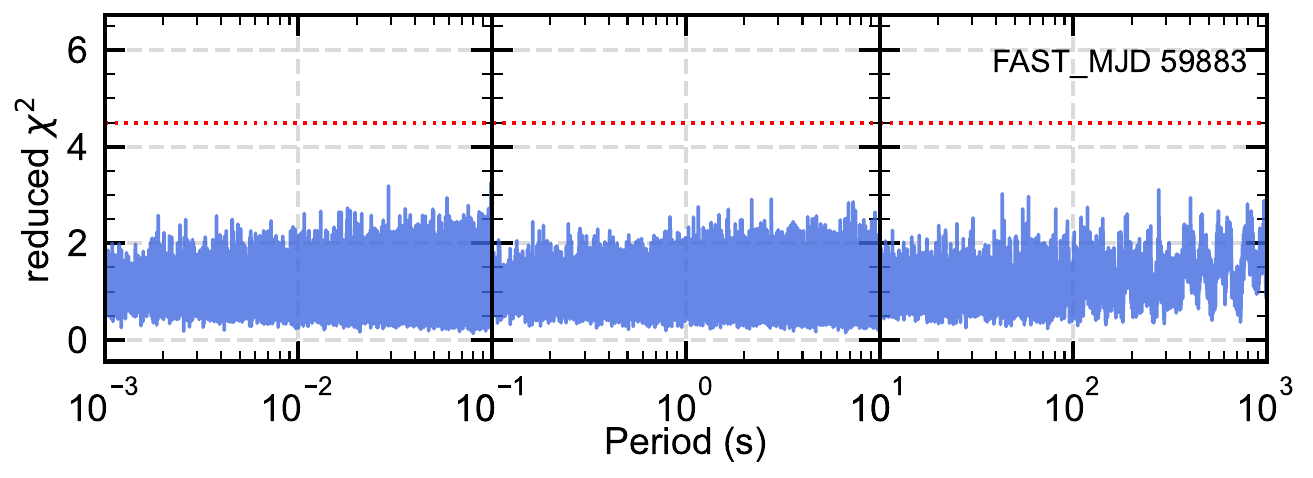}
   \includegraphics[width=0.32\textwidth]{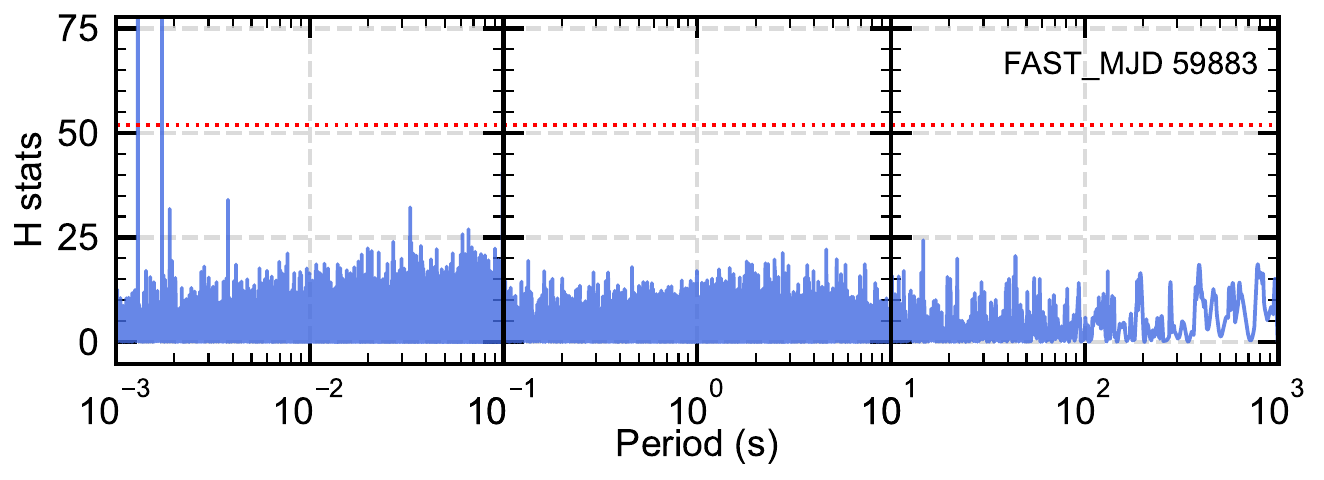}
   \includegraphics[width=0.32\textwidth]{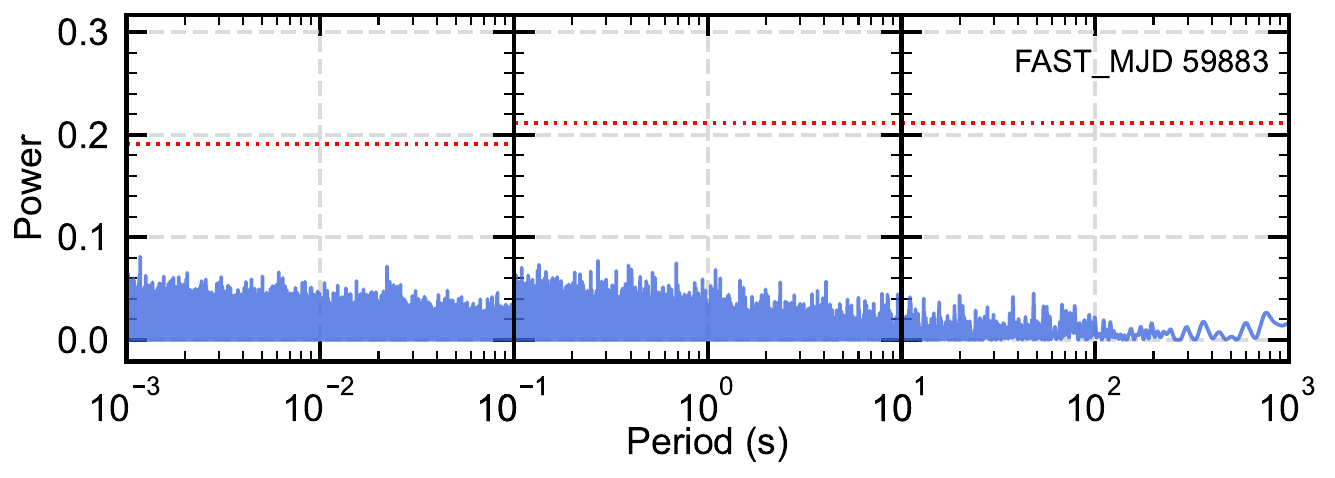}
   \caption{\footnotesize The period search results based on the FAST Dataset \#9
   of FRB 20220912A. The three columns from left to right correspond to the three search methods,
   i.e. the phase folding algorithm, the H-test, and the Lomb-Scargle periodogram, respectively.
   The three rows from top to bottom correspond to the three selected days (MJD 59880, MJD 59883
   and MJD 59901). In the cases of phase folding and the H-test methods, a horizontal dotted
   line is plotted to mark the p-value of $10^{-9}$. In the cases of Lomb-Scargle periodogram,
   a horizontal dotted line indicating a FAP level of $10^{-9}$ is plotted. No clear evidence
   of periodicity is found in these plots. Note that in the middle panels,
   the peak structures in the 1 ms--2 ms
   range are fake signals due to the limited timing accuracy. }
   \label{Fig10}
\end{figure*}

We have used all the three methods mentioned above to analyze the
periodicity. In the phase folding algorithm, the number of bins is
taken as $n = 20$. In the H-test method, the maximum number of
harmonics is $m_{\max}=20$. In the method of the Lomb-Scargle
periodogram, the peak flux of each burst is used as a necessary
input for the intensity. The results of Datasets \#8 and \#9 are
shown in Figures ~\ref{Fig9} and ~\ref{Fig10}, respectively. For
the phase folding method and the H-test method, a horizontal
dotted line is plotted to mark the p-value of $10^{-9}$
(corresponding to a gaussian significance of 6.11$\sigma$). For
the Lomb-Scargle periodogram, a horizontal dotted line showing a
FAP level of $10^{-9}$ is also plotted. Again in the 1--20 ms range, some extremely high peaks could be observed, which
are all harmonics of 0.864 ms ($10^{-8}$ d). They are fake signals
due to limited timing accuracy of observations. In each plot, we
have also examined by hand the phase-folded histograms at the
positions that correspond to the relatively high peaks in the
$\chi^2$/H-statistic/Lomb-Scargle periodograms, but found that no
marked features appear in the histograms. From Figures ~\ref{Fig9}
and ~\ref{Fig10}, we conclude that no periodicity exceeding the
confidence level is found in the whole range of 0.001--1000 s for
FRB 20220912A.

To summarize, we have carefully examined the four active repeating
FRB sources (FRBs 20121102A, 20200120E, 20201124A, and 20220912A).
All the three methods widely used for periodicity analysis are
applied on them. For these four repeaters, no evidence is found
supporting the existence of any periodicity in the whole range of
0.001--1000 s. The null results are consistent across different
datasets. It indicates that regardless of the physical origin of
the FRBs, their periodicity is very likely destroyed or smeared by
the external environment, such as a strongly magnetized medium, or
by the precession of the source itself.


\section{Period searches based on brightness features}
\label{sect:analysis2}
FRBs are highly variable. Their properties vary from event to
event. They have different pulse widths, peak flux, fluence and
pulse profiles, and the emission is in different frequency ranges.
The differences may reflect the different physical conditions of
the central engine. Theoretically, even for the same repeater, the
bursts might be triggered by different mechanisms and could be
emitted at various regions. Therefore, it is possible that some
subclasses of FRBs may show a periodical behavior. In this
section, we explore the periodicity of FRBs based on their
properties. In earlier researches of repeating FRBs, the small
sample size severely limits the robustness of previous attempts at
period searching on subclasses of bursts. However, with the
explosive growth of the observational data in recent years, it is
possible for us to group the sample into subsets and examine their
nature in great detail.

As a first step, we have tried to use some simple parameters such
as the pulse width, peak flux or fluence of each burst as the
weighting factor, and repeated the above analysis. Still, no clear
evidence pointing toward any periodicity is found.

We then consider the brightness temperature as the criteria.
Recent studies by \cite{xiao2022hints} suggest that
FRBs can be divided into two subsets according to their brightness
temperatures. The brightness temperature of an FRB is defined by
equaling the observed power to the luminosity of a blackbody,
which gives:
\begin{equation}
   T_{\mathrm{B}}=F_{\nu} d_{\mathrm{A}}^{2} / 2 \pi k(\nu\cdot W)^{2},
\end{equation}
where $F_{\nu}$ is the flux density, $\nu$ is the emission
frequency and $W$ is the pulse width. Note that $d_{\mathrm{A}}$
here is the angular diameter distance, which is different for
different repeaters.

In calculating the brightness temperature, when the detailed
frequency information is unavailable for some bursts, the central
frequency of the telescope is used. A flat $\Lambda$-CDM cosmology
with $H_0 = 67.4$ km~s$^{-1}$~Mpc$^{-1}$ and $\Omega_{\rm m} =
0.315$ is adopted \citep{aghanim2020planck} to calculate the
distance ($d_{\mathrm{A}}$). For FRB 20121102A at a redshift of $z
= 0.19273$ \citep{tendulkar2017hosta}, we get the distance as
$d_{\mathrm{A}}$ = 0.682 Gpc. For FRB 20201124A at $z = 0.09795$
\citep{xu2022fast}, we have $d_{\mathrm{A}}$ = 0.386 Gpc. For FRB
20220912A at $z = 0.0771$ \citep{ravi2023Deep}, we have
$d_{\mathrm{A}}$ = 0.311 Gpc.

\begin{figure*}
   \centering
   \includegraphics[width=1\textwidth]{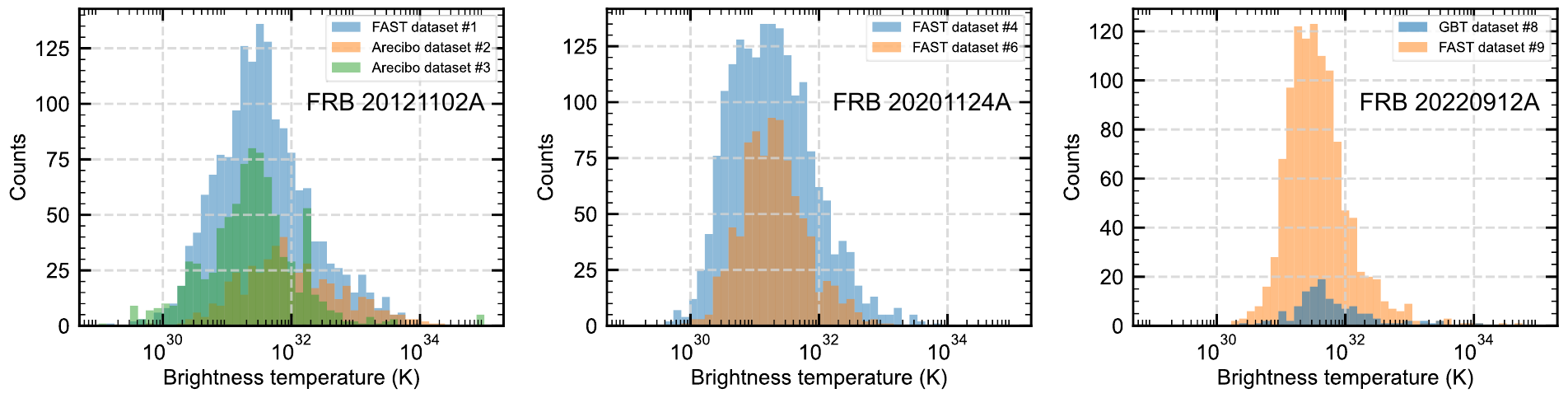}
   \caption{\footnotesize The brightness temperature distribution of the bursts
   from the three repeaters of FRB 20121102A (FAST Dataset \#1, Arecibo Datasets \#2, \#3),
   FRB 20201124A (FAST Datasets \#4, \#6), and FRB 20220912A (GBT Dataset \#8, FAST Dataset \#9).}
   \label{Fig11}
\end{figure*}

We then divide the FRBs into subgroups according to their
brightness temperatures and conduct period searches based on the
subgroups. Still, we continue to use the single-day search
strategy and exclude bursts with waiting times less than 0.1 s in searches for periods larger than 0.1 s. To
ensure that the number of bursts in each subset is not too few,
our analysis is conducted only for the three repeaters with enough
bursts, i.e. FRB 20121102A (Arecibo Datasets
\#1, \#2, FAST Dataset \#3), FRB 20201124A (FAST Datasets \#4, \#6), and FRB
20220912A (GBT Dataset \#8, FAST Dataset \#9). The brightness
temperature distribution of bursts from each repeater generally
follows a log-normal distribution (as shown in Figure
~\ref{Fig11}). The bursts are divided into two subgroups, i.e.
high $T_\mathrm{B}$ group and low $T_\mathrm{B}$ group. The
criteria temperature is taken to ensure that nearly equal numbers
of events are included in each sub-sample. Similar periodicity
analysis as described in Section 4 is then applied to each subset.

\begin{figure*}
   \centering
   \includegraphics[width=0.32\textwidth]{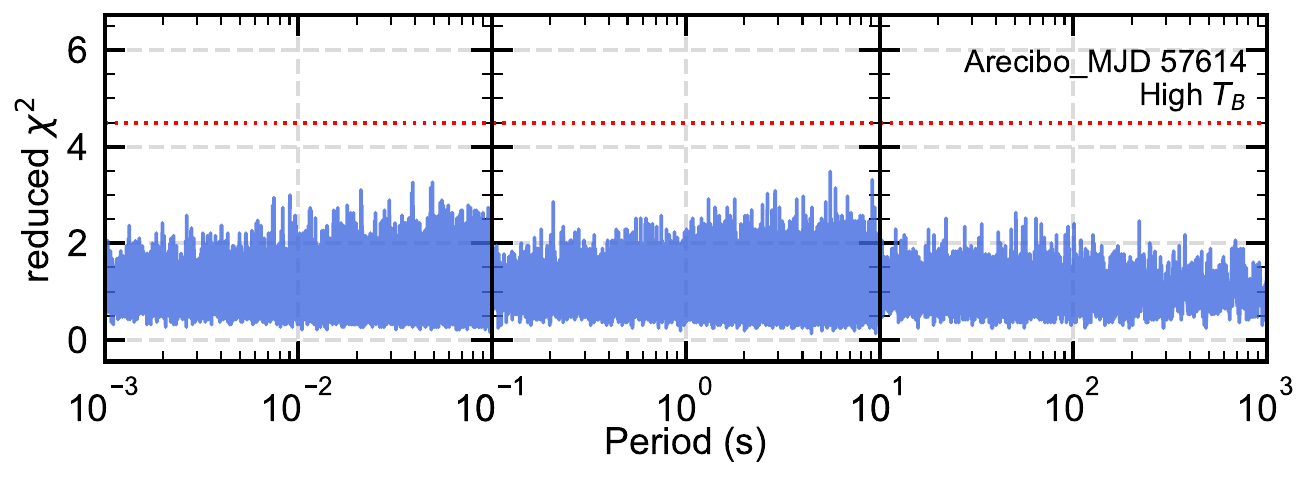}
   \includegraphics[width=0.32\textwidth]{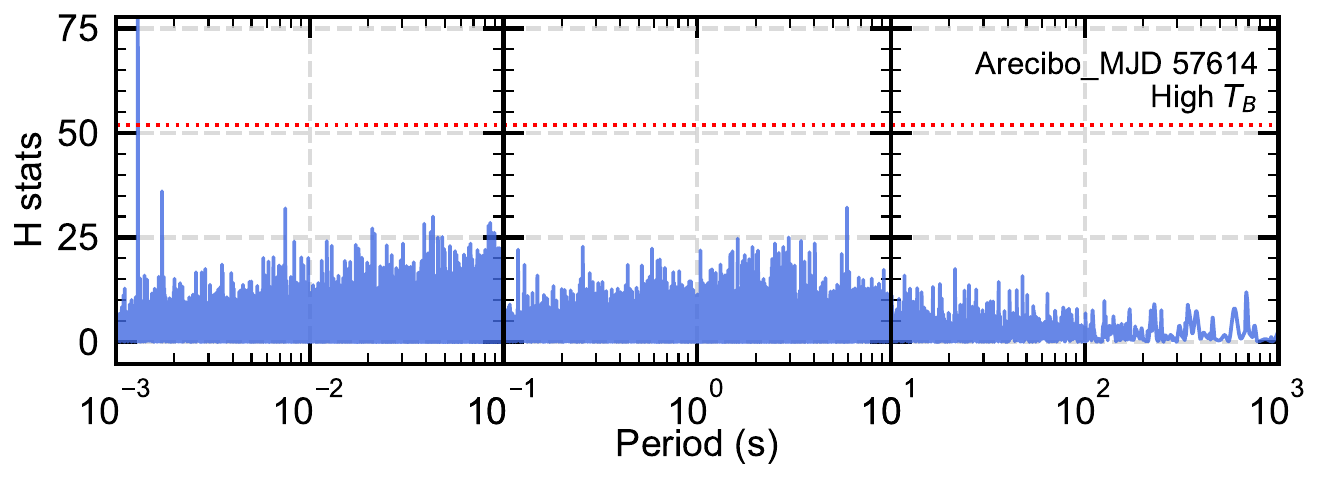}
   \includegraphics[width=0.32\textwidth]{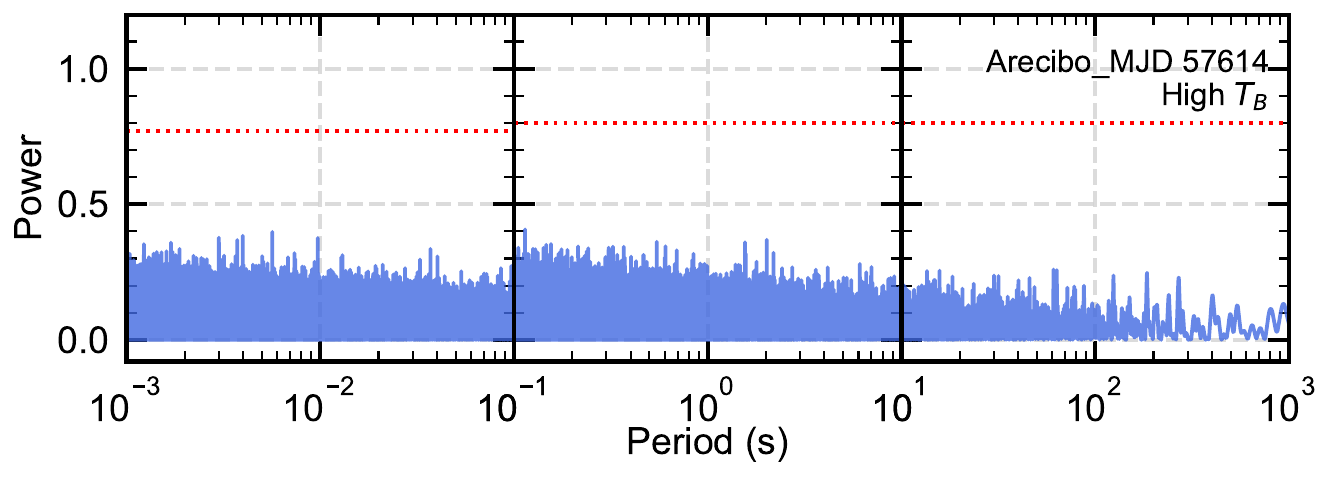}
   \includegraphics[width=0.32\textwidth]{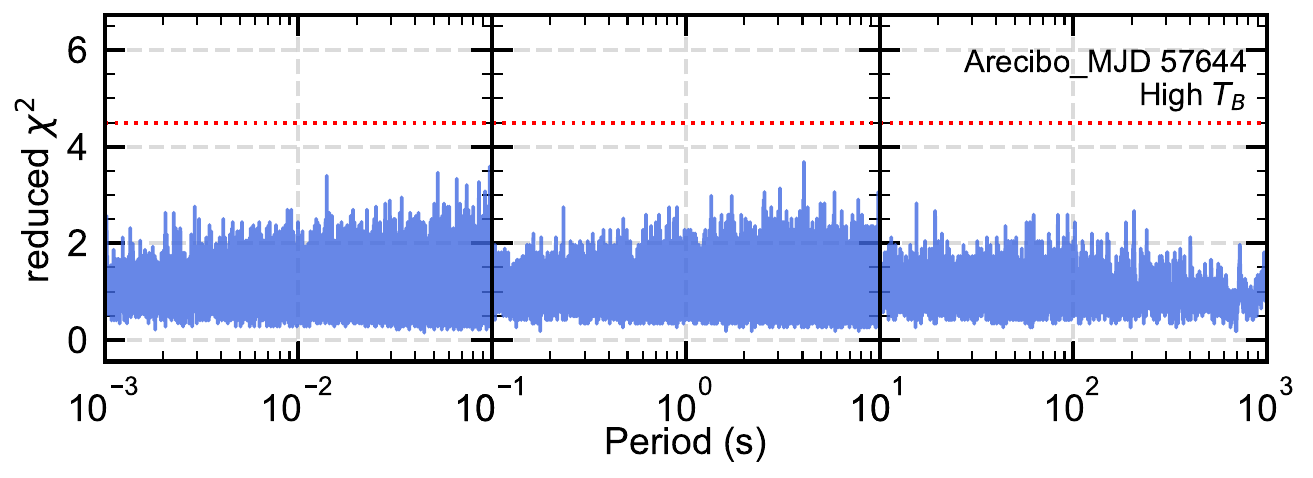}
   \includegraphics[width=0.32\textwidth]{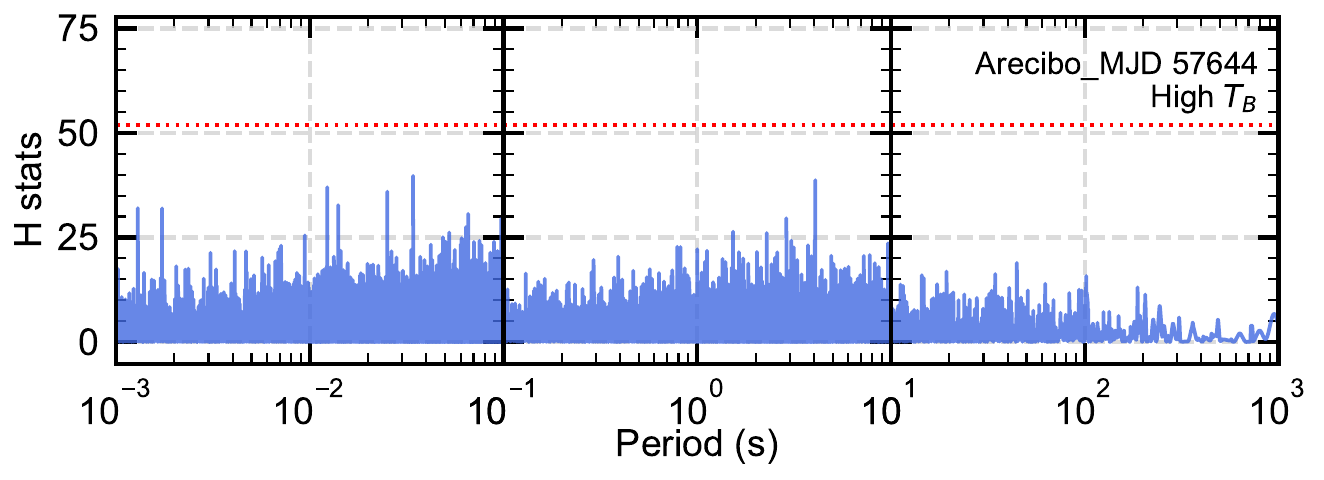}
   \includegraphics[width=0.32\textwidth]{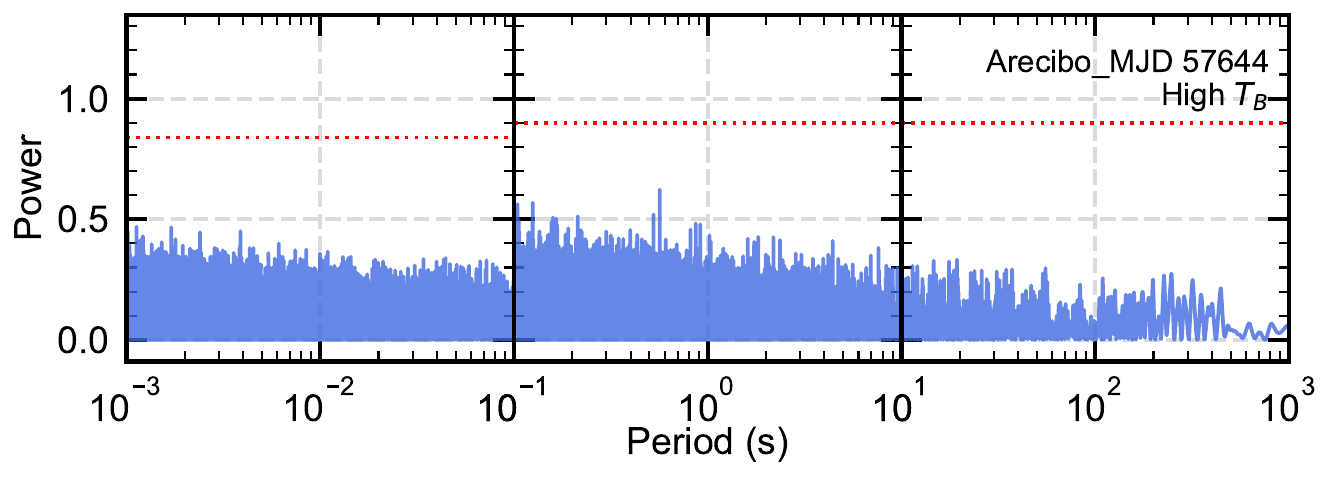}
   \includegraphics[width=0.32\textwidth]{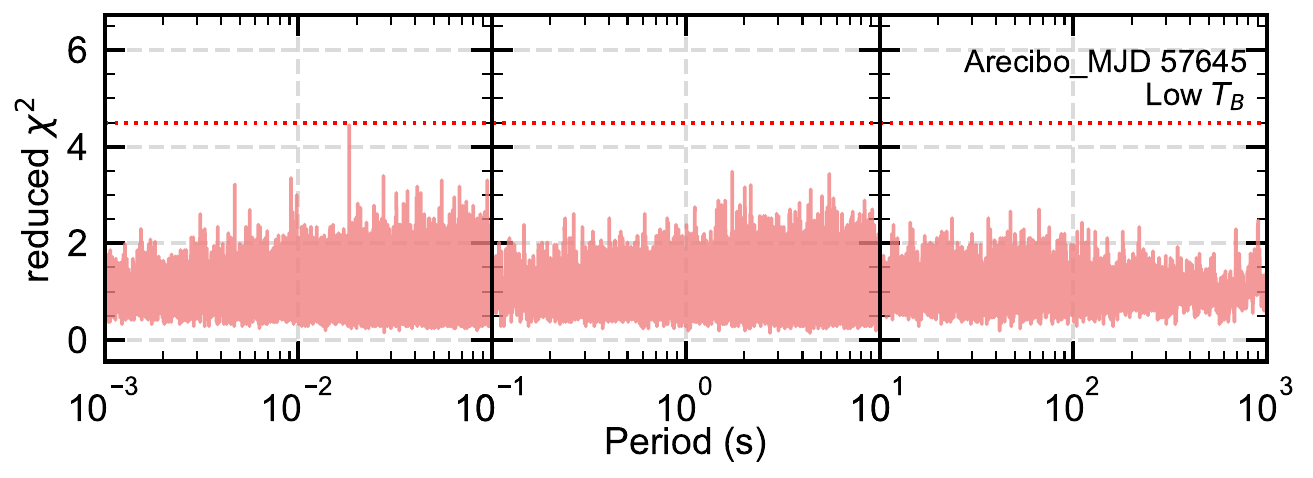}
   \includegraphics[width=0.32\textwidth]{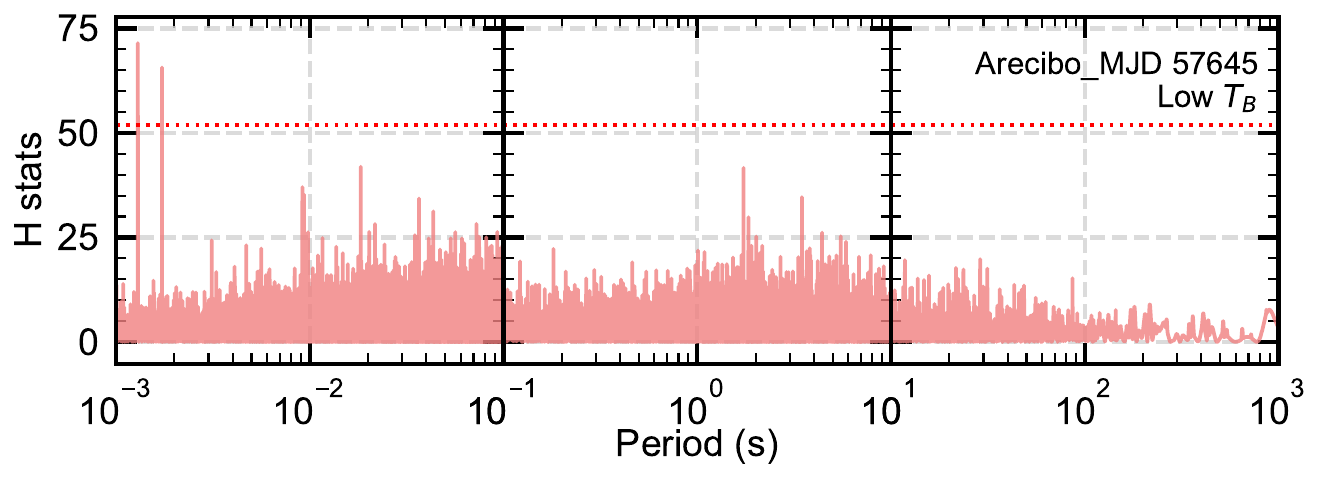}
   \includegraphics[width=0.32\textwidth]{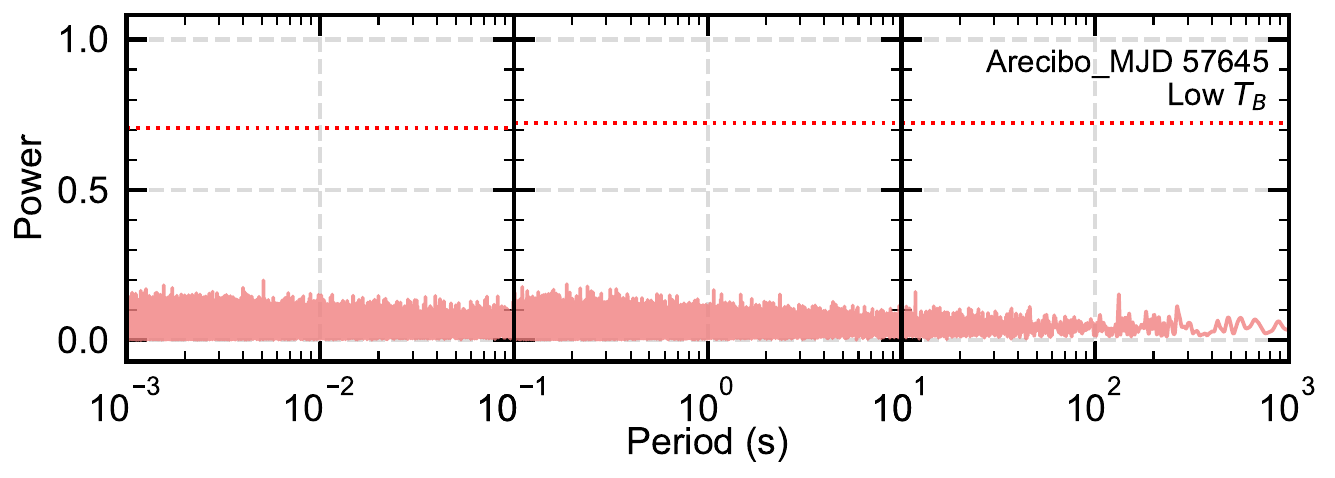}
   \includegraphics[width=0.32\textwidth]{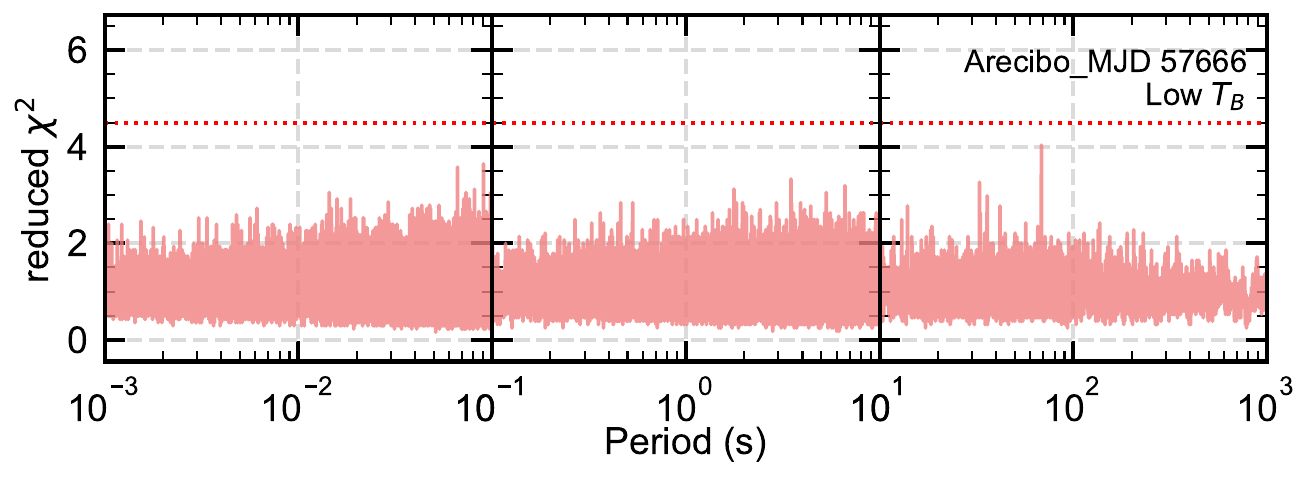}
   \includegraphics[width=0.32\textwidth]{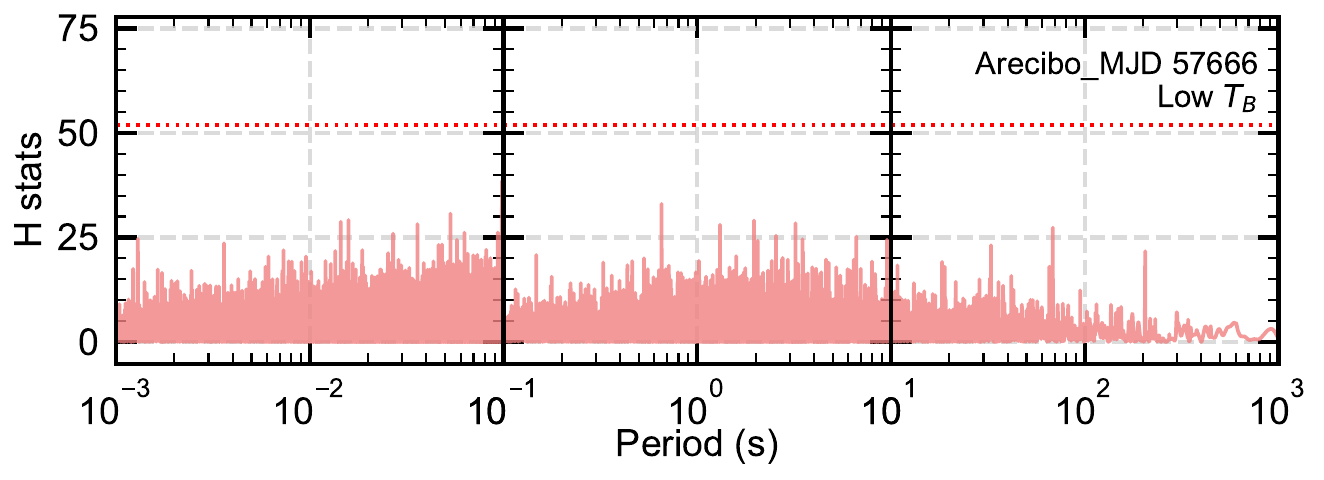}
   \includegraphics[width=0.32\textwidth]{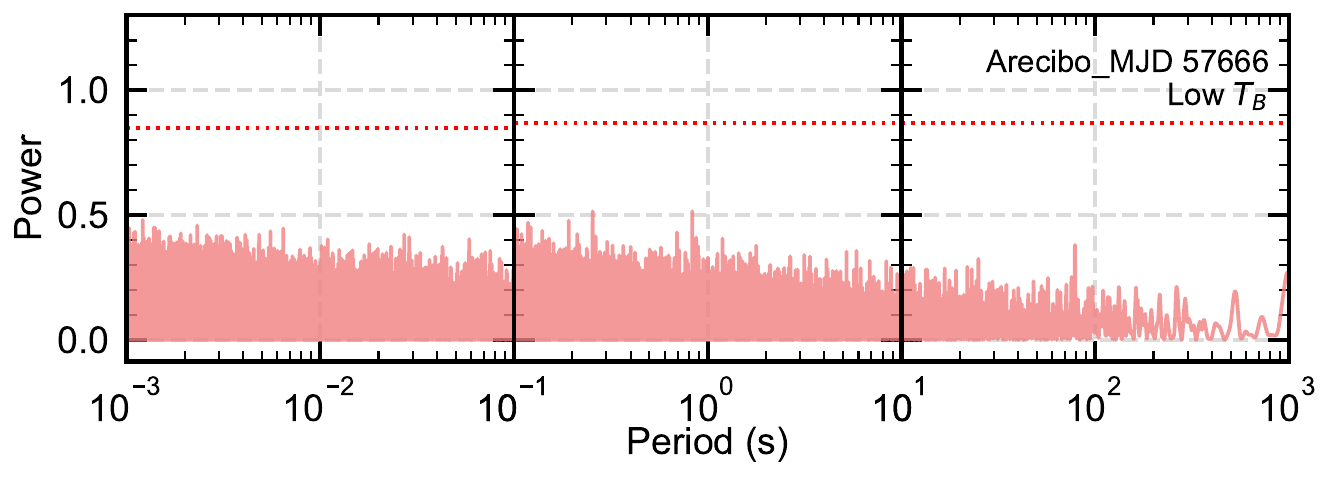}
   \caption{\footnotesize The period search results based on the Arecibo dataset \#1
   of FRB 20121102A, considering the brightness temperature of the bursts.
   The three columns from left to right correspond to the three search methods, i.e.
   the phase folding algorithm, the H-test, and the Lomb-Scargle periodogram, respectively.
   The rows from top to bottom correspond to high temperature or low temperature events
   on selected days: high $T_\mathrm{B}$ bursts on MJD 57614 and MJD 57644; and
   low $T_\mathrm{B}$ bursts on MJD 57645 and MJD 57666. In the cases of phase folding and the H-test methods,
   a horizontal dotted line is plotted to mark the p-value of $10^{-9}$. In the cases of
   Lomb-Scargle periodogram, a horizontal dotted line indicating a FAP level of $10^{-9}$ is plotted.
   No clear evidence of periodicity is found in these plots. Note that in the middle panels,
   the peak structures in the 1 ms--2 ms
   range are fake signals due to the limited timing accuracy.  }
   \label{Fig12}
\end{figure*}

For the Arecibo Dataset \#1 of FRB 20121102A, the high
$T_\mathrm{B}$ subset consists of bursts on two days: 40 bursts on
MJD 57614, of which 37 bursts have a waiting time > 0.1 s; and 33
bursts on MJD 57644, of which 27 bursts have a waiting time > 0.1
s. The low $T_\mathrm{B}$ subset of the Arecibo Dataset \#1 also
consists of bursts on two days: 48 bursts on MJD 57645, of which
46 bursts have a waiting time > 0.1 s; and 32 bursts on MJD 57666,
of which 30 bursts have a waiting time > 0.1 s. The results are
shown in Figure ~\ref{Fig12}.

\begin{figure*}
   \centering
   \includegraphics[width=0.32\textwidth]{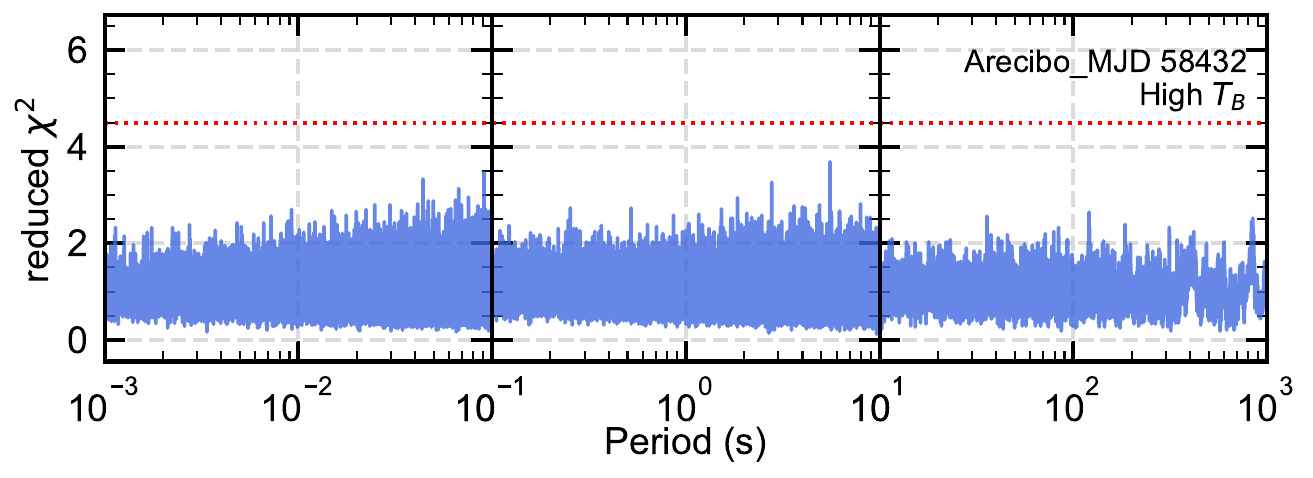}
   \includegraphics[width=0.32\textwidth]{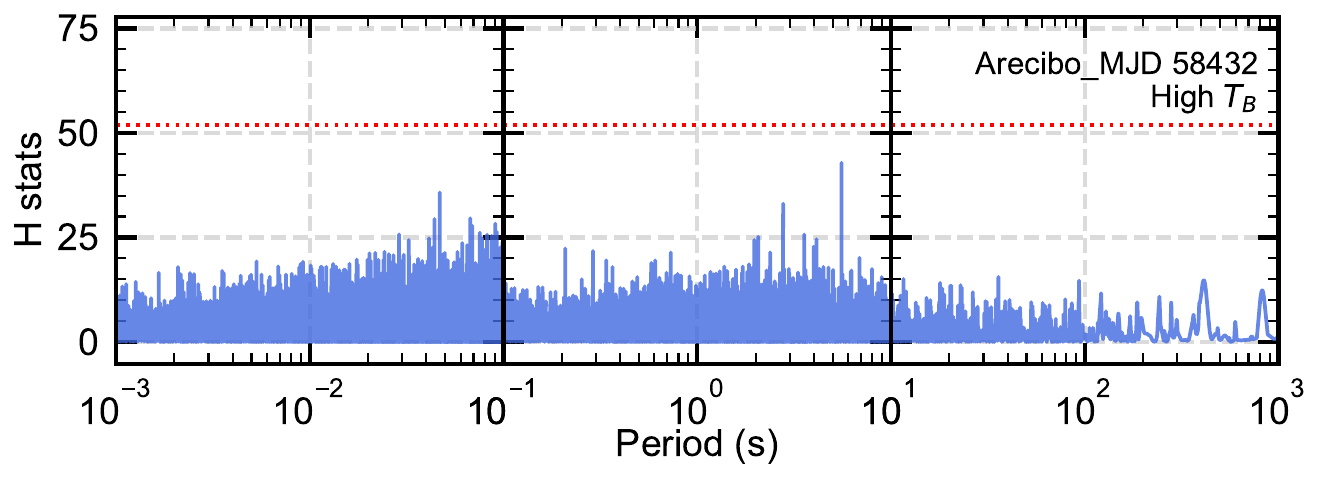}
   \includegraphics[width=0.32\textwidth]{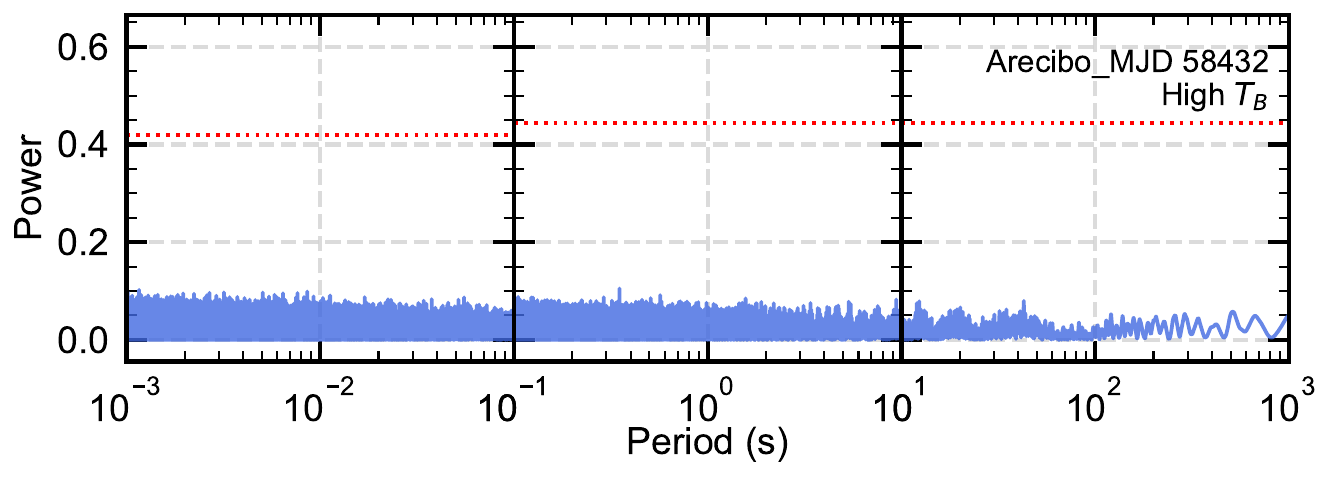}
   \includegraphics[width=0.32\textwidth]{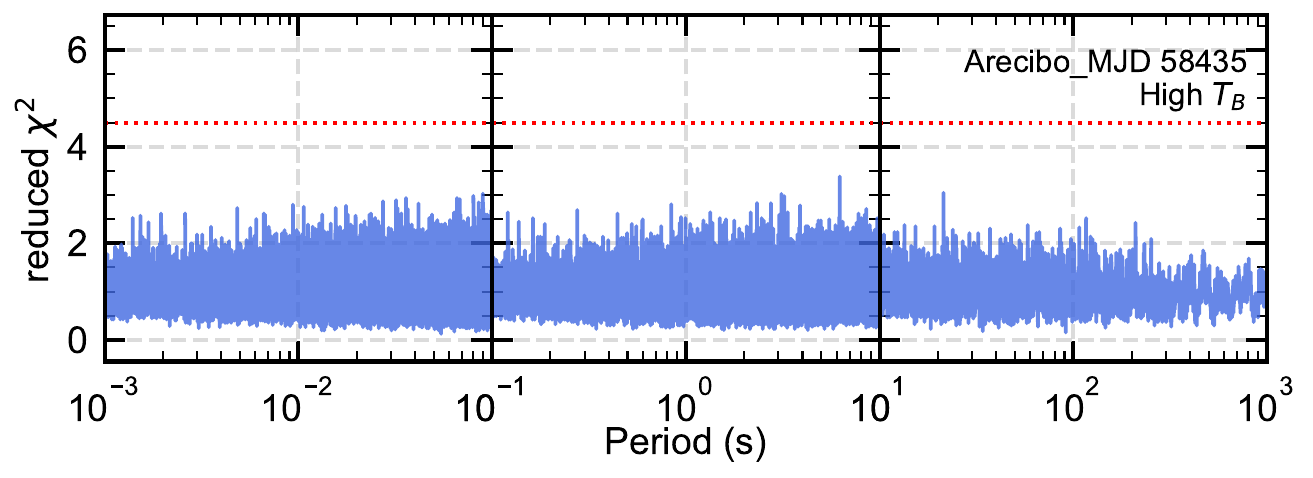}
   \includegraphics[width=0.32\textwidth]{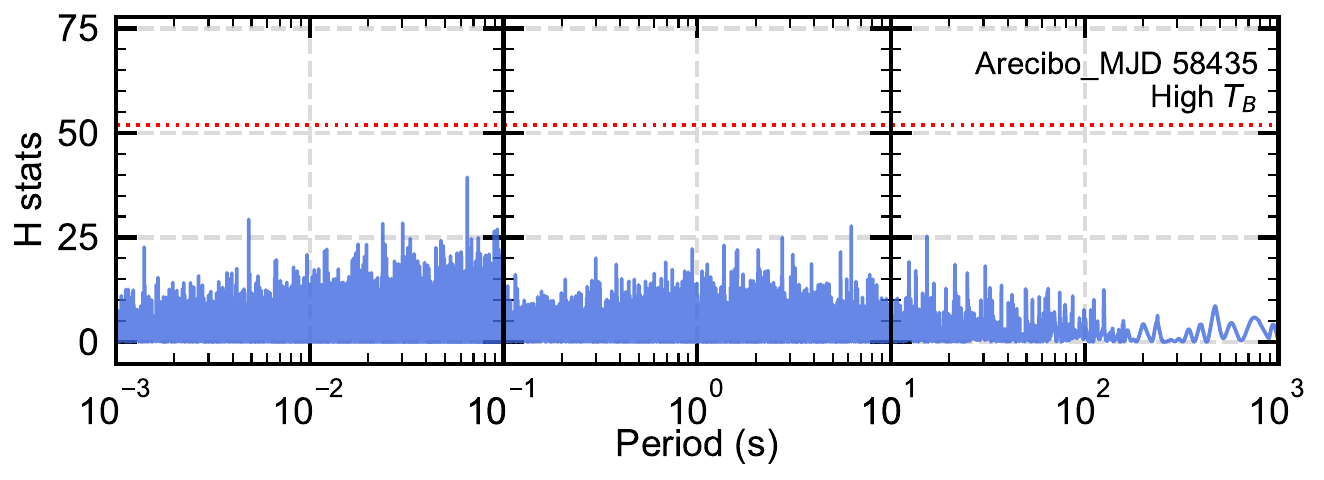}
   \includegraphics[width=0.32\textwidth]{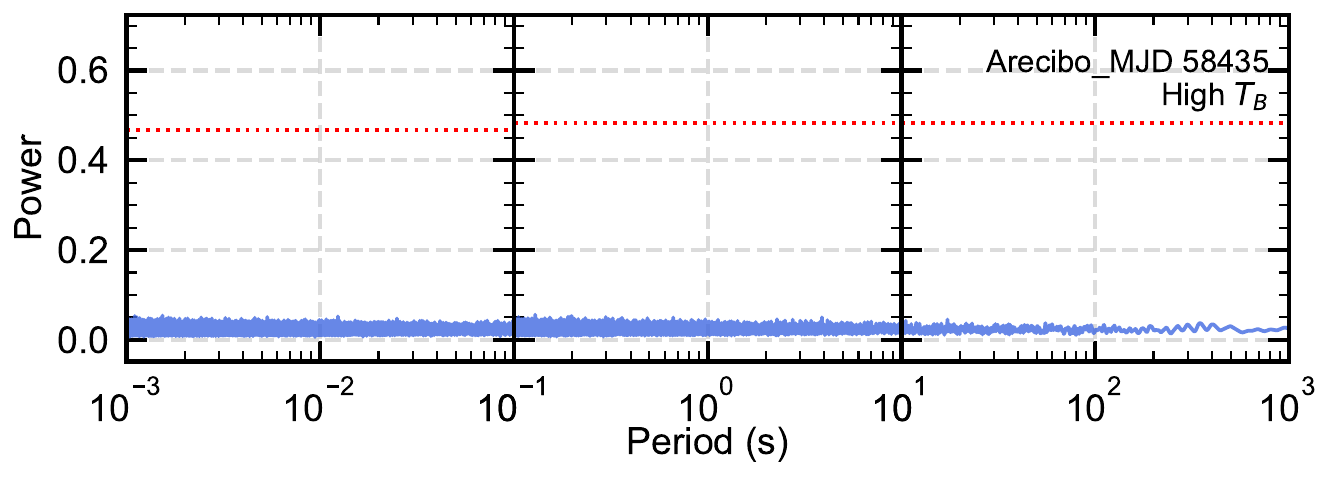}
   \includegraphics[width=0.32\textwidth]{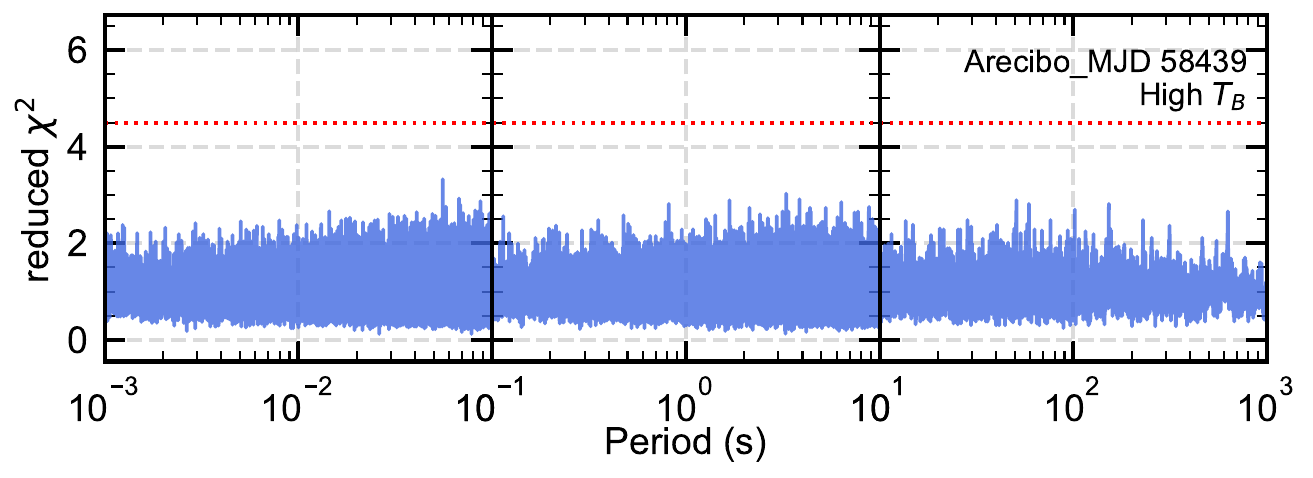}
   \includegraphics[width=0.32\textwidth]{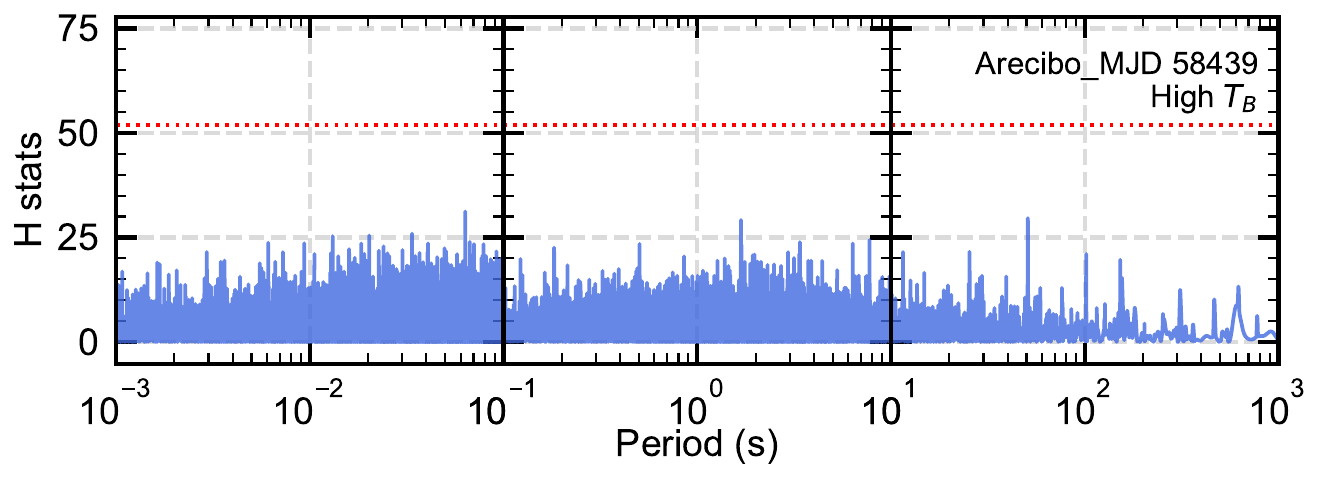}
   \includegraphics[width=0.32\textwidth]{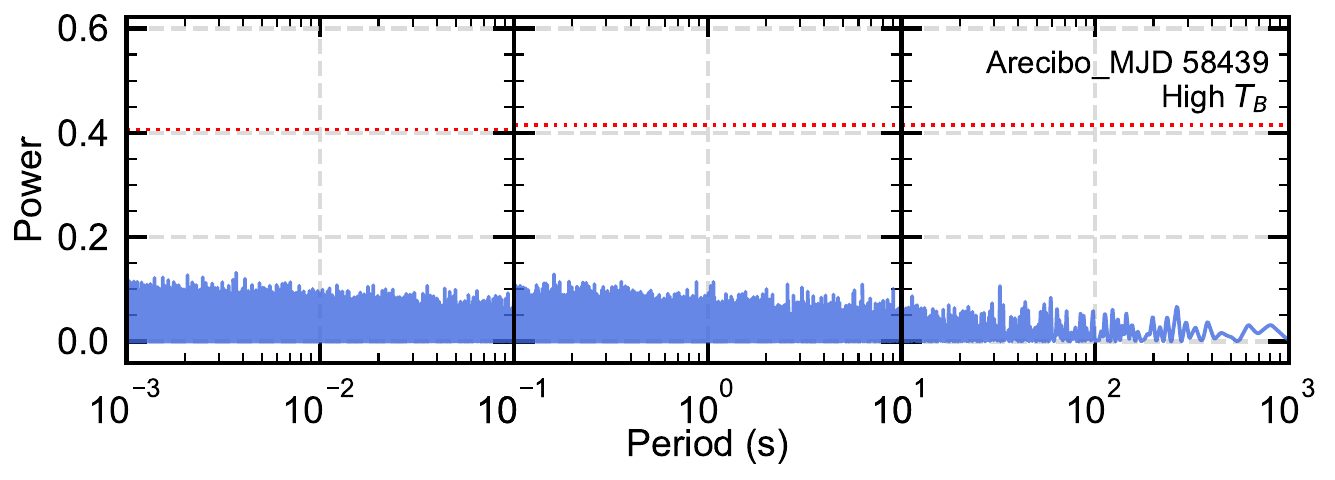}
   \includegraphics[width=0.32\textwidth]{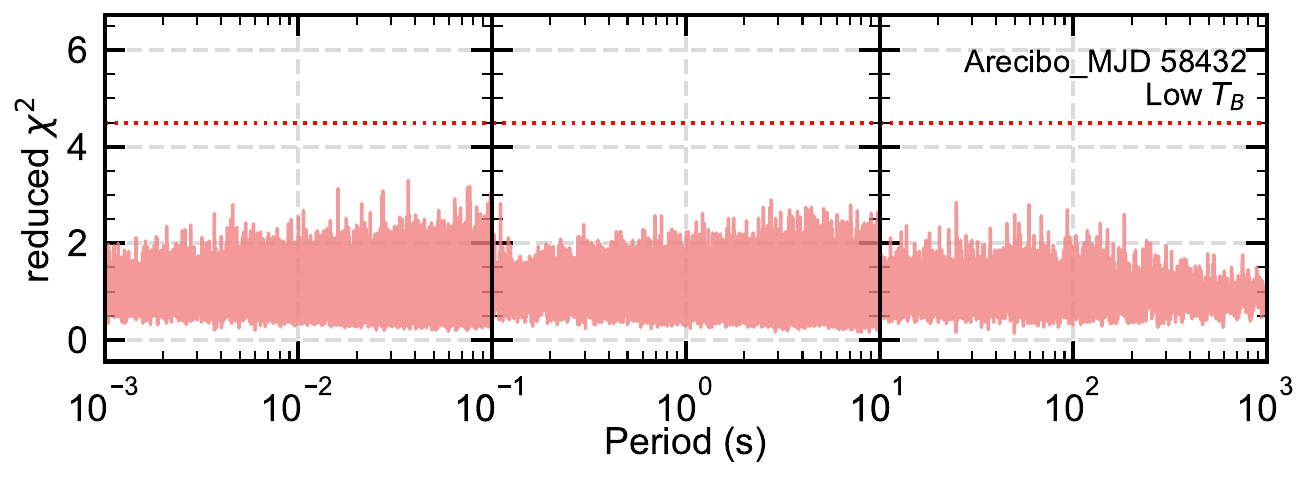}
   \includegraphics[width=0.32\textwidth]{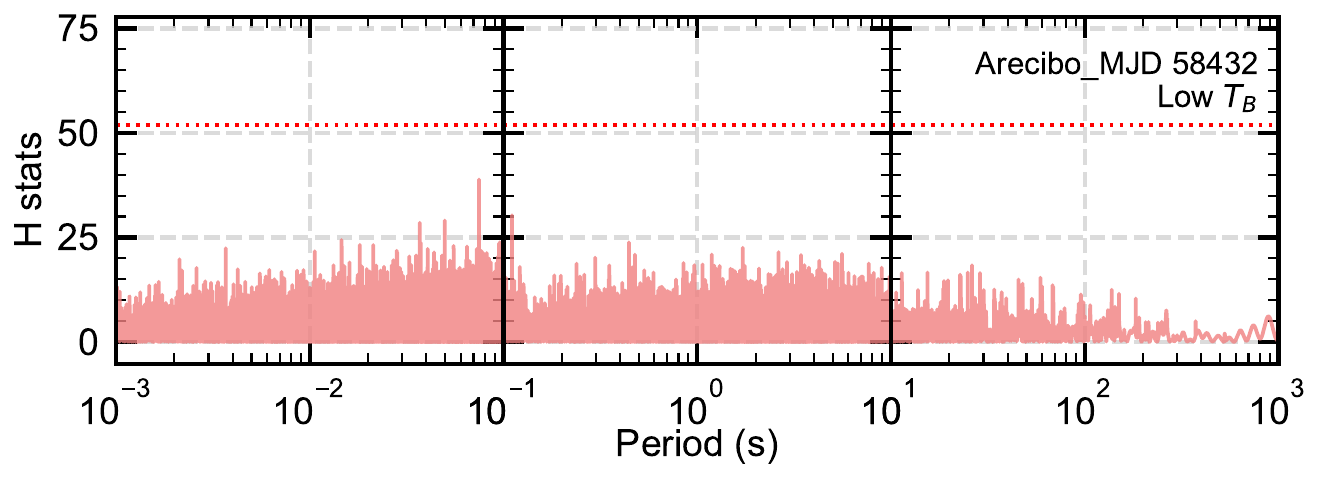}
   \includegraphics[width=0.32\textwidth]{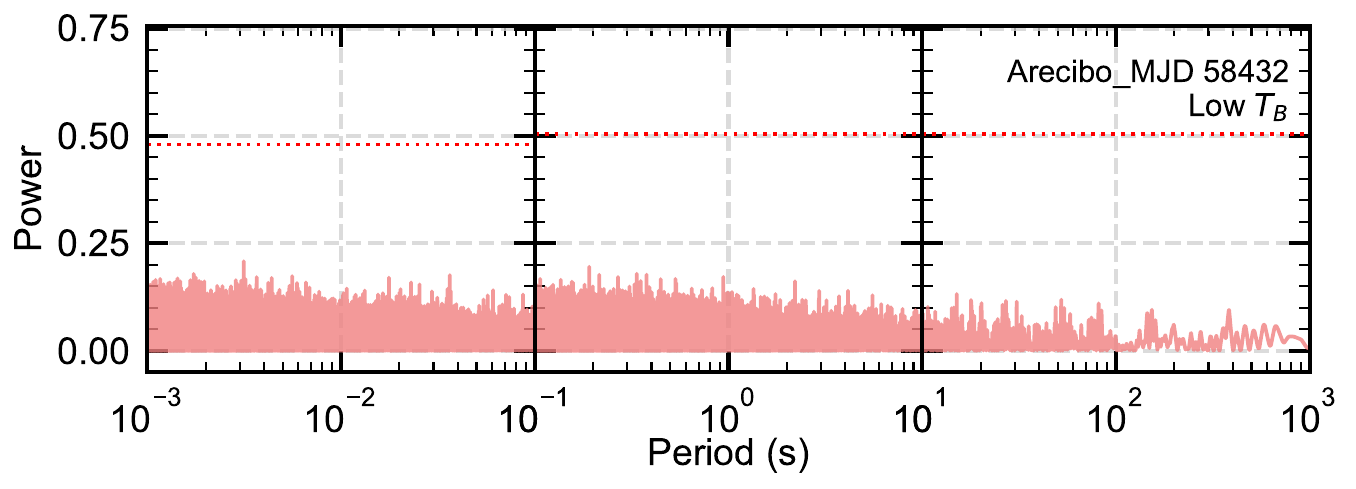}
   \includegraphics[width=0.32\textwidth]{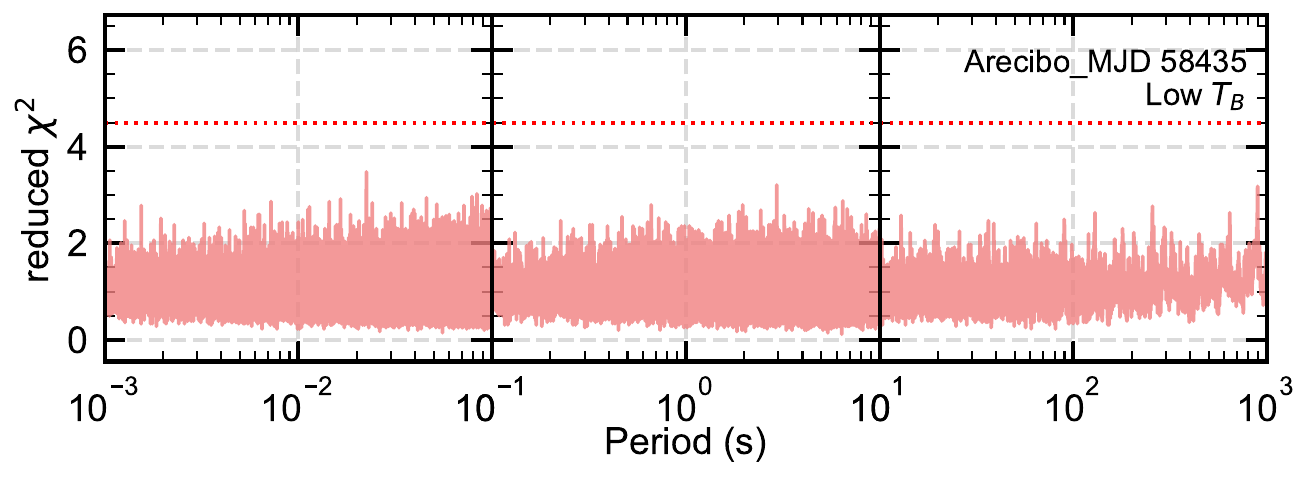}
   \includegraphics[width=0.32\textwidth]{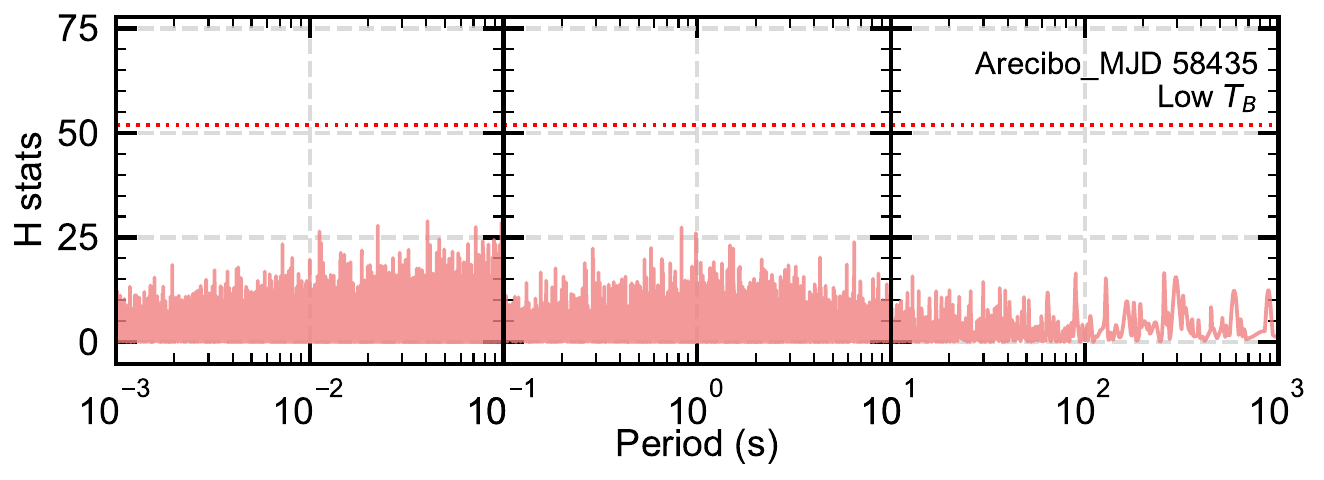}
   \includegraphics[width=0.32\textwidth]{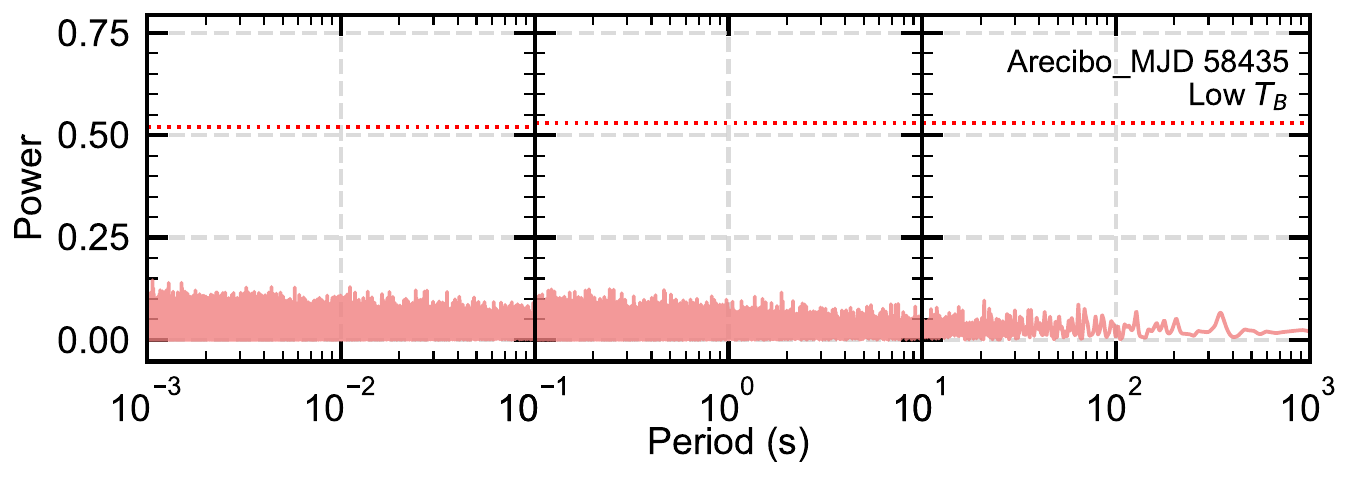}
   \includegraphics[width=0.32\textwidth]{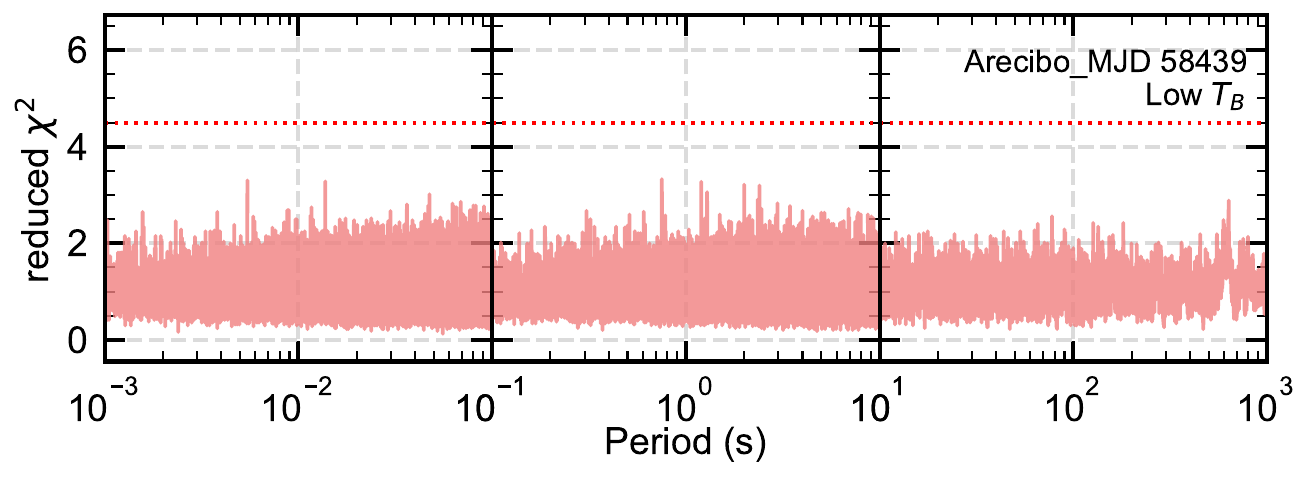}
   \includegraphics[width=0.32\textwidth]{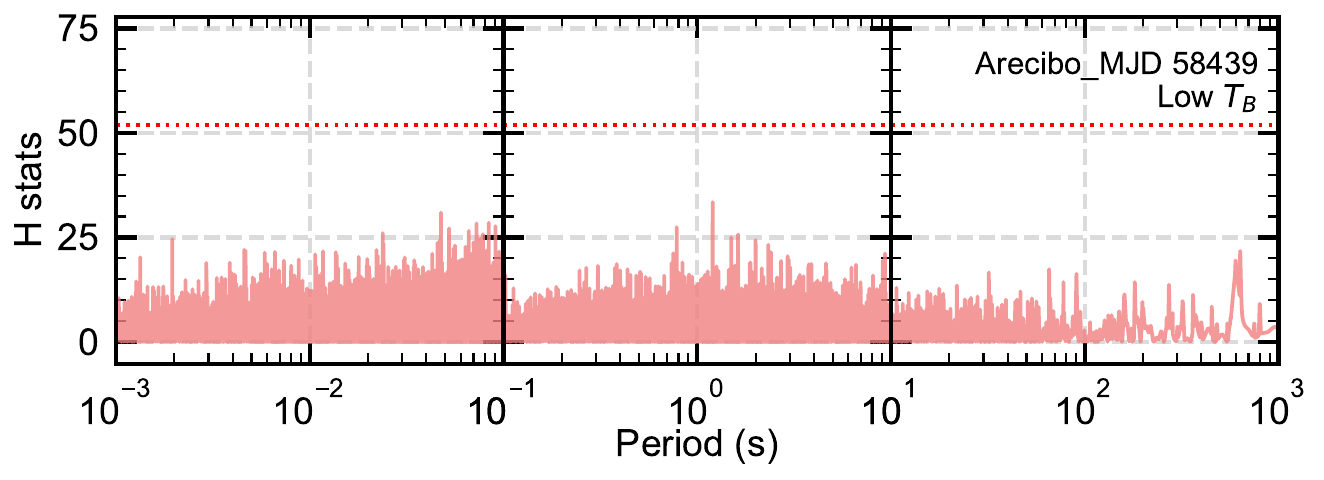}
   \includegraphics[width=0.32\textwidth]{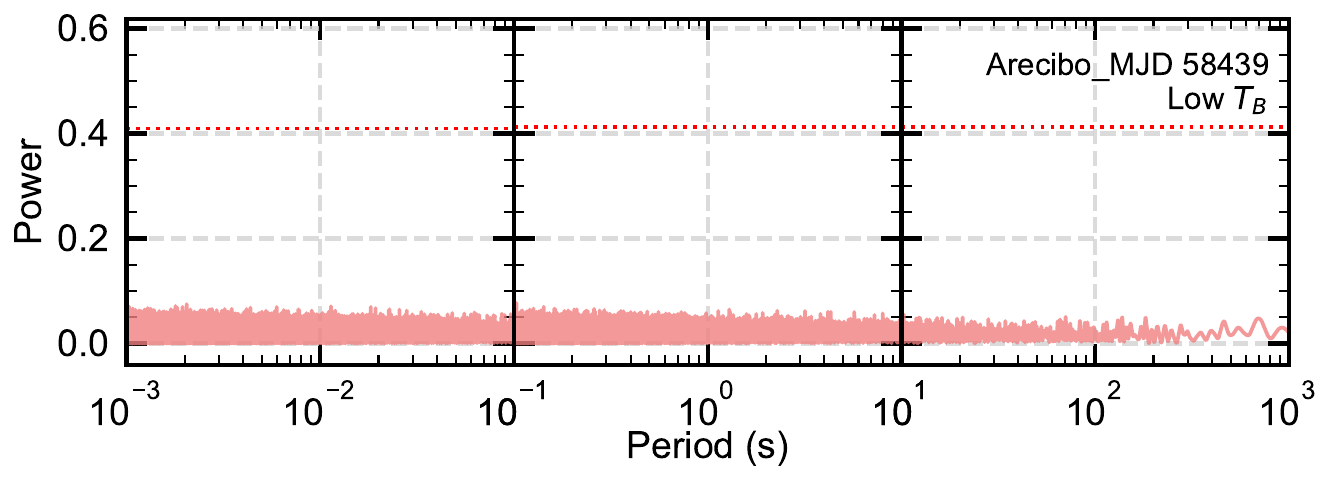}
   \caption{\footnotesize The period search results based on the Arecibo Dataset \#2
   of FRB 20121102A, considering the brightness temperature of the bursts.
   The three columns from left to right correspond to the three search methods,
   i.e. the phase folding algorithm, the H-test, and the Lomb-Scargle periodogram, respectively.
   The rows from top to bottom correspond to high temperature or low temperature events
   on selected days: high $T_\mathrm{B}$ bursts on MJD 58432, MJD 58435 and MJD 58439;  and
   low $T_\mathrm{B}$ bursts on MJD 58432, MJD 58435 and MJD 58439. In the cases of phase
   folding and the H-test methods, a horizontal dotted line is plotted to mark
   the p-value of $10^{-9}$. In the cases of Lomb-Scargle periodogram,
   a horizontal dotted line indicating a FAP level of $10^{-9}$ is plotted.
   No clear evidence of periodicity is found in these plots.}
   \label{Fig13}
\end{figure*}

For the Arecibo Dataset \#2 of FRB 20121102A, the high
$T_\mathrm{B}$ subset consists of bursts on three days: 107 bursts
on MJD 58432, of which 99 bursts have a waiting time > 0.1 s; 92
bursts on MJD 58435, of which 88 bursts have a waiting time > 0.1
s; and 111 bursts on MJD 58439, of which 108 bursts have a waiting
time > 0.1 s. The low $T_\mathrm{B}$ subset of the Arecibo Dataset
\#2 also consists of bursts on three days: 89 bursts on MJD 58432,
of which 83 bursts have a waiting time > 0.1 s; 79 bursts on MJD
58435, of which 77 bursts have a waiting time > 0.1 s; and 110
bursts on MJD 58439, of which 109 bursts have a waiting time > 0.1
s. The results are shown in Figure ~\ref{Fig13}.

\begin{figure*}
   \centering
   \includegraphics[width=0.32\textwidth]{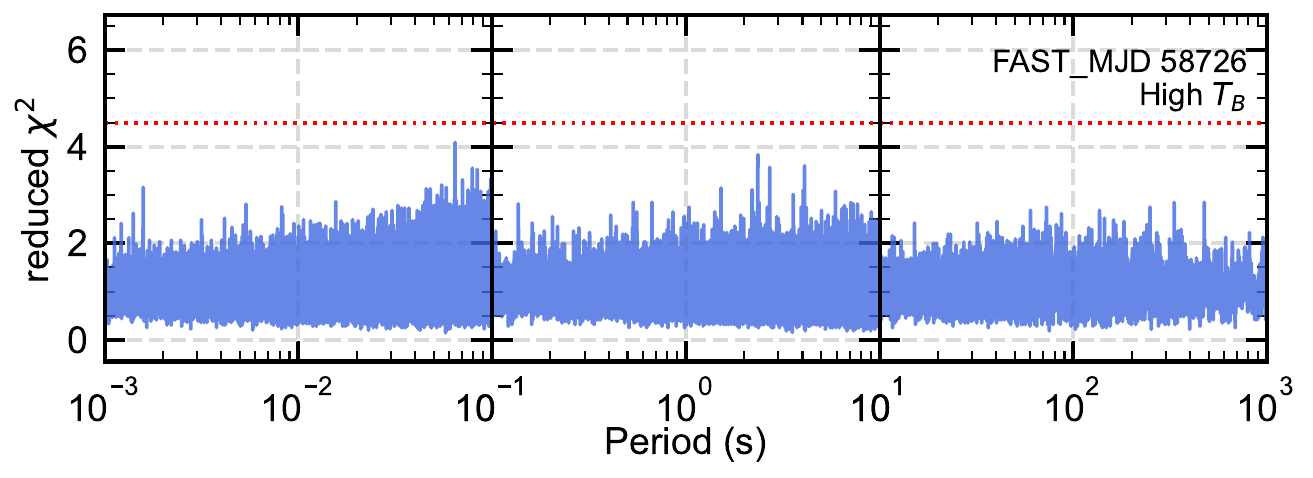}
   \includegraphics[width=0.32\textwidth]{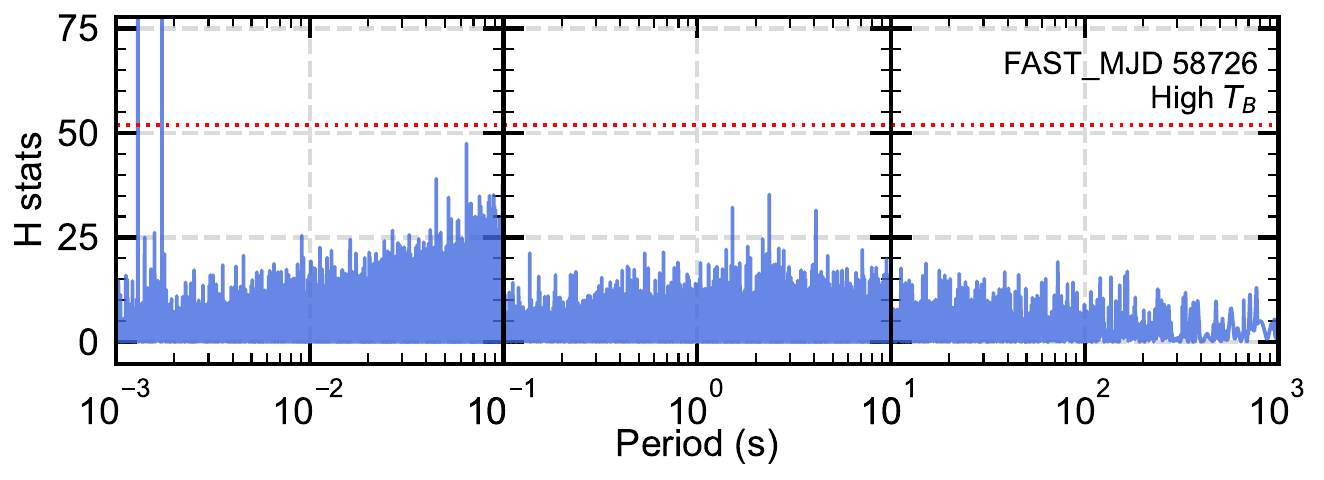}
   \includegraphics[width=0.32\textwidth]{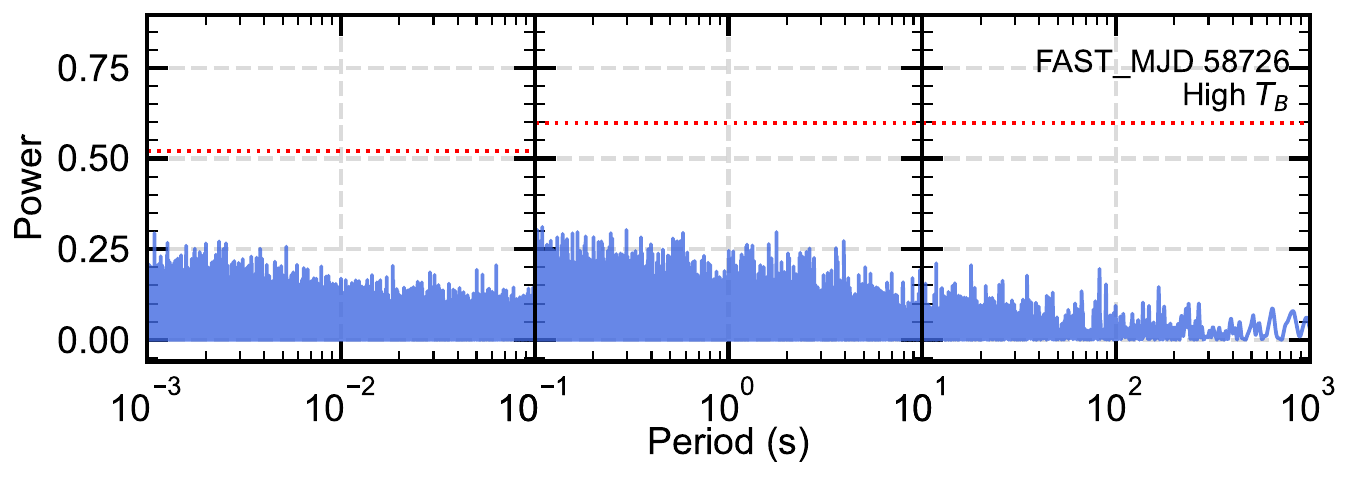}
   \includegraphics[width=0.32\textwidth]{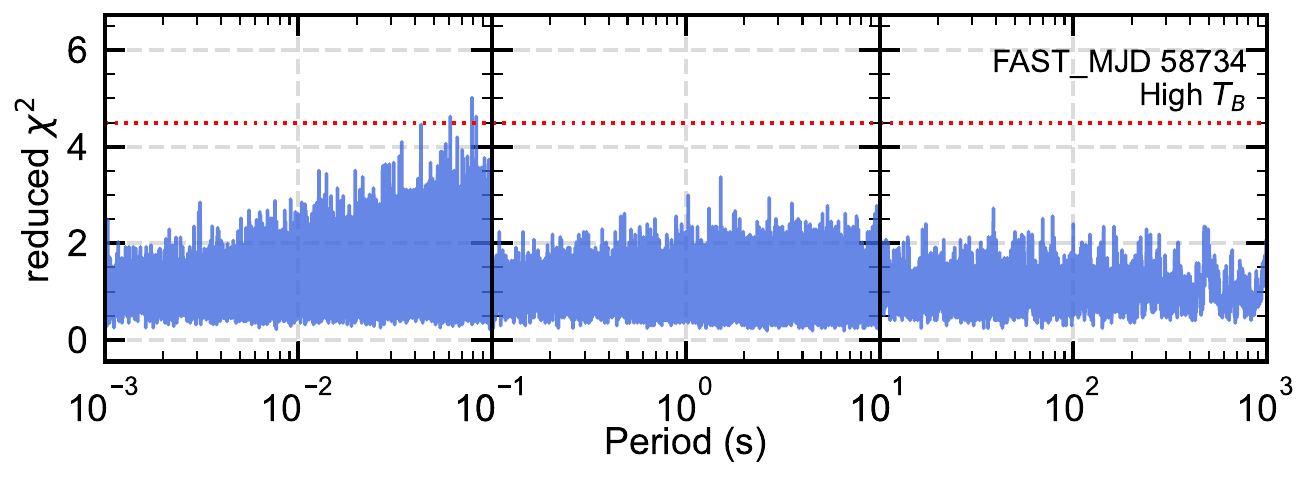}
   \includegraphics[width=0.32\textwidth]{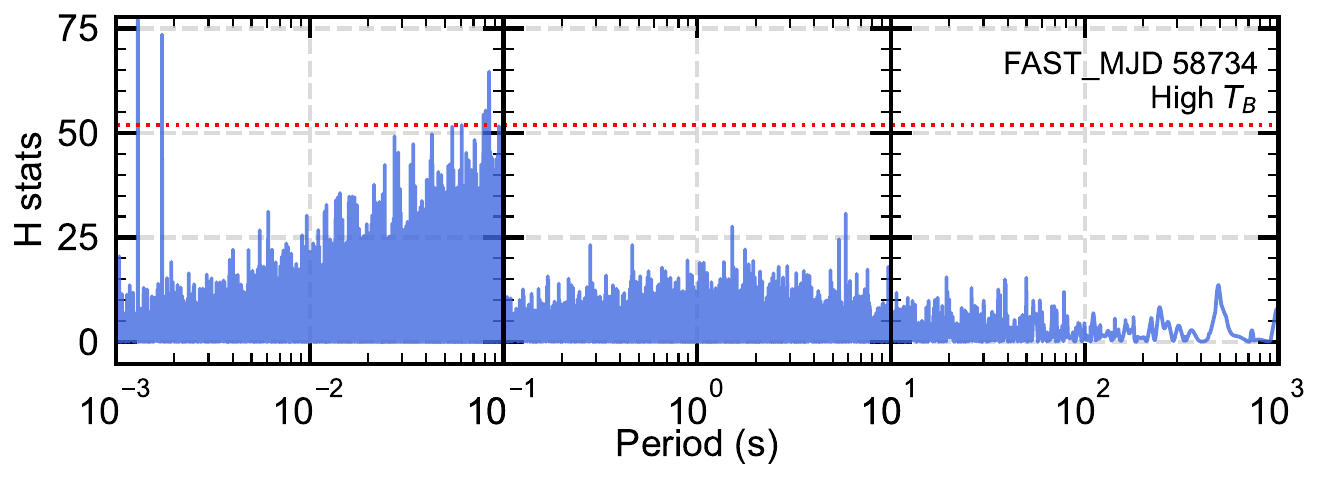}
   \includegraphics[width=0.32\textwidth]{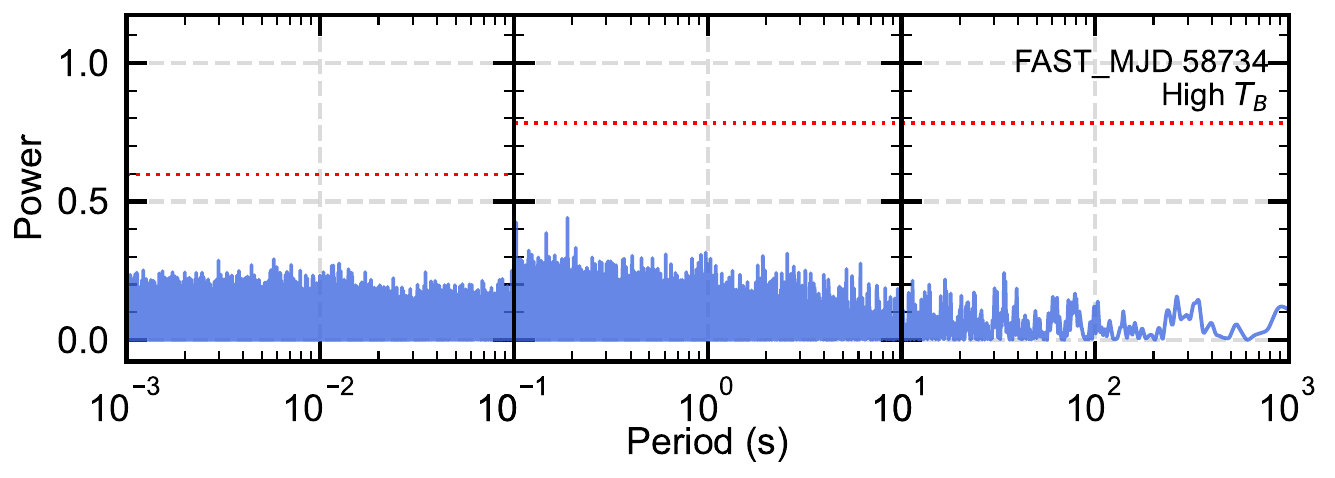}
   \includegraphics[width=0.32\textwidth]{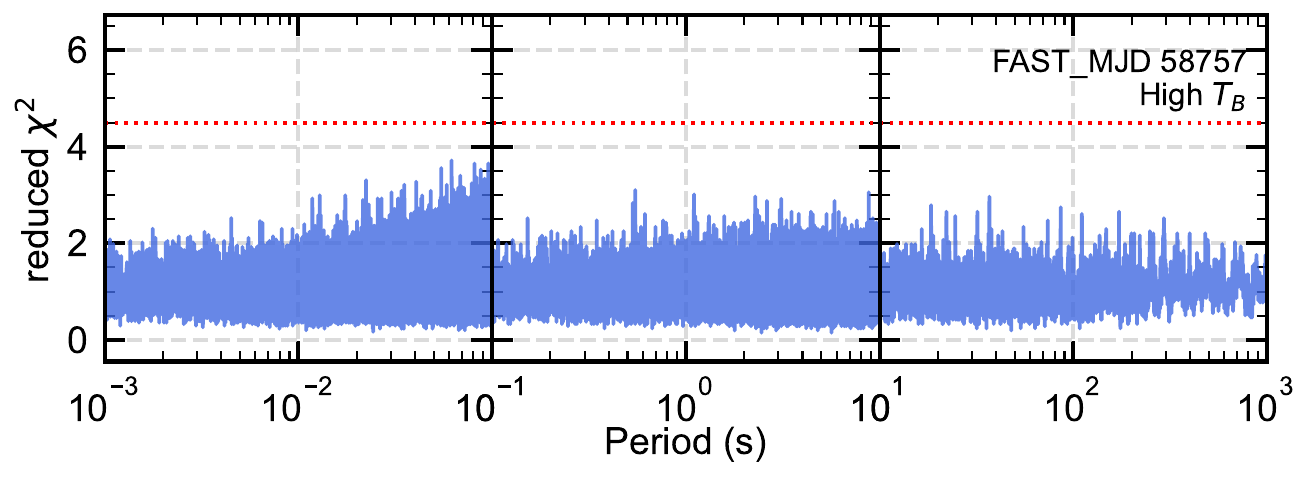}
   \includegraphics[width=0.32\textwidth]{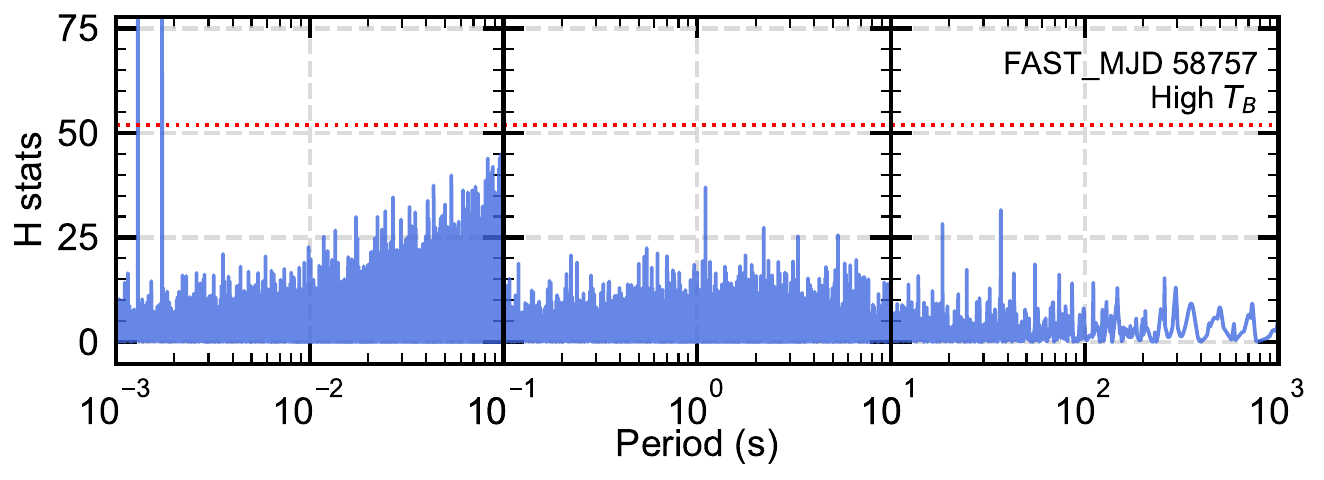}
   \includegraphics[width=0.32\textwidth]{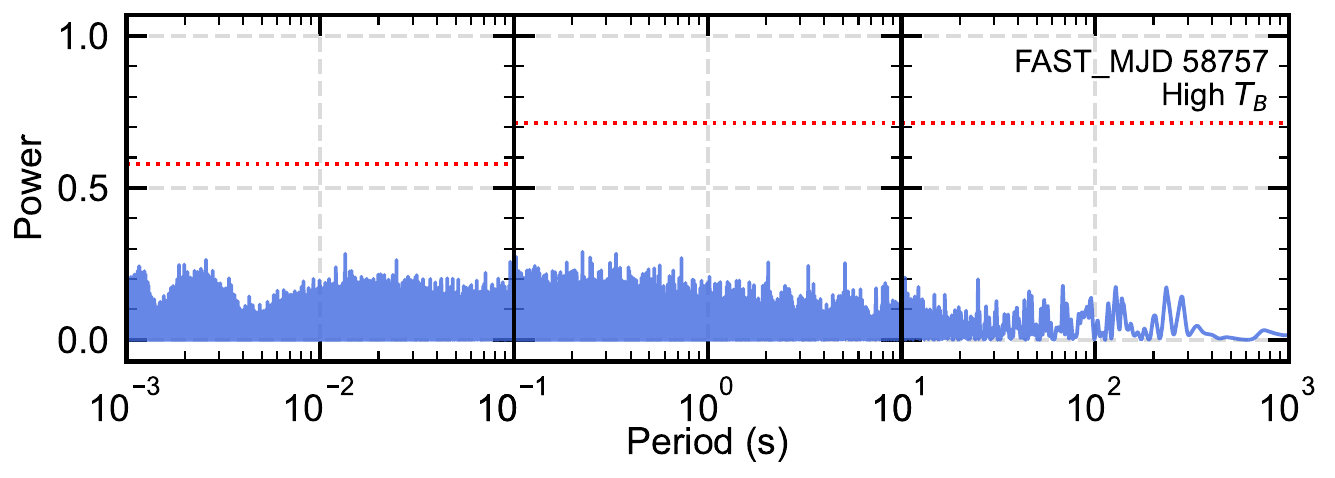}
   \includegraphics[width=0.32\textwidth]{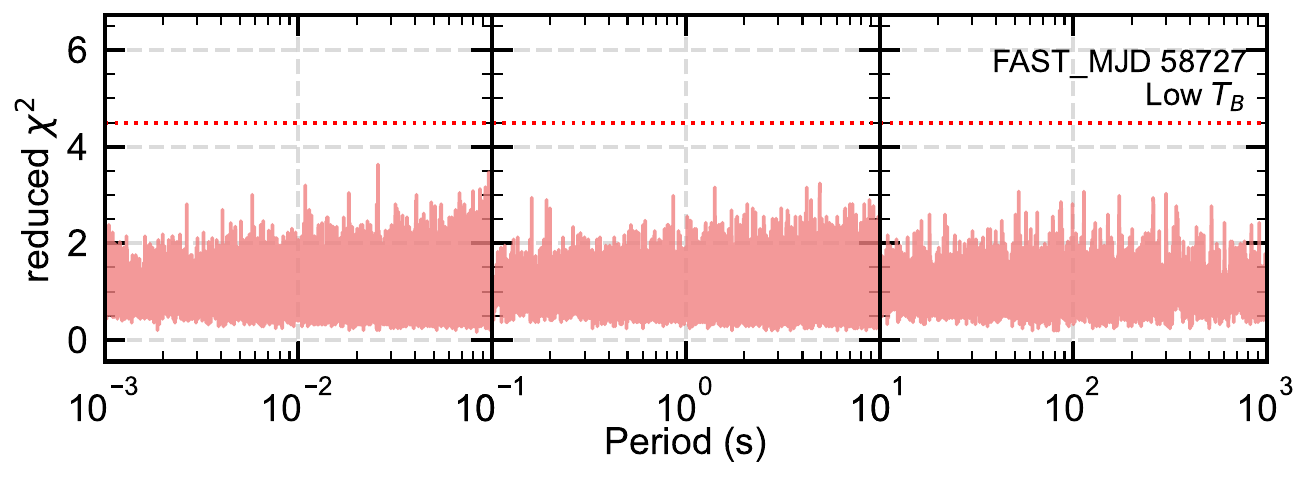}
   \includegraphics[width=0.32\textwidth]{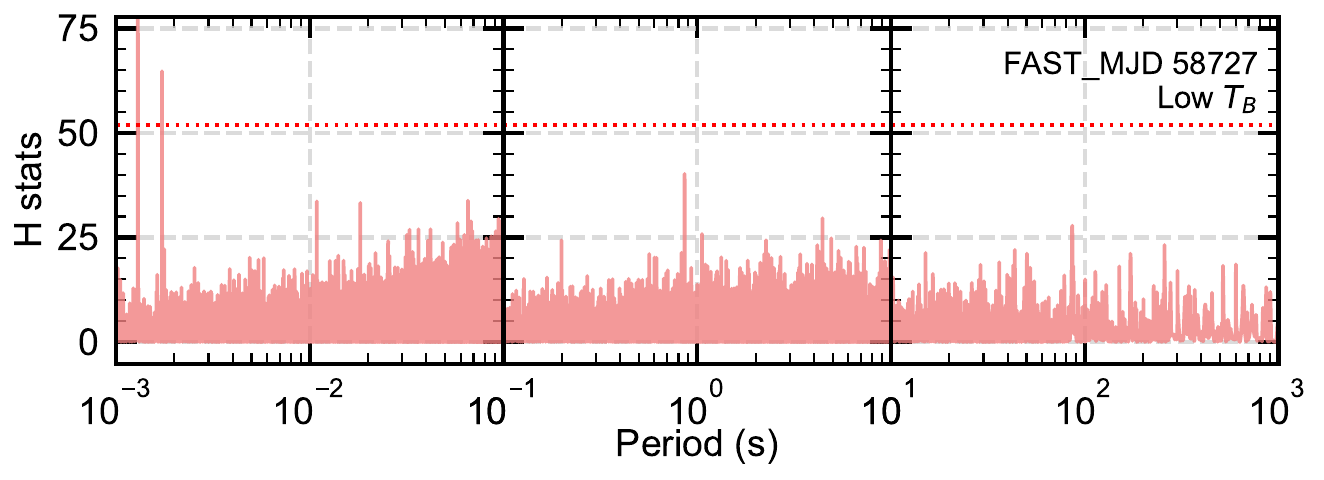}
   \includegraphics[width=0.32\textwidth]{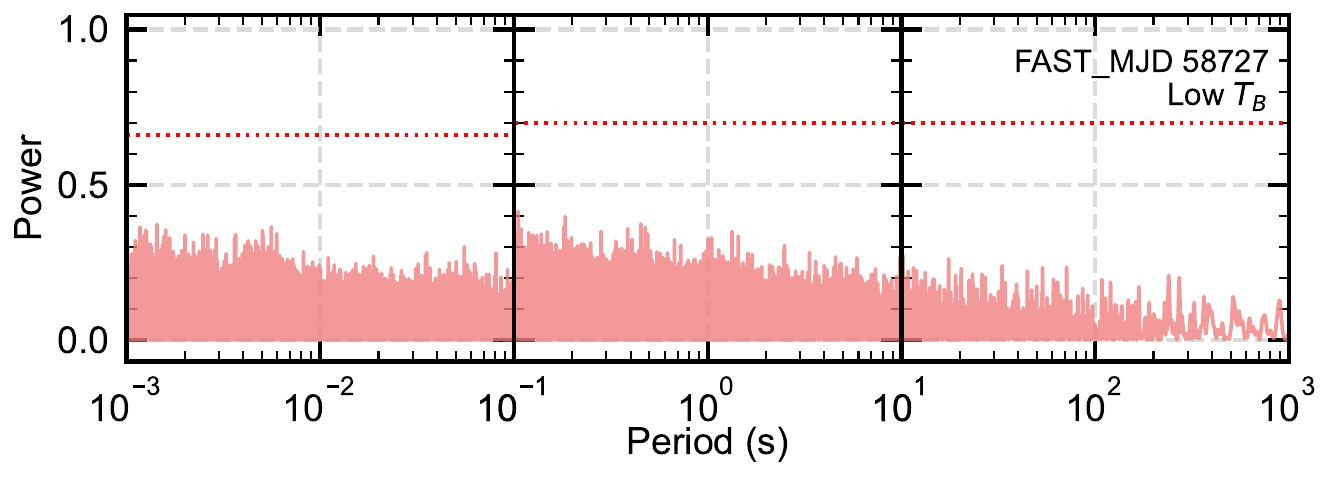}
   \includegraphics[width=0.32\textwidth]{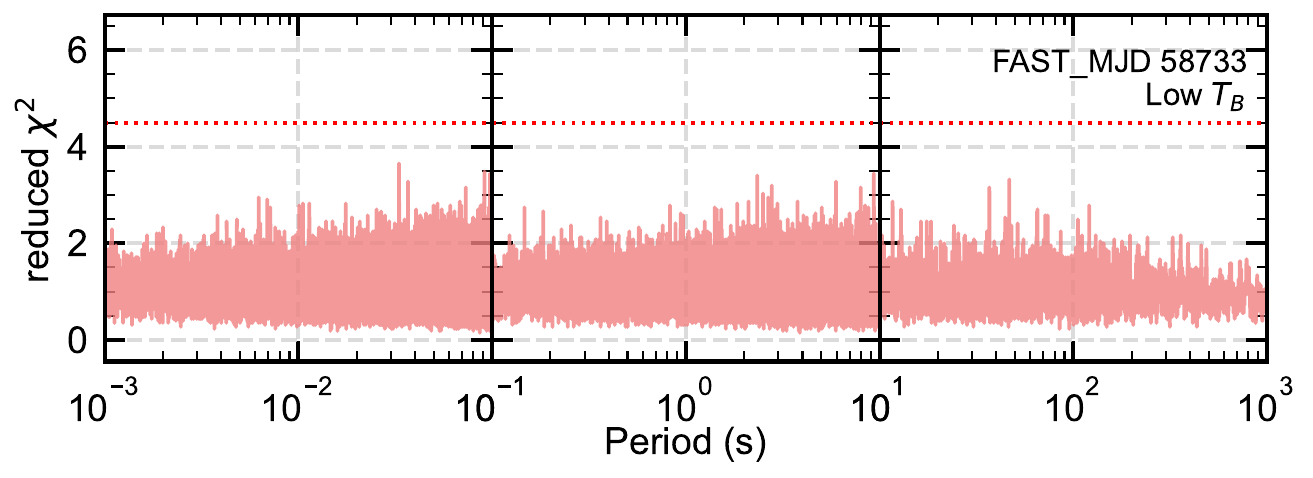}
   \includegraphics[width=0.32\textwidth]{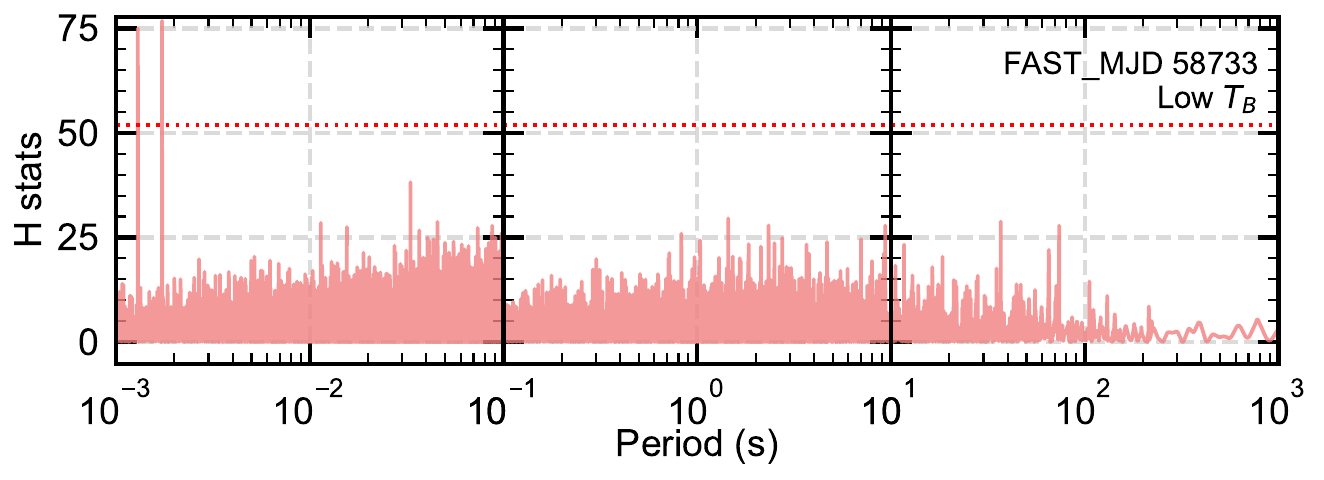}
   \includegraphics[width=0.32\textwidth]{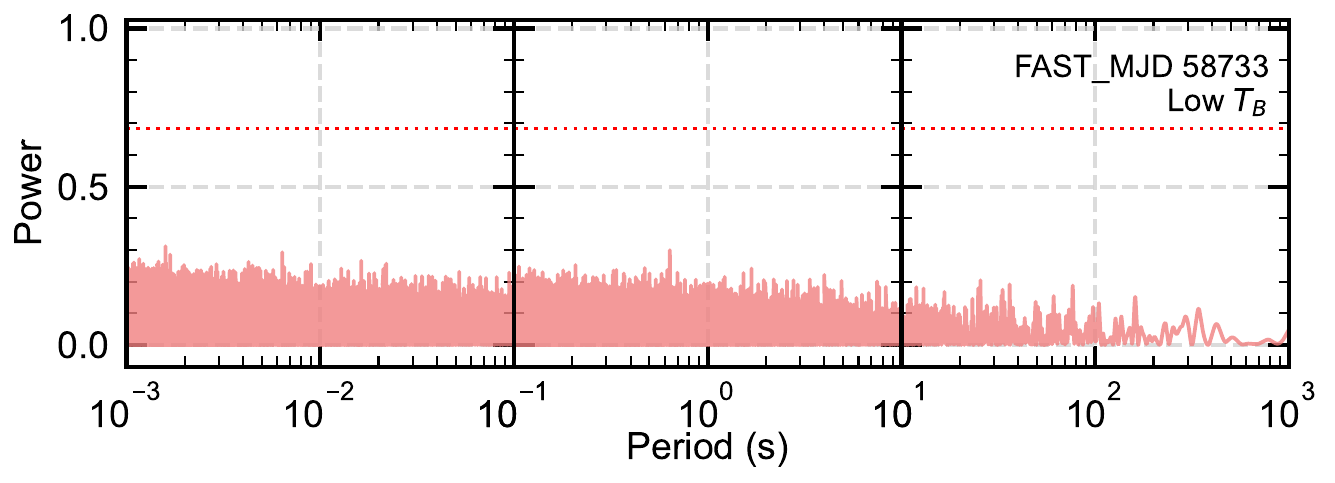}
   \includegraphics[width=0.32\textwidth]{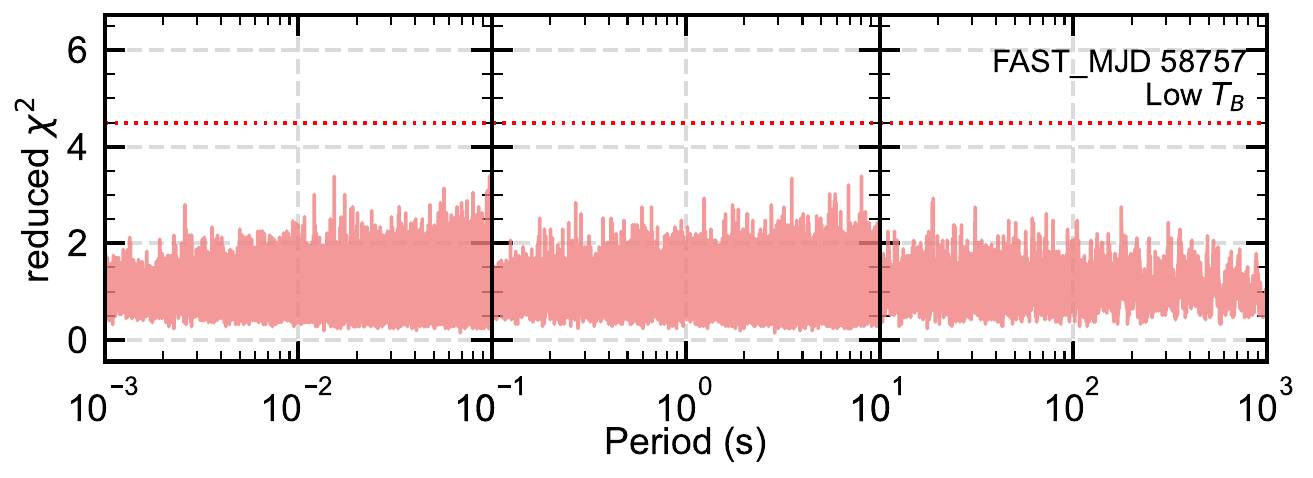}
   \includegraphics[width=0.32\textwidth]{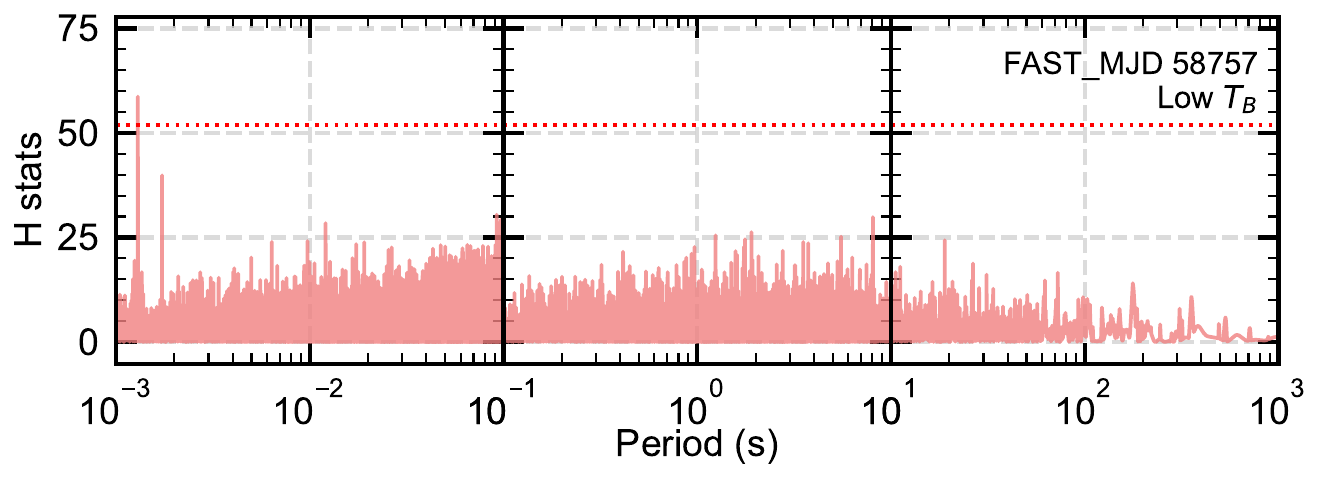}
   \includegraphics[width=0.32\textwidth]{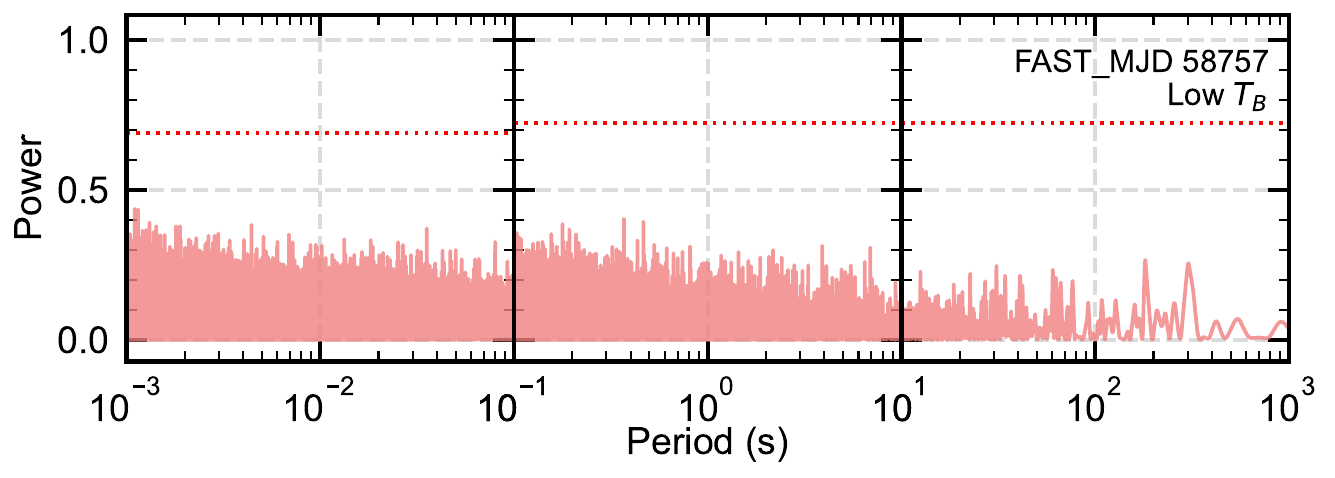}
   \caption{\footnotesize The period search results based on the FAST dataset \#3
   of FRB 20121102A, considering the brightness temperature of the bursts.
   The three columns from left to right correspond to the three search methods, i.e.
   the phase folding algorithm, the H-test, and the Lomb-Scargle periodogram, respectively.
   The rows from top to bottom correspond to high temperature or low temperature events
   on selected days: high $T_\mathrm{B}$ bursts on MJD 58726, MJD 58734 and MJD 58757; and
   low $T_\mathrm{B}$ bursts on MJD 58727, MJD 58733 and MJD 58757.
   In the cases of phase folding and the H-test methods, a horizontal dotted line is
   plotted to mark the p-value of $10^{-9}$. In the cases of Lomb-Scargle periodogram,
   a horizontal dotted line indicating a FAP level of $10^{-9}$ is plotted. No clear
   evidence of periodicity is found in these plots. Note that in the middle panels,
   the peak structures in the 1 ms--2 ms
   range are fake signals due to the limited timing accuracy. }
   \label{Fig14}
\end{figure*}

For the FAST Dataset \#3 of FRB 20121102A, the high
$T_\mathrm{B}$ subset consists of bursts on three days: 79 bursts
on MJD 58726, of which 64 bursts have a waiting time > 0.1 s; 64
bursts on MJD 58734, of which 39 bursts have a waiting time > 0.1
s; and 67 bursts on MJD 58757, of which 47 bursts have a waiting
time > 0.1 s. The low $T_\mathrm{B}$ subset of the FAST Dataset
\#3 also consists of bursts on three days: 54 bursts on MJD 58727,
of which 49 bursts have a waiting time > 0.1 s; 51 bursts on MJD
58733, all with the waiting time > 0.1 s; and 50 bursts on MJD
58757, of which 46 bursts have a waiting time > 0.1 s. The
numerical results are shown in Figure ~\ref{Fig14}.

In the 1--2 ms range, several extremely high peaks could
be observed, which are all harmonics of 0.864 ms ($10^{-8}$ d).
They are fake signals due to limited timing accuracy of
observations. In each plot, we have also examined by hand the
phase-folded histograms at the positions that correspond to the
relatively high peaks in the $\chi^2$/H-statistic/Lomb-Scargle
periodograms, but found that no marked features appear in the
histograms. From Figures ~\ref{Fig12}, ~\ref{Fig13}, and
~\ref{Fig14}, we can clearly see that no periodicity significantly
exceeding the confidence level is found in the whole range of
0.001--1000 s either in high $T_\mathrm{B}$ or in low
$T_\mathrm{B}$ bursts from FRB 20121102A.

\begin{figure*}
   \centering
   \includegraphics[width=0.32\textwidth]{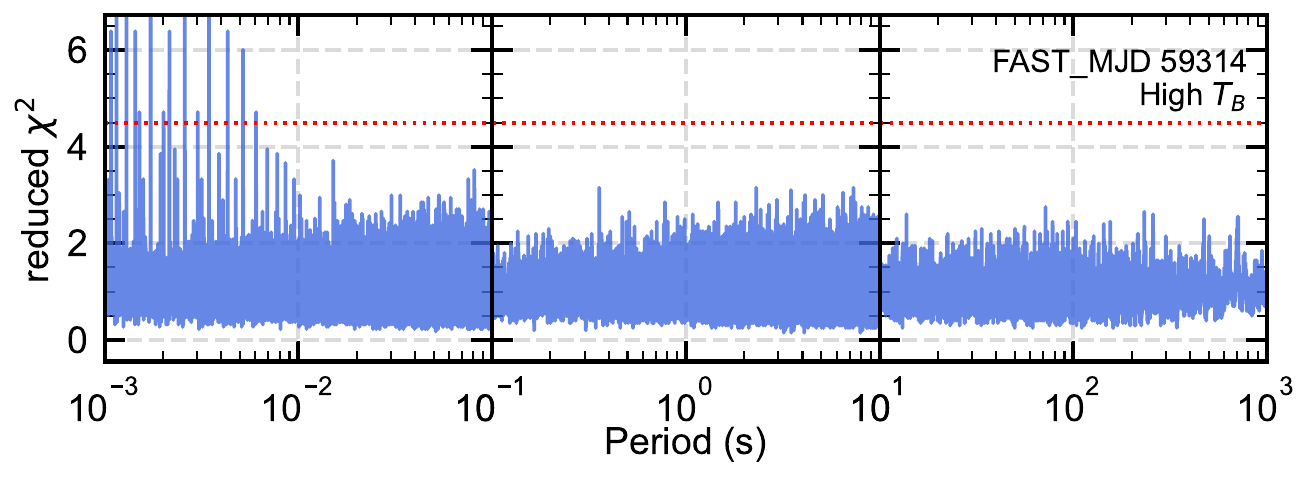}
   \includegraphics[width=0.32\textwidth]{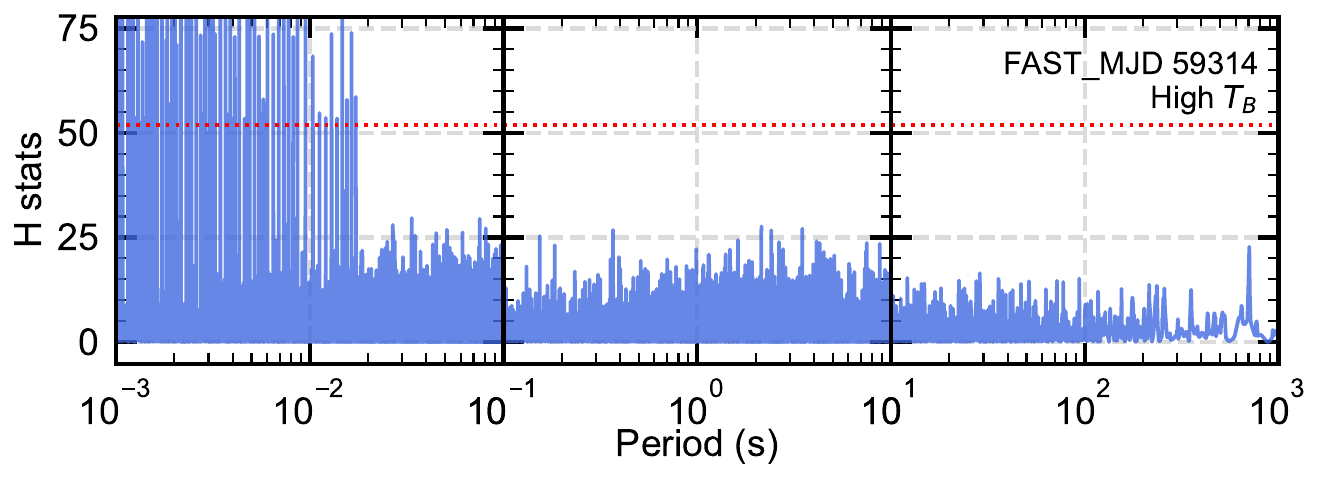}
   \includegraphics[width=0.32\textwidth]{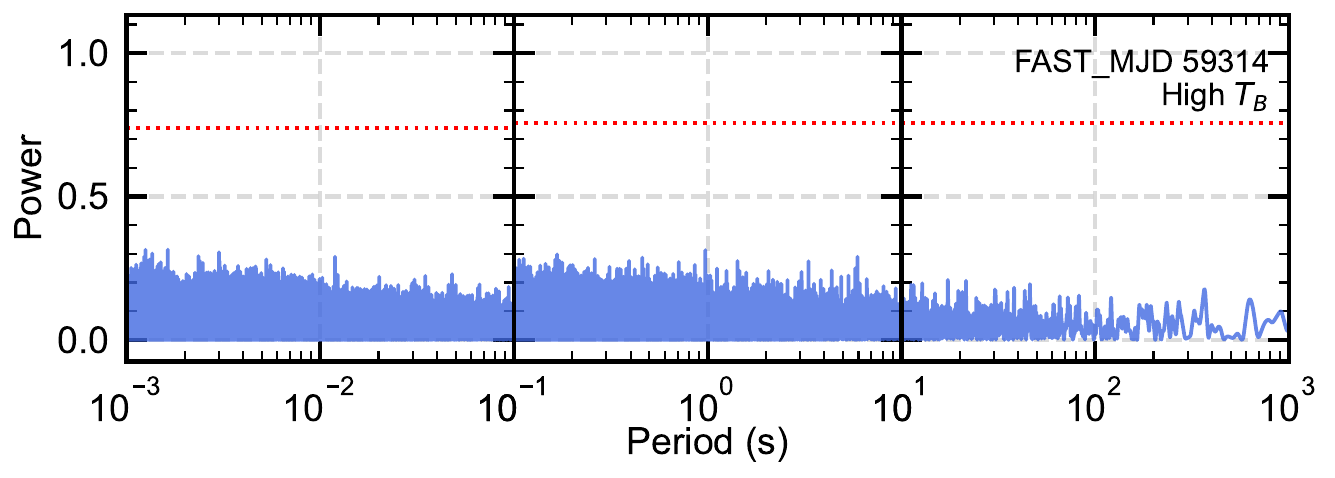}
   \includegraphics[width=0.32\textwidth]{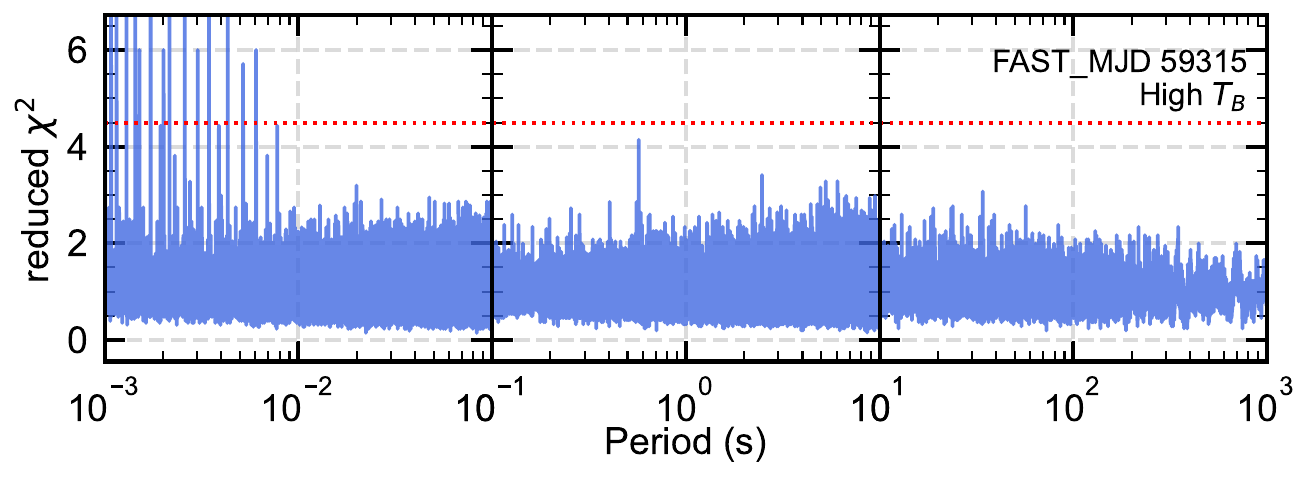}
   \includegraphics[width=0.32\textwidth]{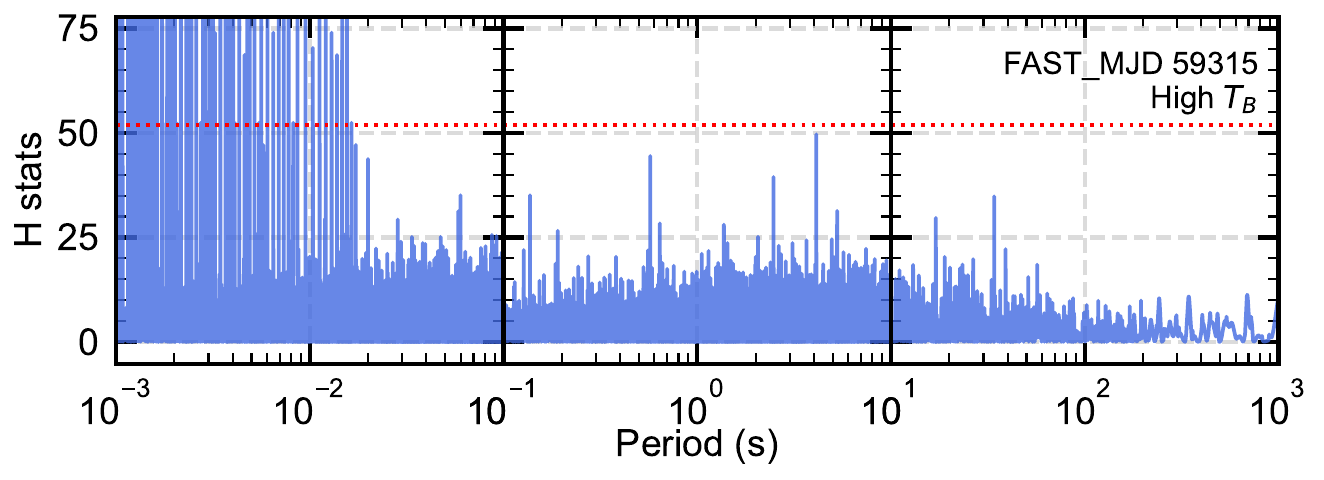}
   \includegraphics[width=0.32\textwidth]{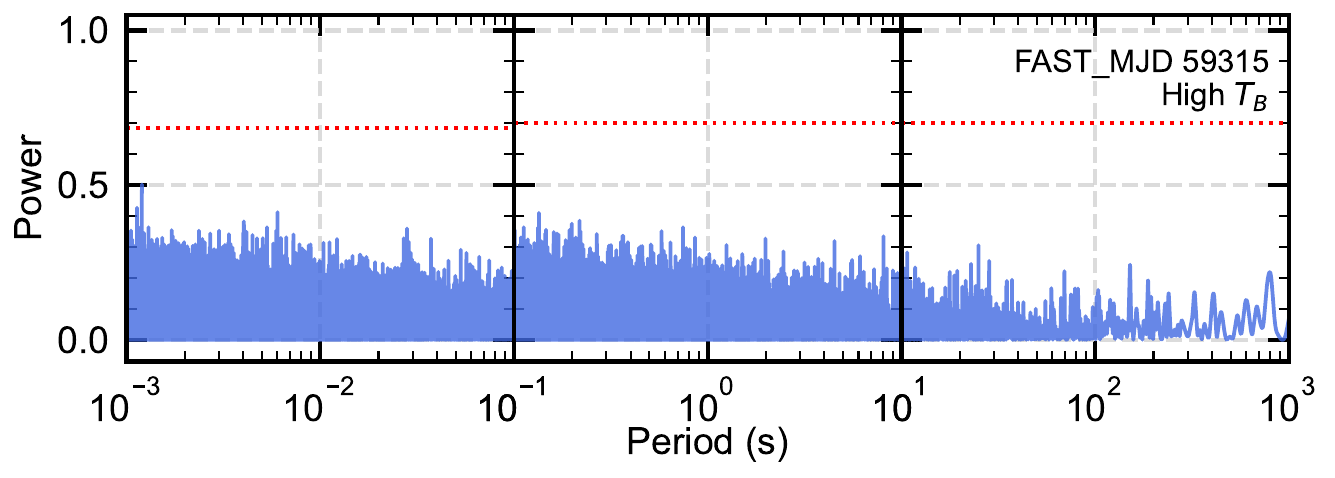}
   \includegraphics[width=0.32\textwidth]{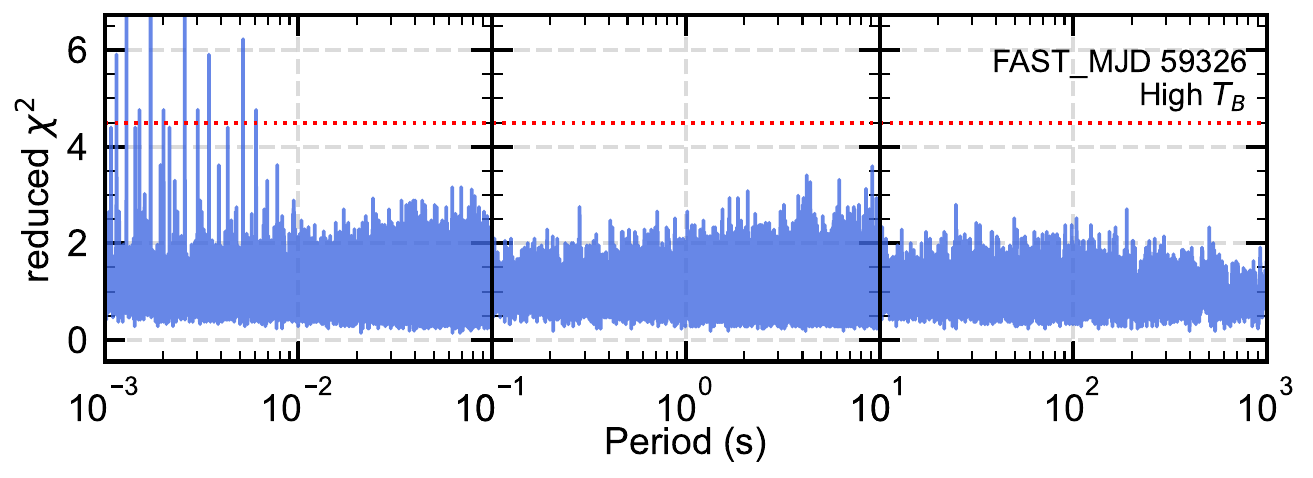}
   \includegraphics[width=0.32\textwidth]{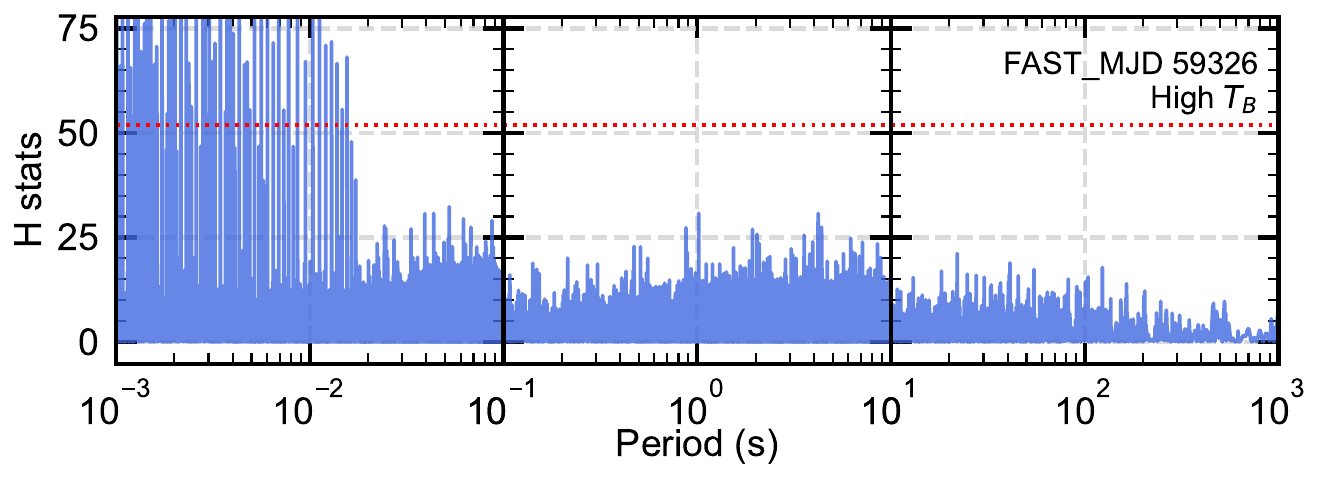}
   \includegraphics[width=0.32\textwidth]{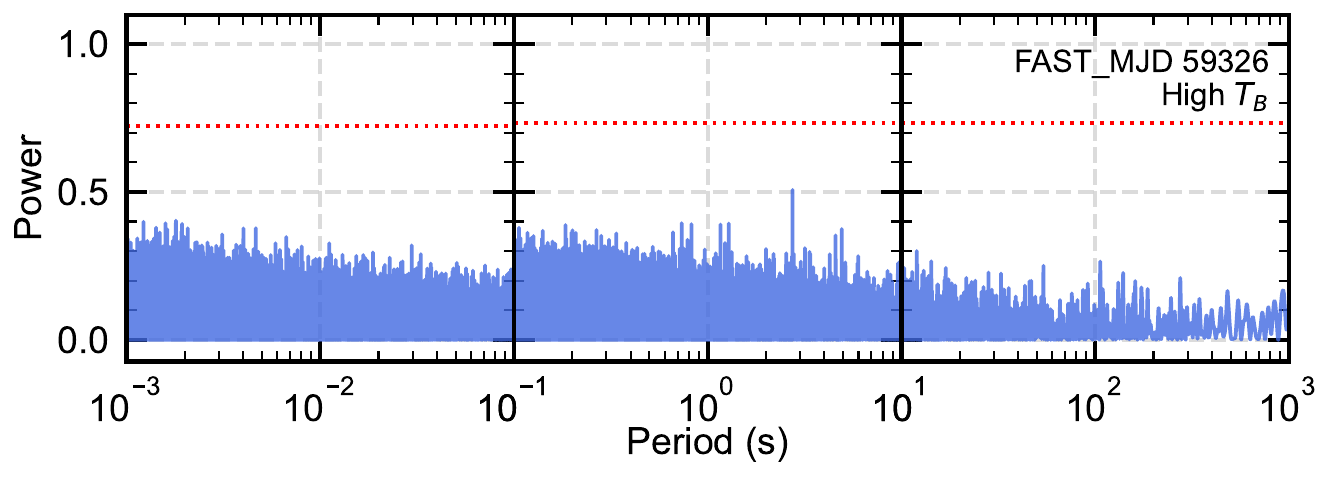}
   \includegraphics[width=0.32\textwidth]{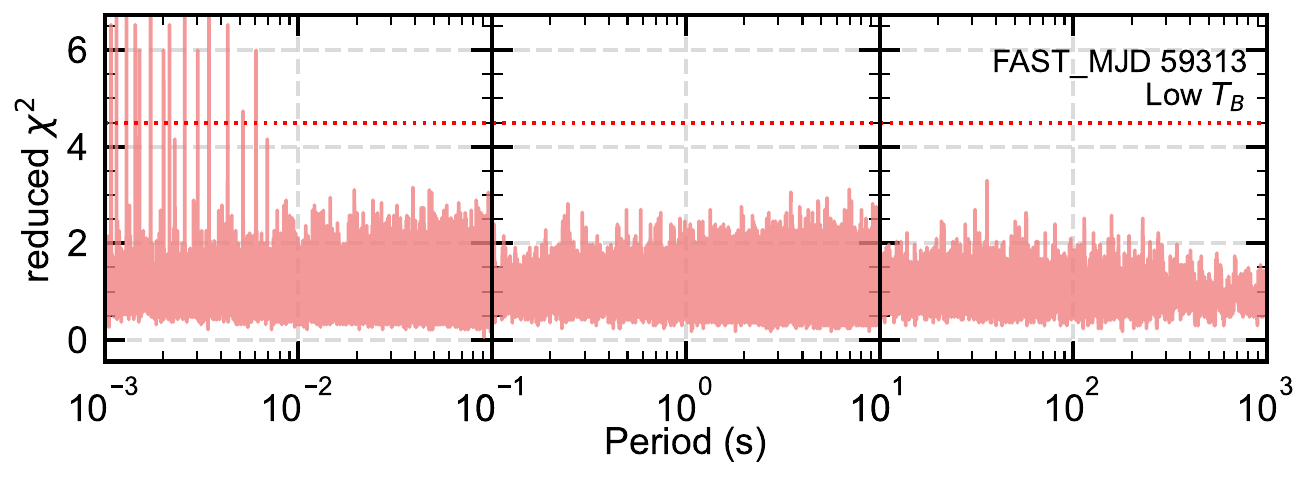}
   \includegraphics[width=0.32\textwidth]{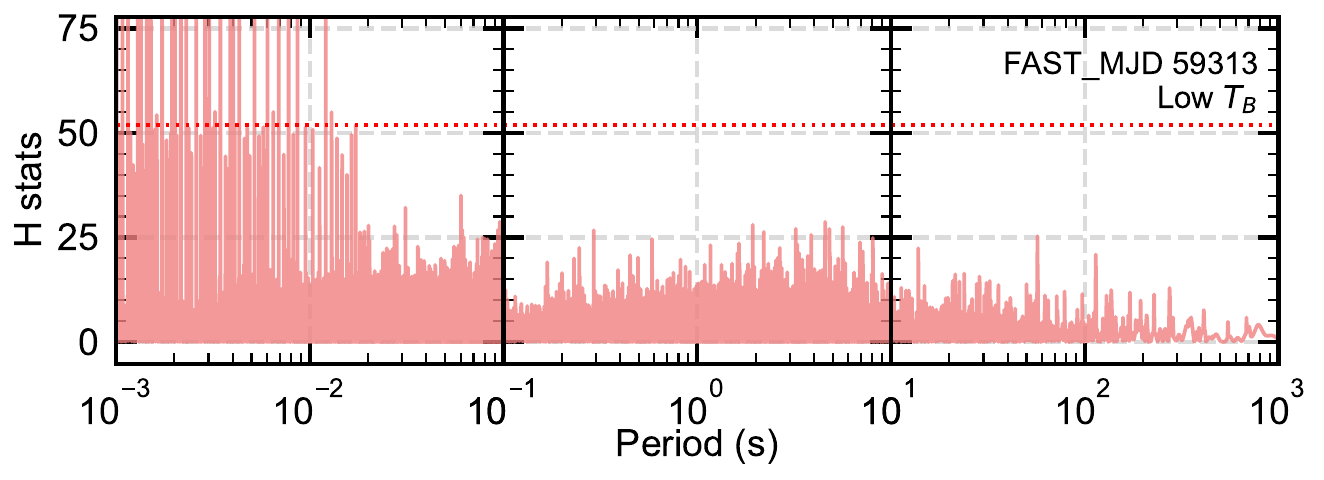}
   \includegraphics[width=0.32\textwidth]{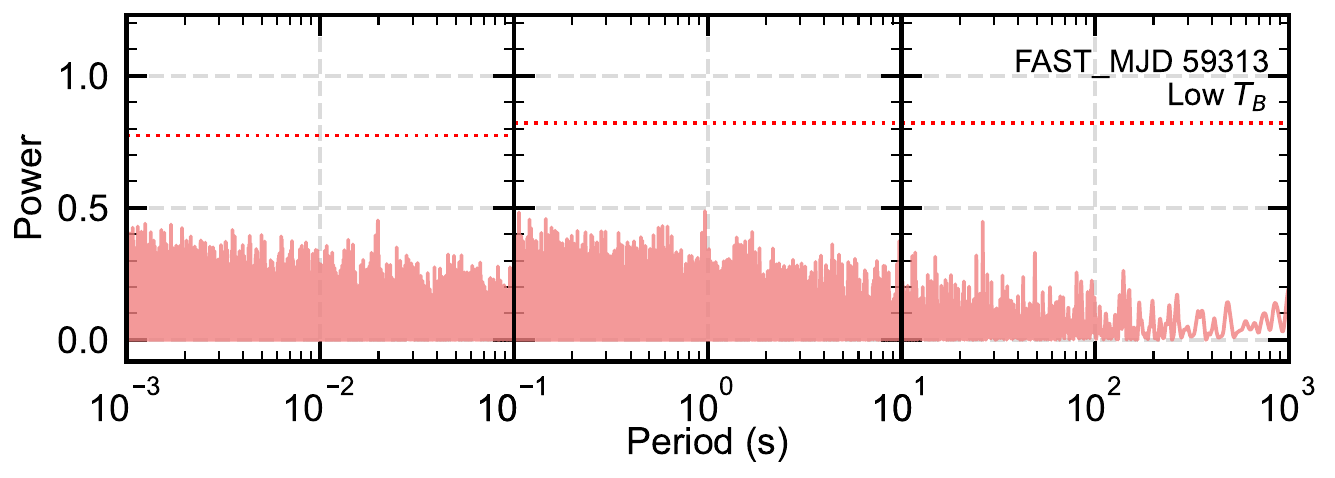}
   \includegraphics[width=0.32\textwidth]{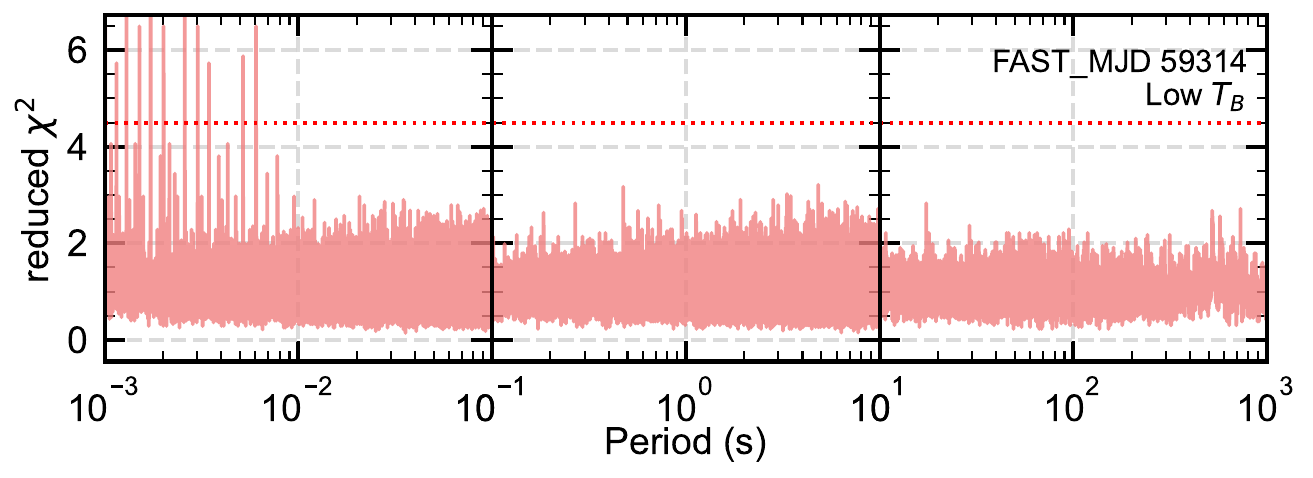}
   \includegraphics[width=0.32\textwidth]{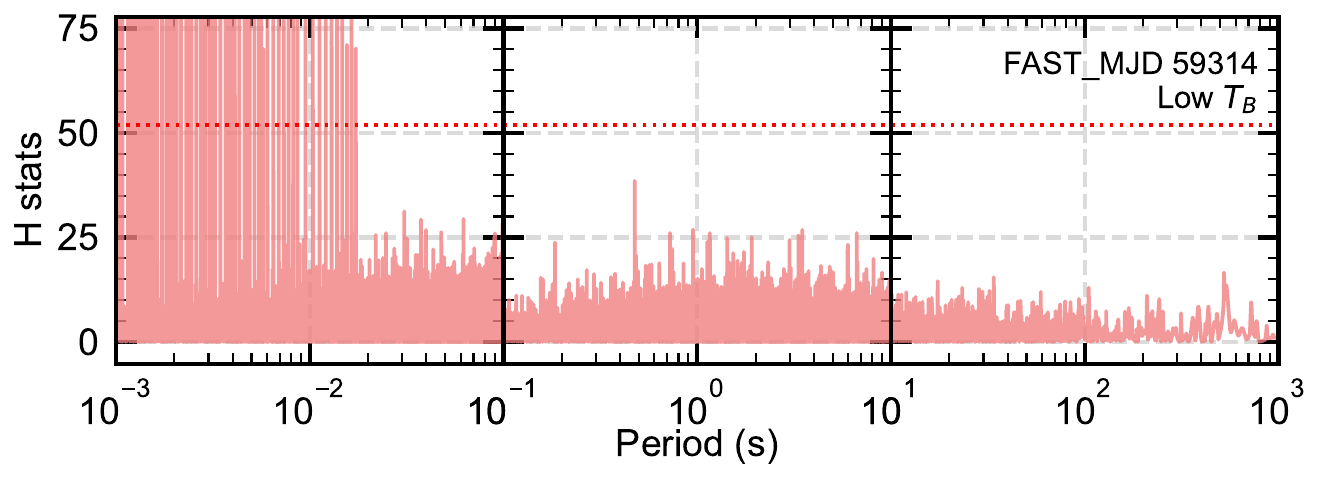}
   \includegraphics[width=0.32\textwidth]{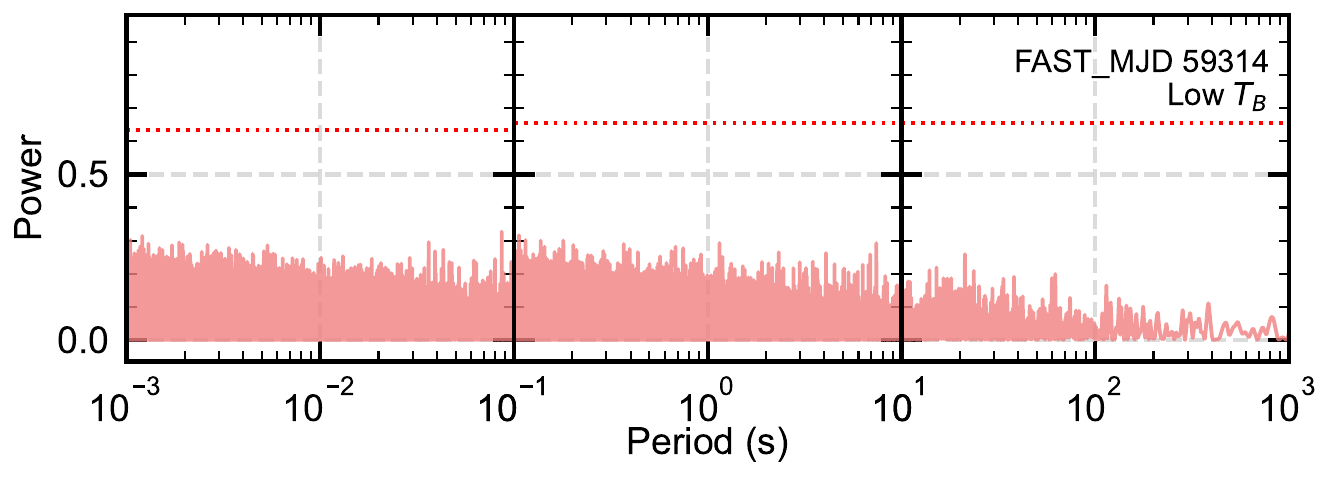}
   \includegraphics[width=0.32\textwidth]{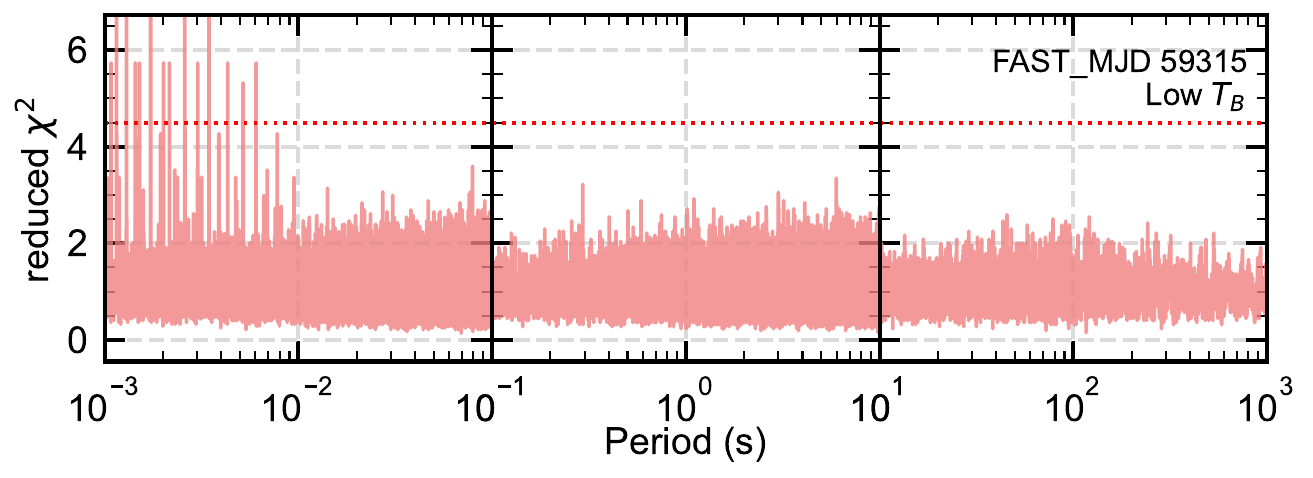}
   \includegraphics[width=0.32\textwidth]{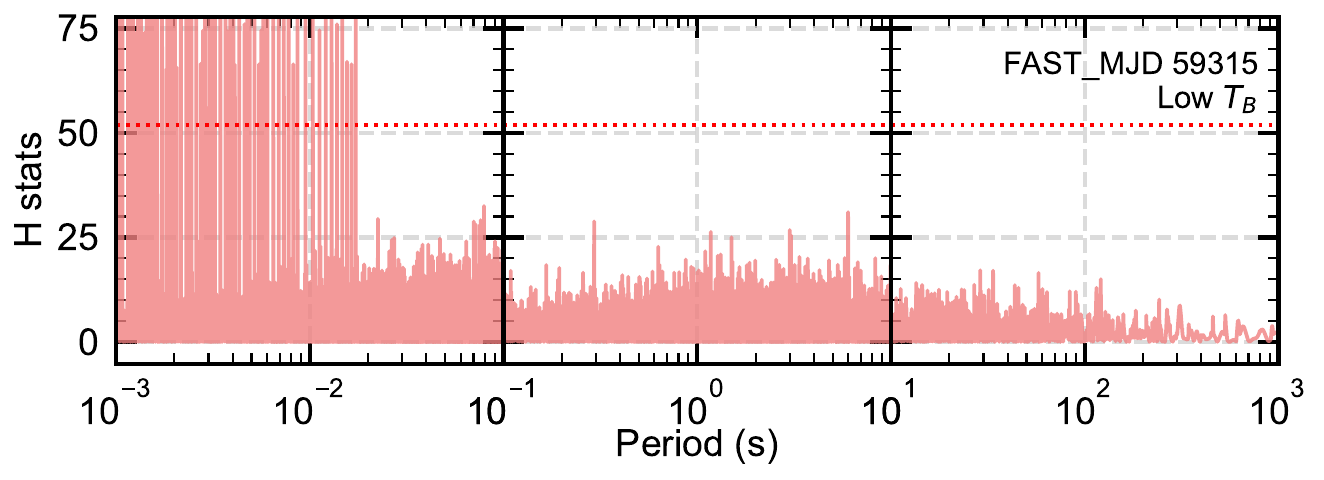}
   \includegraphics[width=0.32\textwidth]{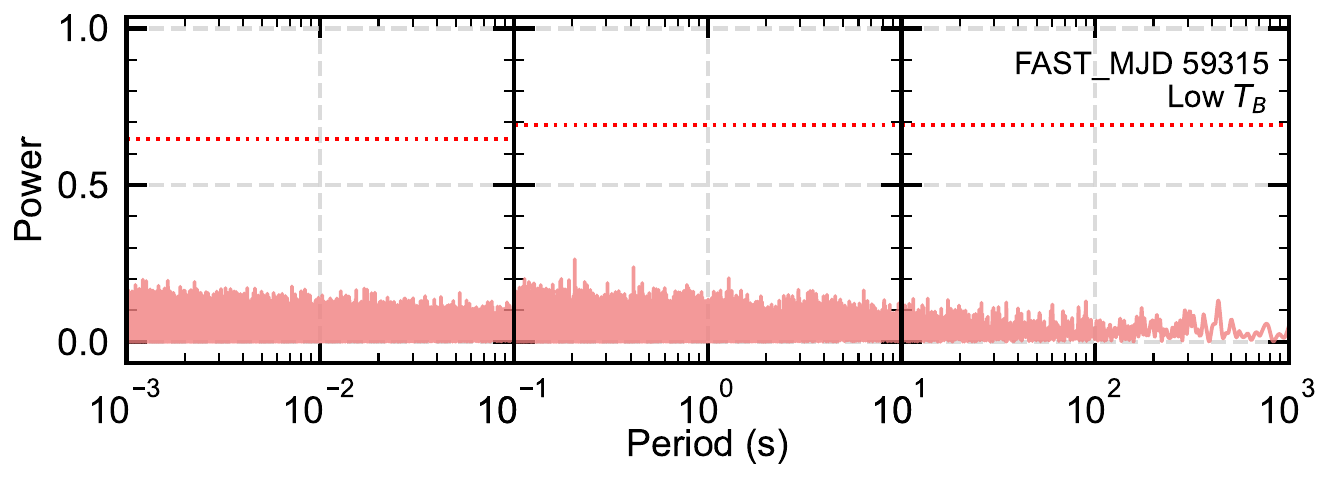}
   \caption{\footnotesize The period search results based the FAST Dataset \#4
   of FRB 20201124A, considering the brightness temperature of the bursts.
   The three columns from left to right correspond to the three search methods,
   i.e. the phase folding algorithm, the H-test, and the Lomb-Scargle periodogram, respectively.
   The rows from top to bottom correspond to high temperature or low temperature events
   on selected days: high $T_\mathrm{B}$ bursts on MJD 59314, MJD 59315 and MJD 59326;
   and low $T_\mathrm{B}$ bursts on MJD 59313, MJD 59314 and MJD 59315. In the cases
   of phase folding and the H-test methods, a horizontal dotted line is plotted to mark
   the p-value of $10^{-9}$. In the cases of Lomb-Scargle periodogram, a horizontal
   dotted line indicating a FAP level of $10^{-9}$ is plotted. No clear evidence
   of periodicity is found in these plots. Note that in the left and middle panels,
   the peak structures in the 1 ms--20 ms
   range are fake signals due to the limited timing accuracy. }
   \label{Fig15}
\end{figure*}

\begin{figure*}
   \centering
   \includegraphics[width=0.32\textwidth]{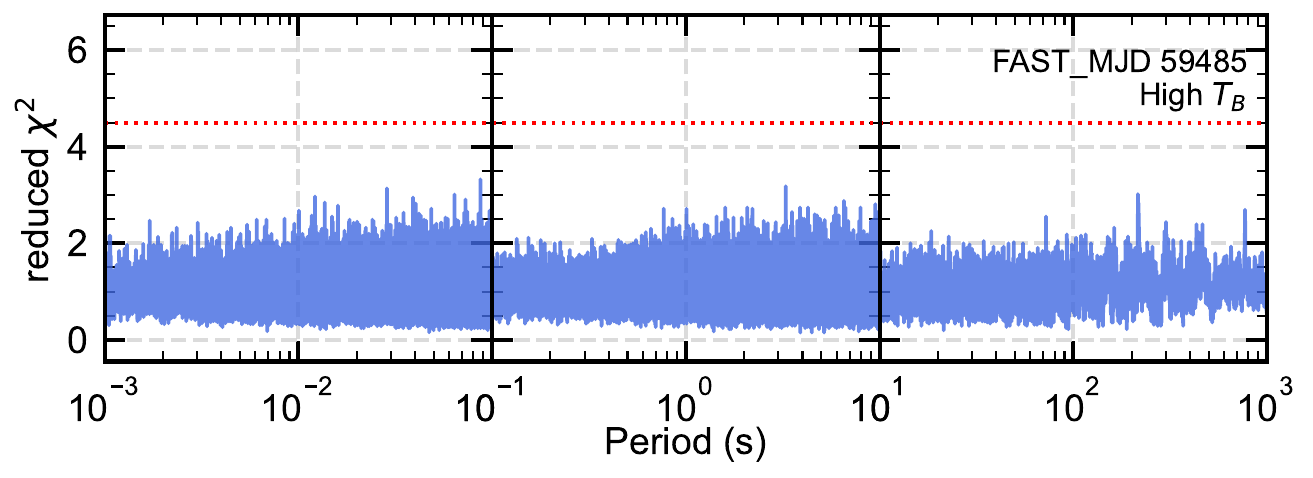}
   \includegraphics[width=0.32\textwidth]{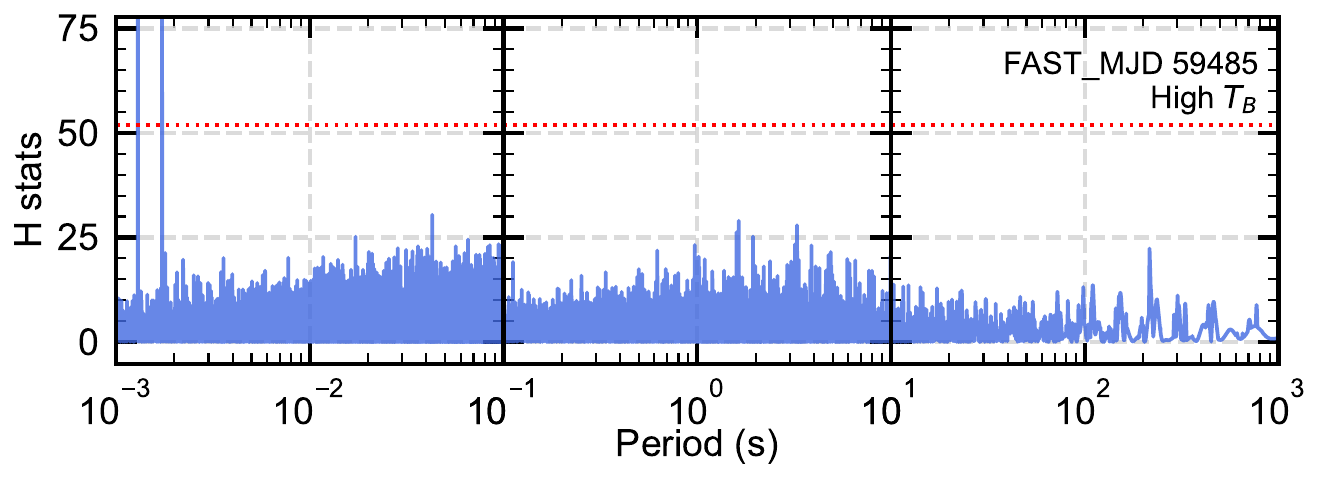}
   \includegraphics[width=0.32\textwidth]{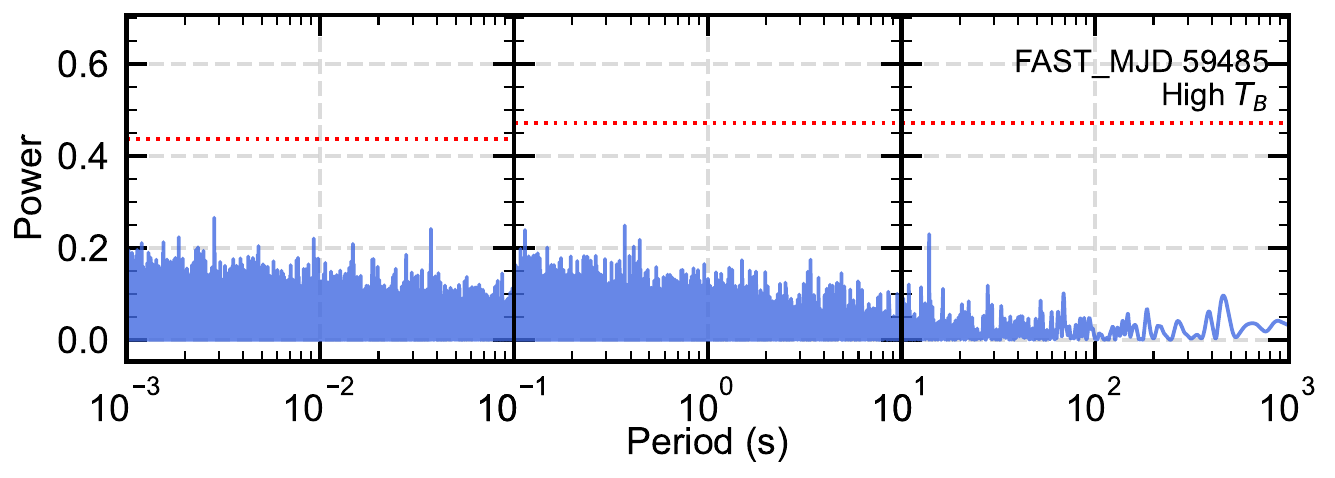}
   \includegraphics[width=0.32\textwidth]{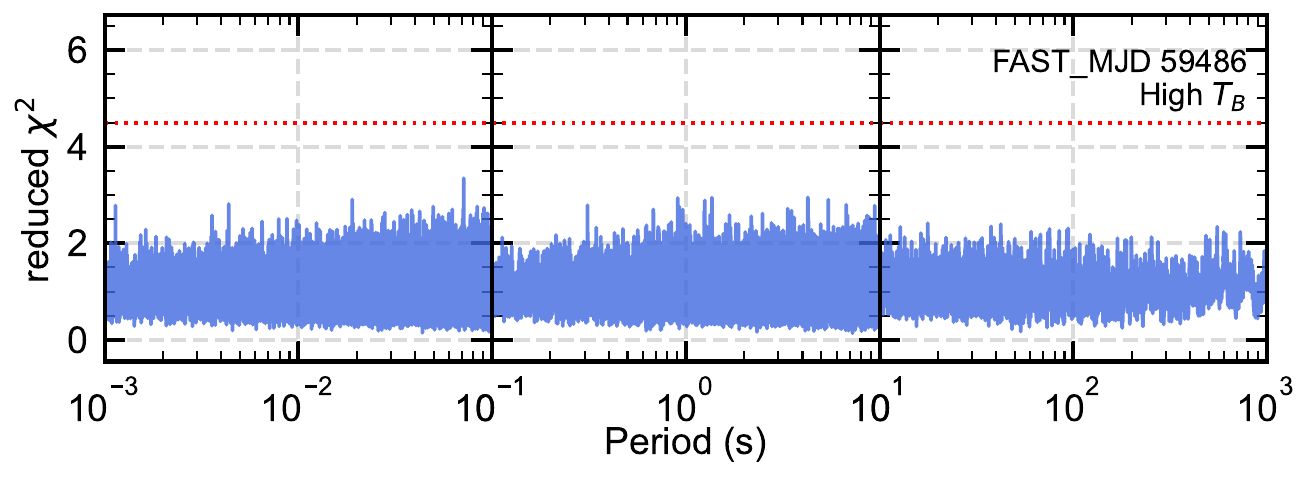}
   \includegraphics[width=0.32\textwidth]{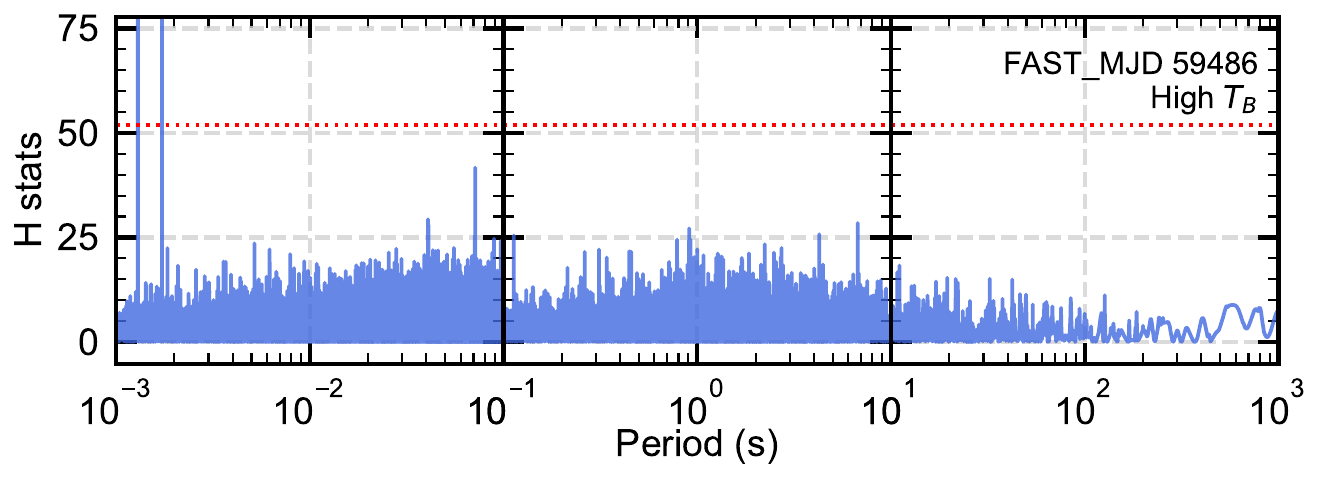}
   \includegraphics[width=0.32\textwidth]{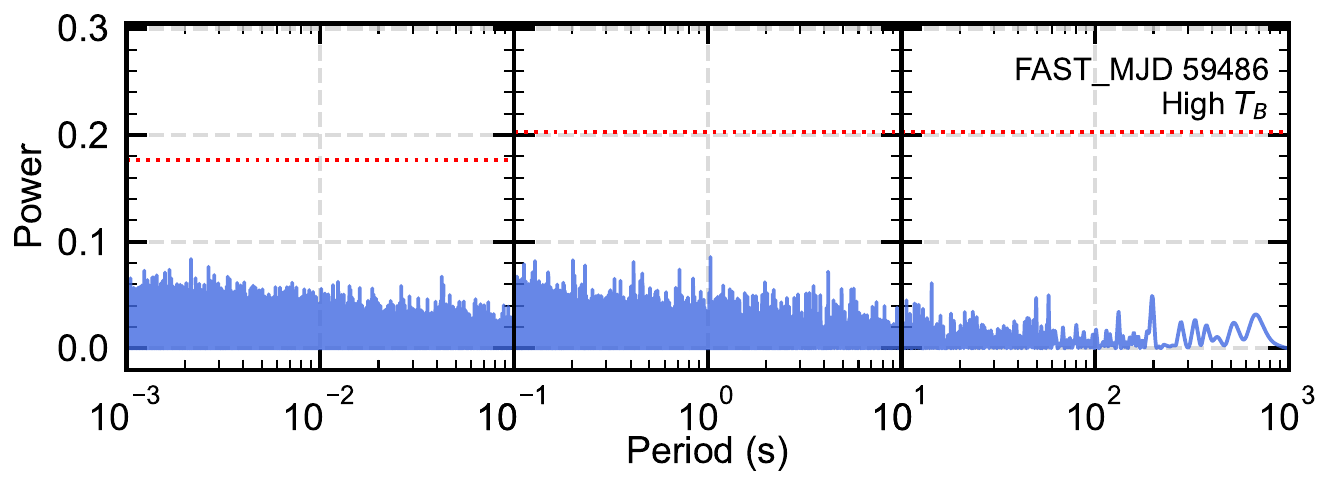}
   \includegraphics[width=0.32\textwidth]{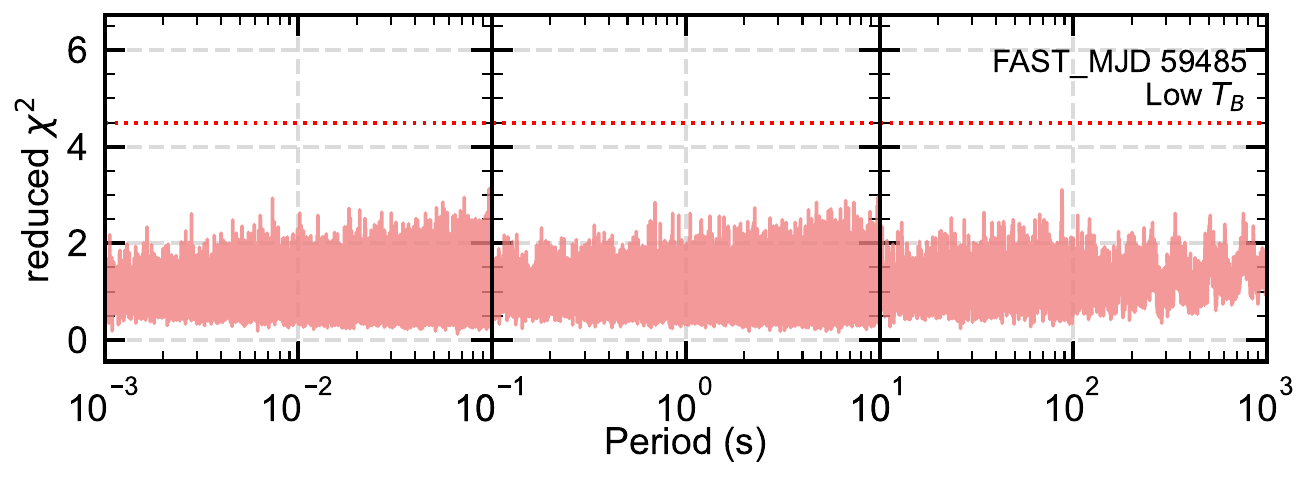}
   \includegraphics[width=0.32\textwidth]{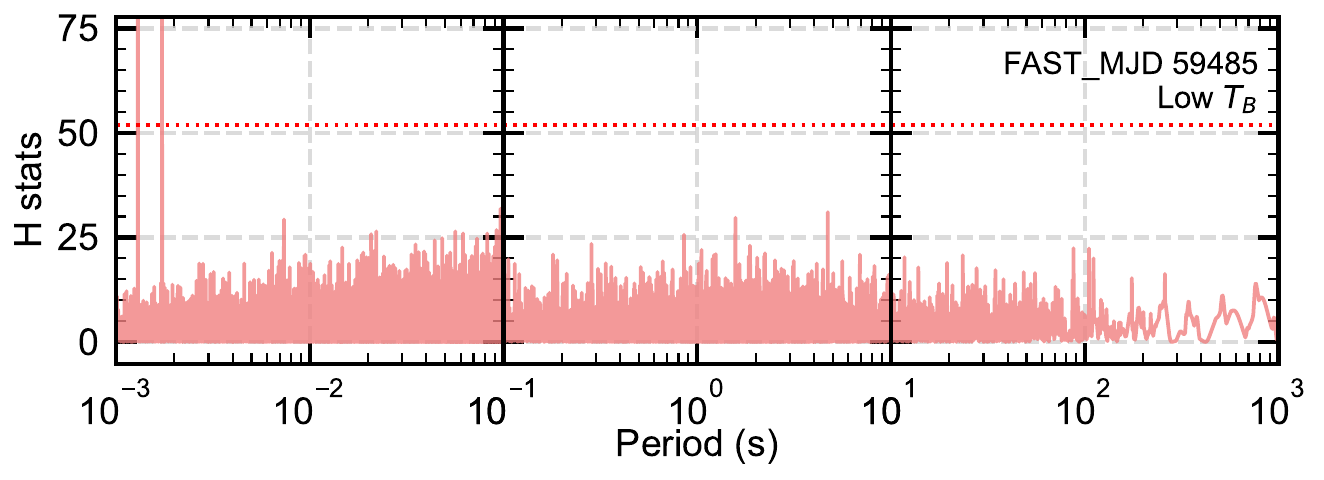}
   \includegraphics[width=0.32\textwidth]{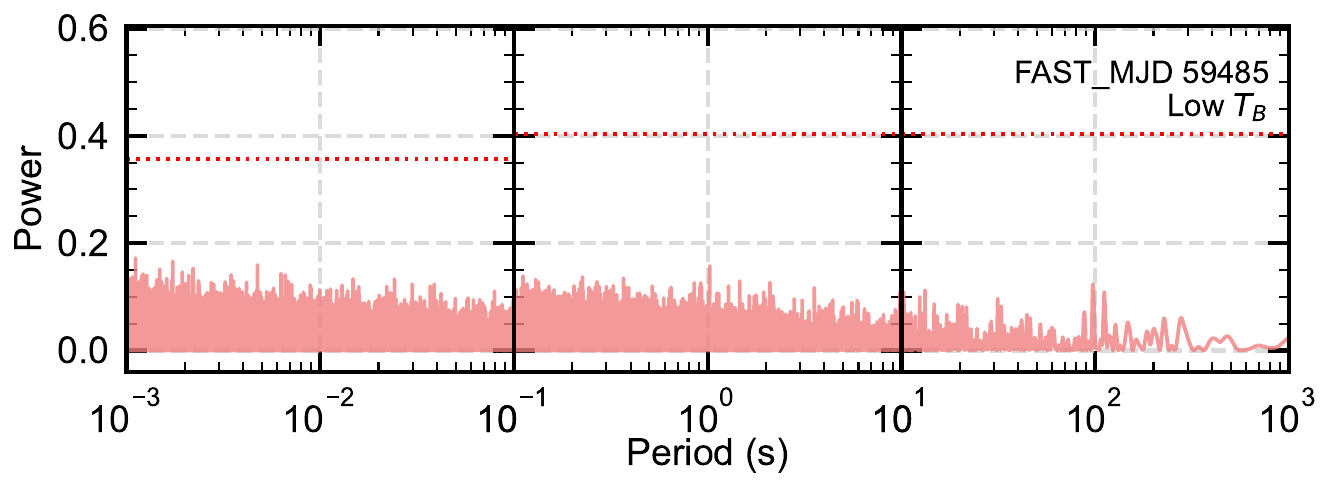}
   \includegraphics[width=0.32\textwidth]{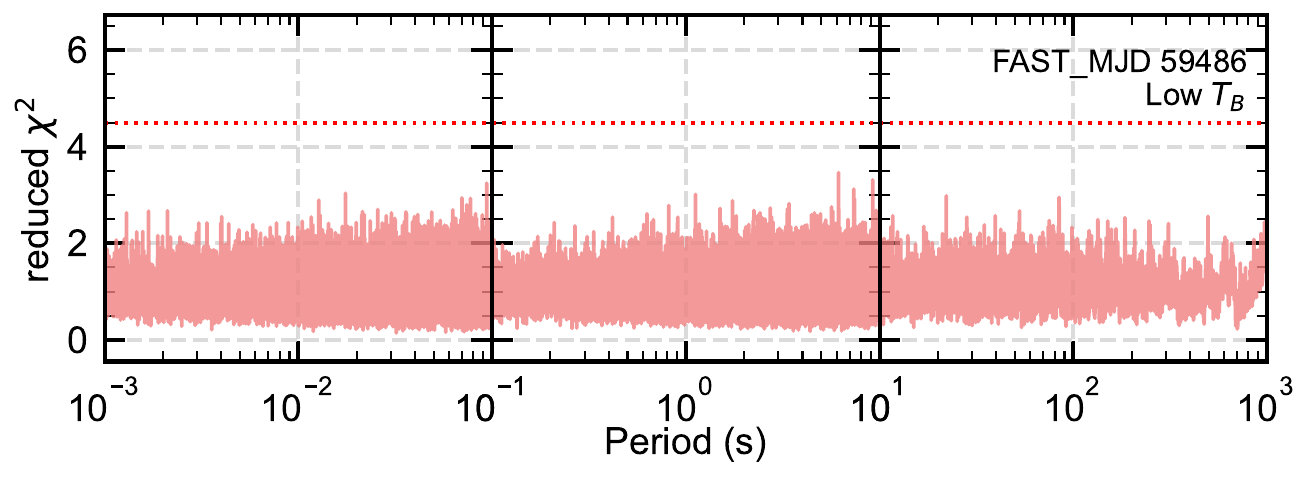}
   \includegraphics[width=0.32\textwidth]{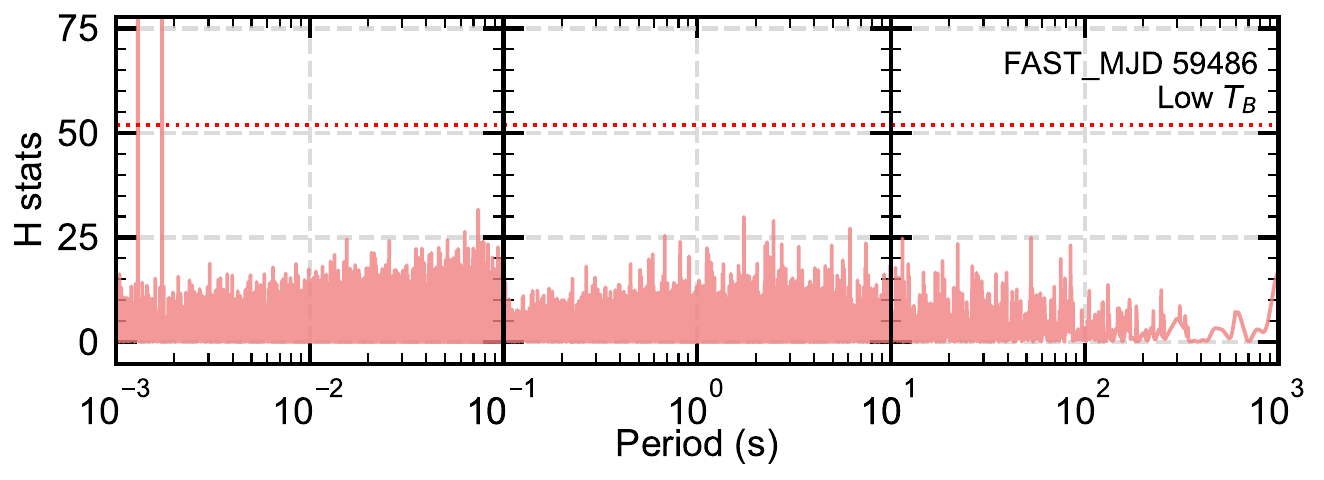}
   \includegraphics[width=0.32\textwidth]{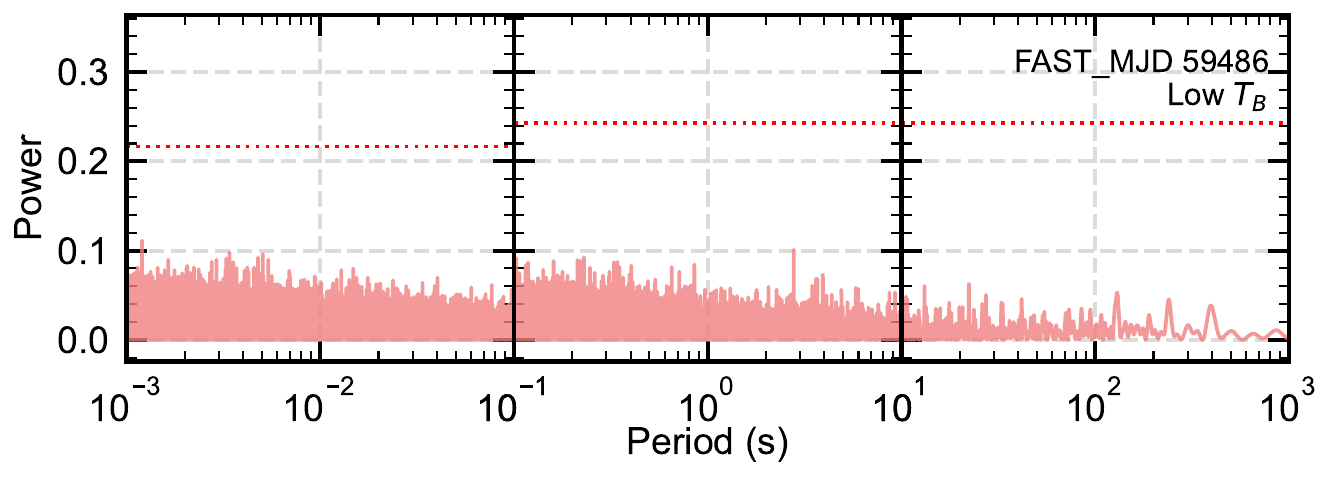}
   \caption{\footnotesize The period search results based on the FAST Dataset \#6 of
   FRB 20201124A, considering the brightness temperature of the bursts. The three columns
   from left to right correspond to the three search methods, i.e. the phase folding algorithm,
   the H-test, and the Lomb-Scargle periodogram, respectively. The rows from top to bottom
   correspond to high temperature or low temperature events on selected days: high $T_\mathrm{B}$
   bursts on MJD 59485 and MJD 59486; and low $T_\mathrm{B}$ bursts on MJD 59485 and MJD 59486.
   In the cases of phase folding and the H-test methods, a horizontal dotted line is plotted
   to mark the p-value of $10^{-9}$. In the cases of Lomb-Scargle periodogram, a horizontal
   dotted line indicating a FAP level of $10^{-9}$ is plotted.
   No clear evidence of periodicity is found in these plots. Note that in the middle panels,
   the peak structures in the 1 ms--2 ms
   range are fake signals due to the limited timing accuracy.  }
   \label{Fig16}
\end{figure*}

For the FAST Dataset \#4 of FRB 20201124A, the high
$T_\mathrm{B}$ subset consists of bursts on three days: 44 bursts
on MJD 59314, of which 42 bursts have a waiting time > 0.1 s; 51
bursts on MJD 59315, of which 49 bursts have a waiting time > 0.1
s; and 46 bursts on MJD 59326, of which 45 bursts have a waiting
time > 0.1 s. The low $T_\mathrm{B}$ subset of the FAST Dataset
\#4 also consists of bursts on three days: 40 bursts on MJD 59313,
of which 35 bursts have a waiting time > 0.1 s; 58 bursts on MJD
59314, of which 55 bursts have a waiting time > 0.1 s; and 56
bursts on MJD 59315, of which 50 bursts have a waiting time > 0.1
s. The results are shown in Figure ~\ref{Fig15}. For the FAST
Dataset \#6 of FRB 20201124A, the high $T_\mathrm{B}$ subset
consists of bursts on two days: 101 bursts on MJD 59485, of which
91 bursts have a waiting time > 0.1 s; and 302 bursts on MJD
59486, of which 258 bursts have a waiting time > 0.1 s. The low
$T_\mathrm{B}$ subset of the FAST Dataset \#6 also consists of
bursts on two days: 131 bursts on MJD 59485, of which 112 bursts
have a waiting time > 0.1 s; and 240 bursts on MJD 59486, of which
210 bursts have a waiting time > 0.1 s. The results are shown in
Figure ~\ref{Fig16}. In the 1--2 ms range and the 1--20 ms range, several extremely high peaks could be observed, which
are all harmonics of 0.864 ms ($10^{-8}$ d). They are fake signals
due to limited timing accuracy of observations. In each plot, we
have also examined by hand the phase-folded histograms at the
positions that correspond to the relatively high peaks in the
$\chi^2$/H-statistic/Lomb-Scargle periodograms, but found that no
marked features appear in the histograms. Additionally, a peak
occurring on one day usually does not appear on other days, which
further excludes its possibility of being a true period. From
Figures ~\ref{Fig15} and ~\ref{Fig16}, we conclude that no
periodicity exceeding the confidence level is found in the whole
range of 0.001--1000 s either in high $T_\mathrm{B}$ or in low
$T_\mathrm{B}$ bursts from FRB 20201124A.

\begin{figure*}
   \centering
   \includegraphics[width=0.32\textwidth]{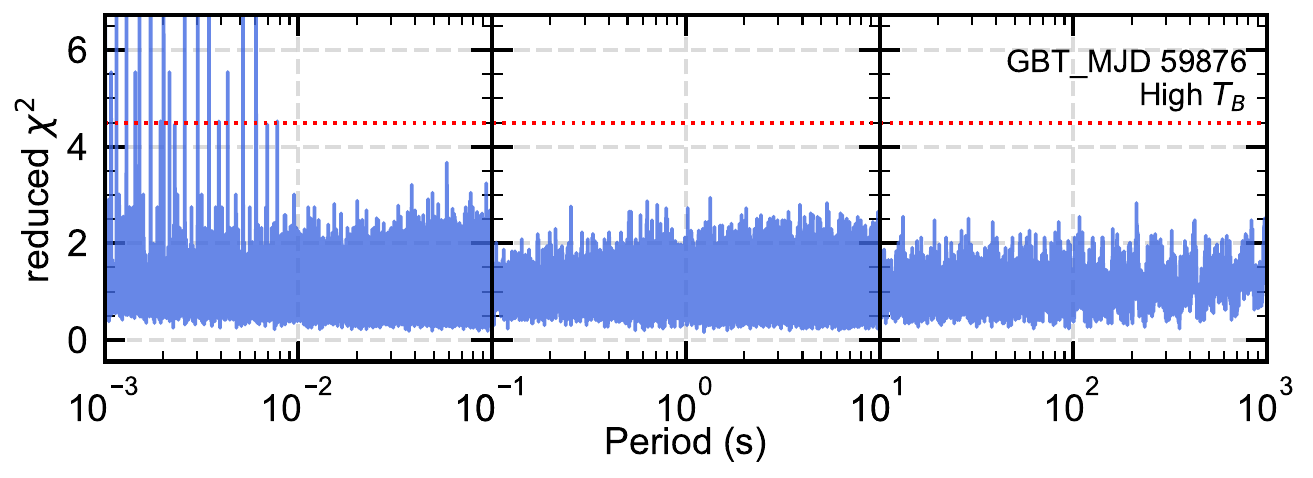}
   \includegraphics[width=0.32\textwidth]{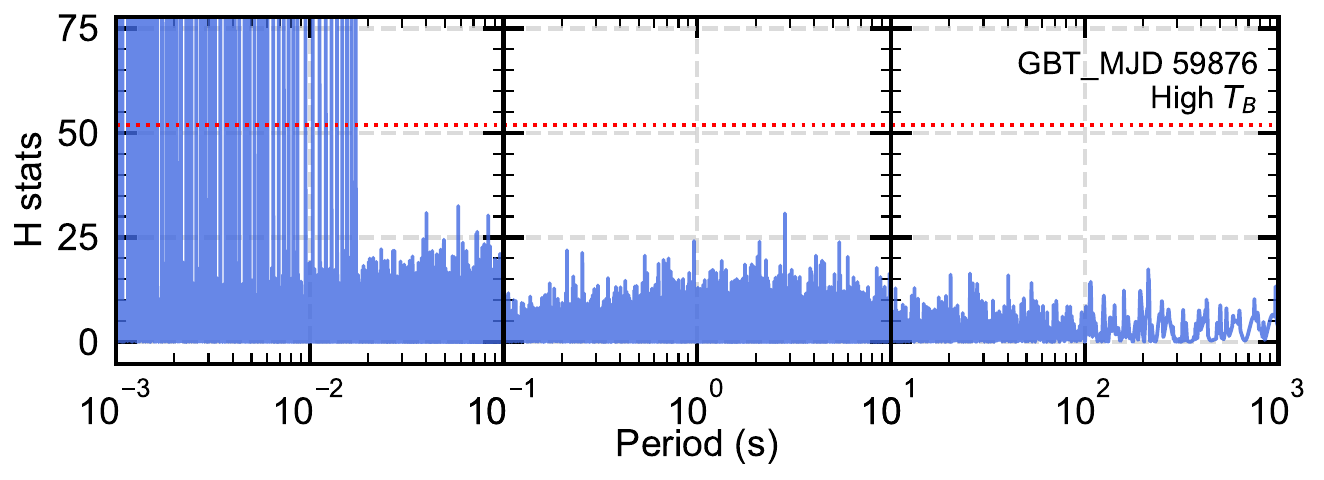}
   \includegraphics[width=0.32\textwidth]{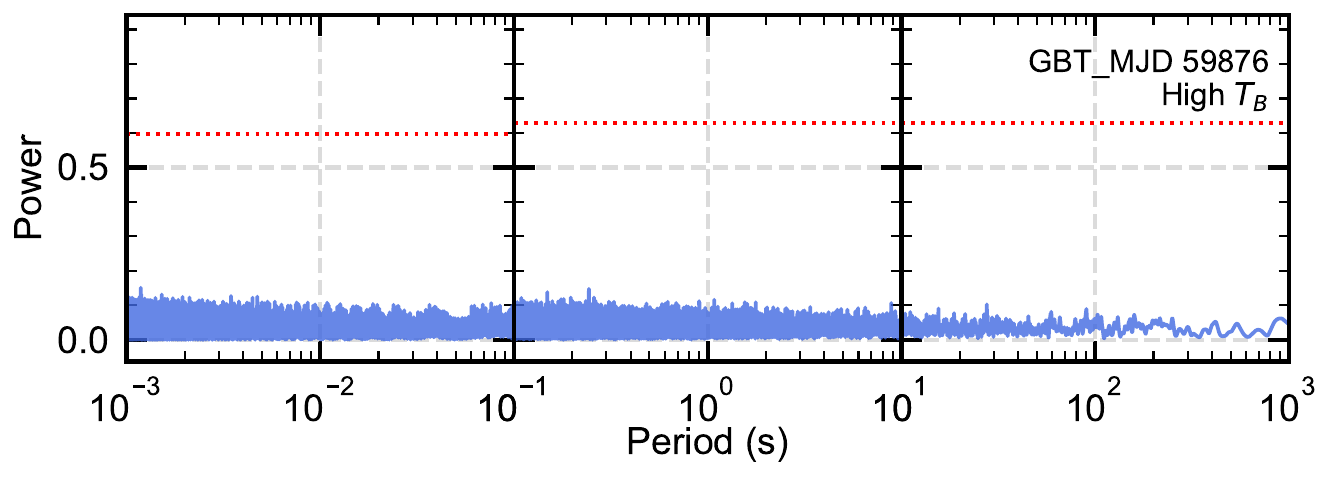}
   \includegraphics[width=0.32\textwidth]{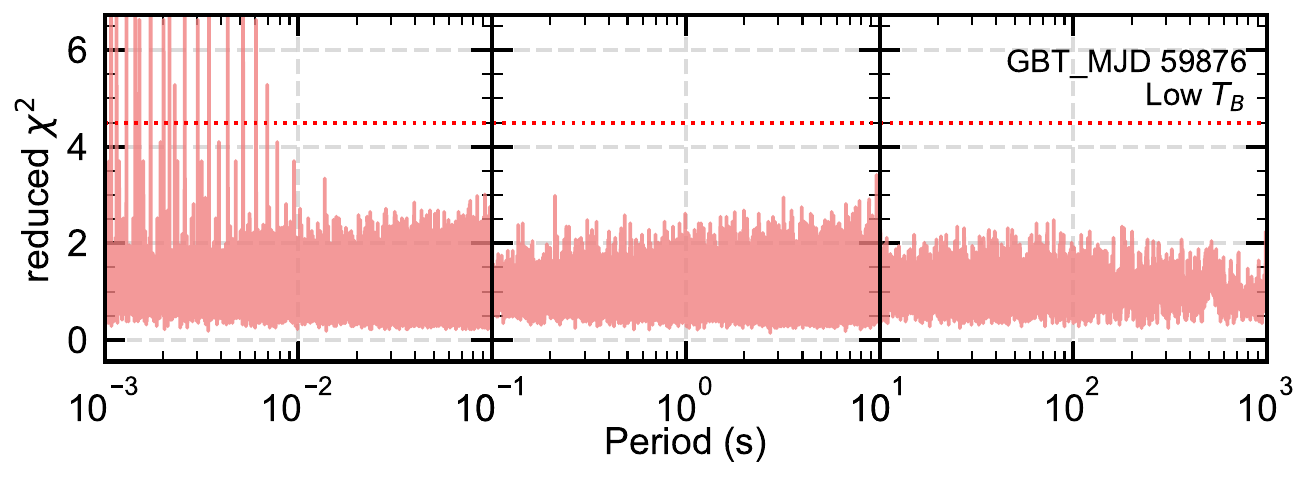}
   \includegraphics[width=0.32\textwidth]{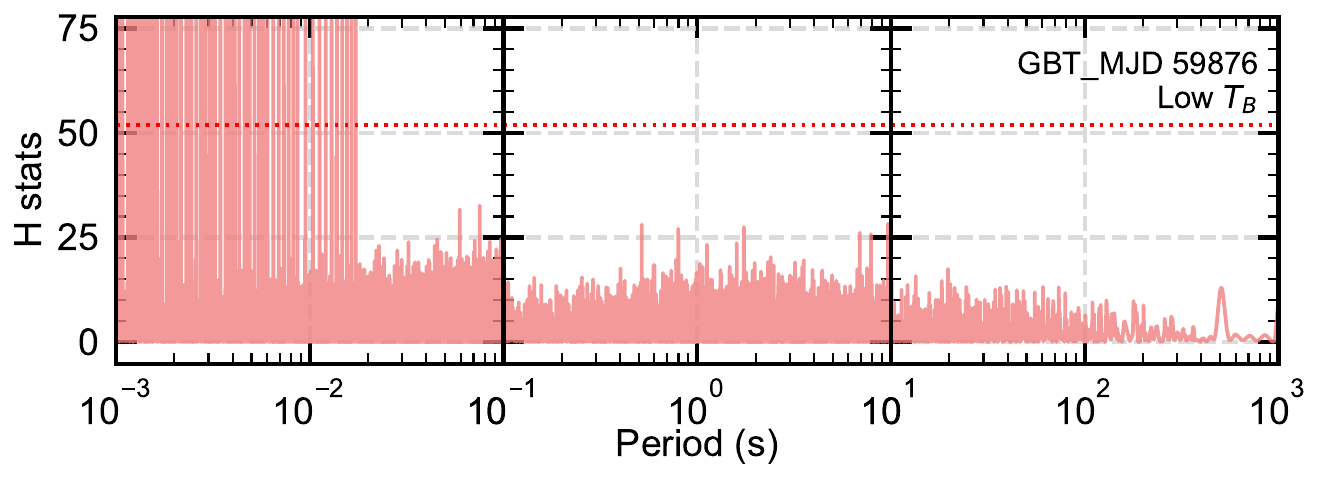}
   \includegraphics[width=0.32\textwidth]{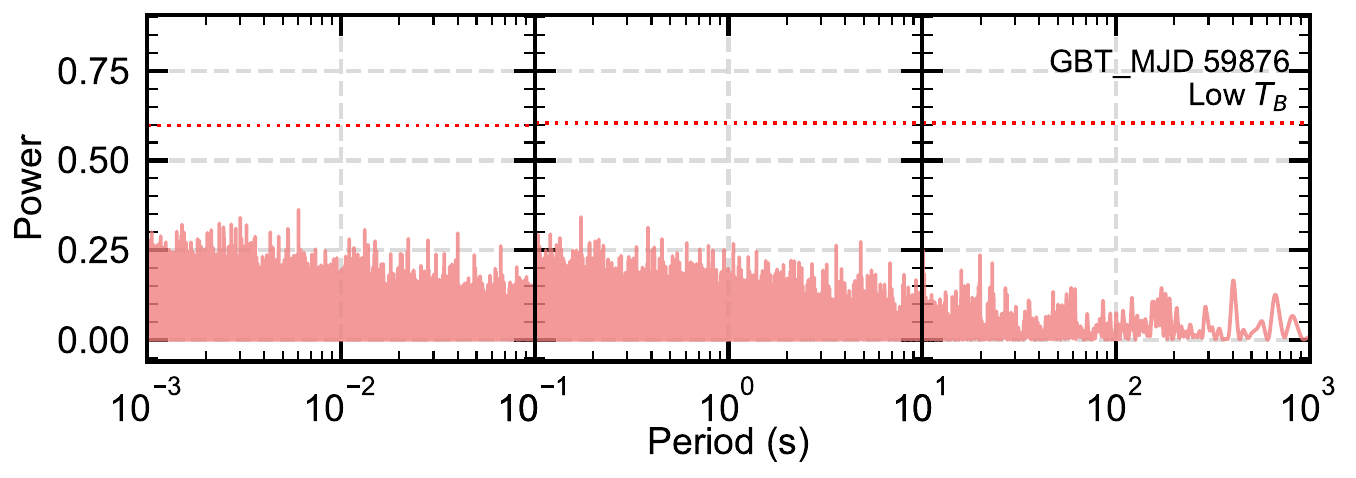}
   \caption{\footnotesize The period search results based on the GBT Dataset \#8 of
   FRB 20220912A, considering the brightness temperature of the bursts. The three columns from left
   to right correspond to the three search methods, i.e. the phase folding algorithm, the H-test,
   and the Lomb-Scargle periodogram, respectively. The rows from top to bottom correspond
   to high temperature or low temperature events on selected days: high $T_\mathrm{B}$
   bursts on MJD 59876 and low $T_\mathrm{B}$ bursts on MJD 59876. In the cases of phase folding
   and the H-test methods, a horizontal dotted line is plotted to mark the p-value of $10^{-9}$.
   In the cases of Lomb-Scargle periodogram, a horizontal dotted line indicating a FAP level
   of $10^{-9}$ is plotted. No clear evidence of periodicity is found in these plots. Note that in the left and middle panels,
   the peak structures in the 1 ms--20 ms
   range are fake signals due to the limited timing accuracy. }
   \label{Fig17}
\end{figure*}

\begin{figure*}
   \centering
   \includegraphics[width=0.32\textwidth]{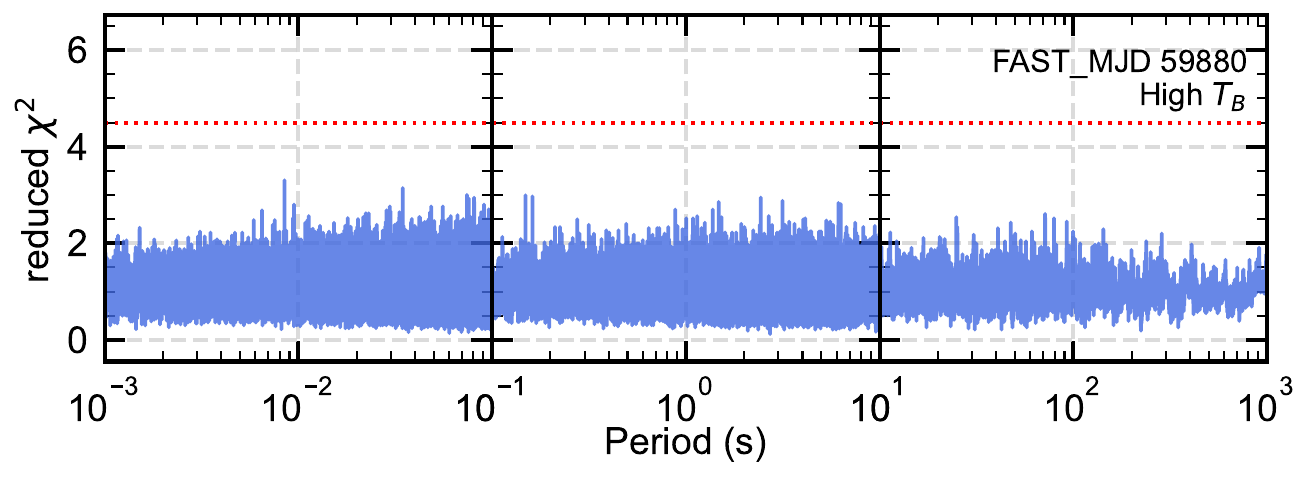}
   \includegraphics[width=0.32\textwidth]{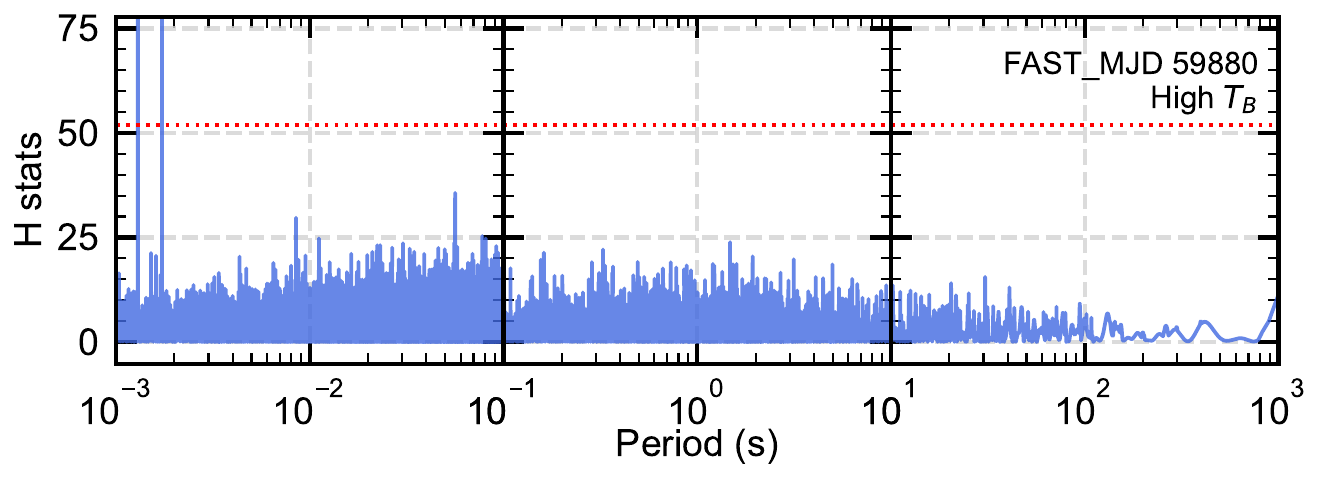}
   \includegraphics[width=0.32\textwidth]{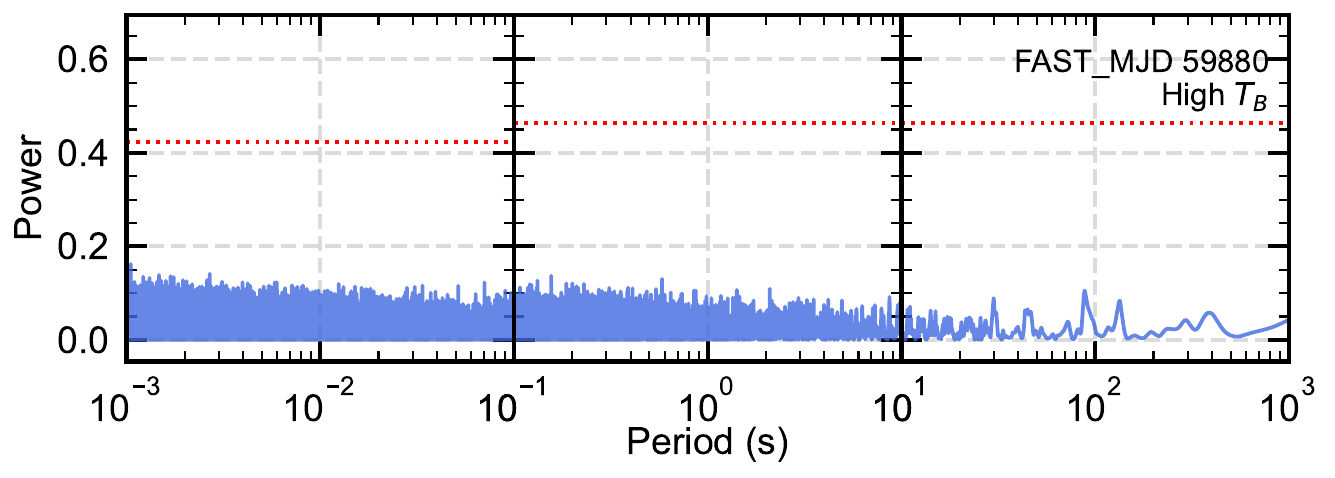}
   \includegraphics[width=0.32\textwidth]{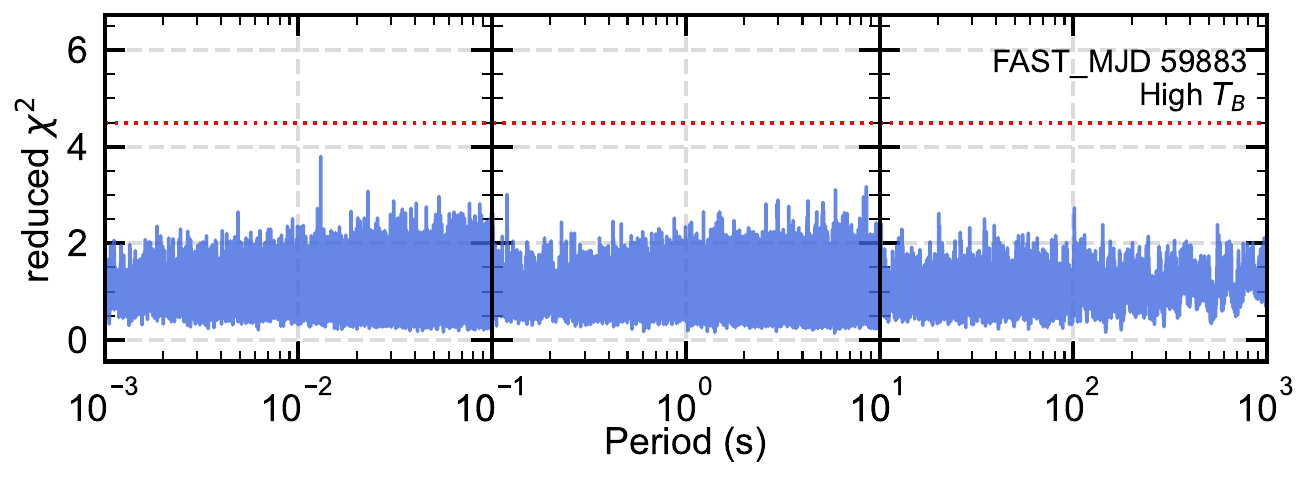}
   \includegraphics[width=0.32\textwidth]{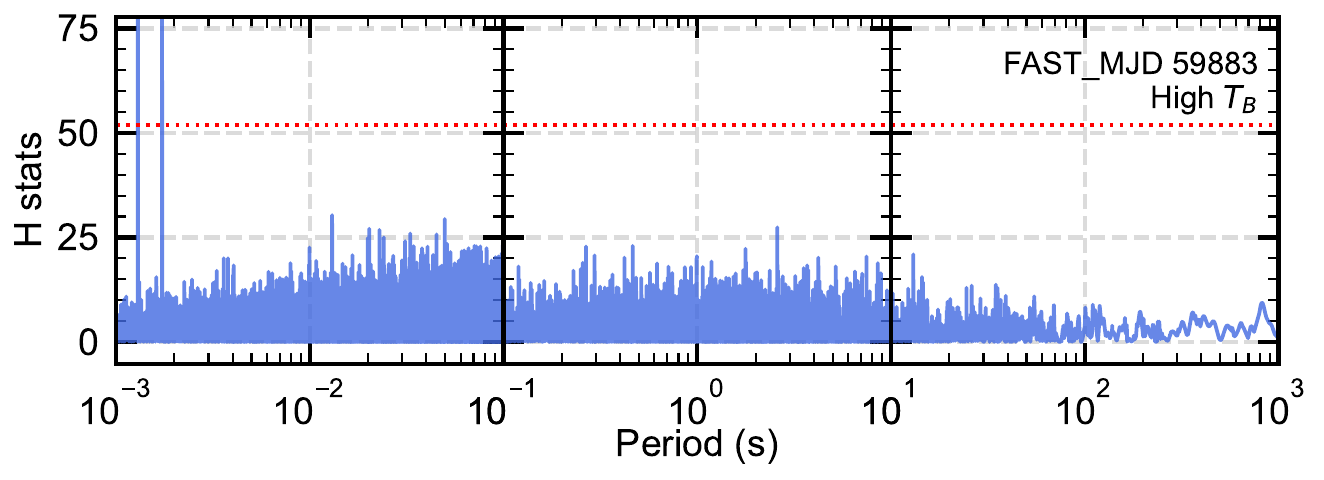}
   \includegraphics[width=0.32\textwidth]{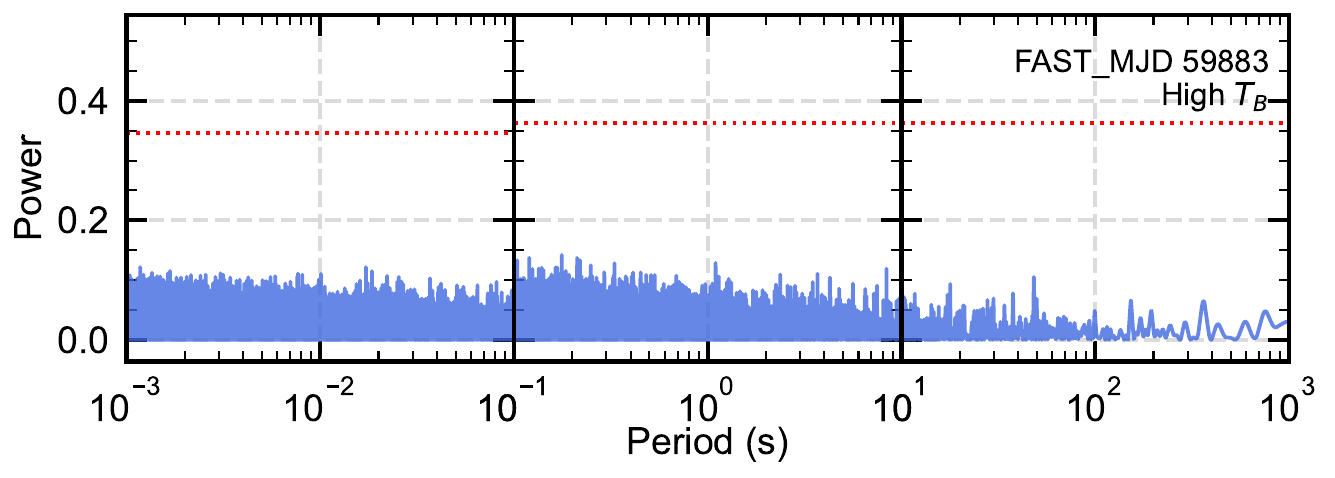}
   \includegraphics[width=0.32\textwidth]{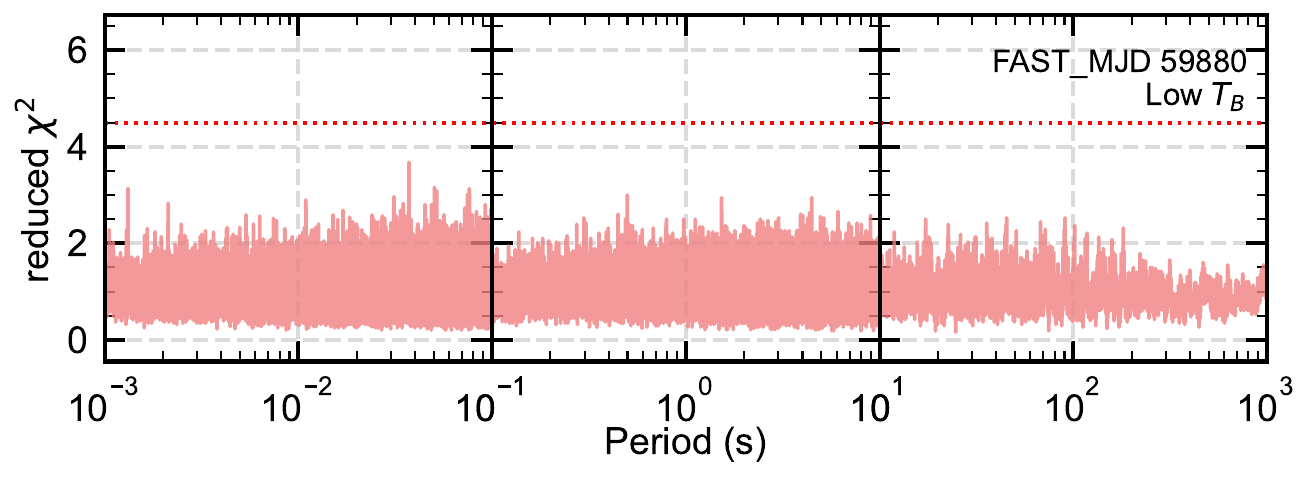}
   \includegraphics[width=0.32\textwidth]{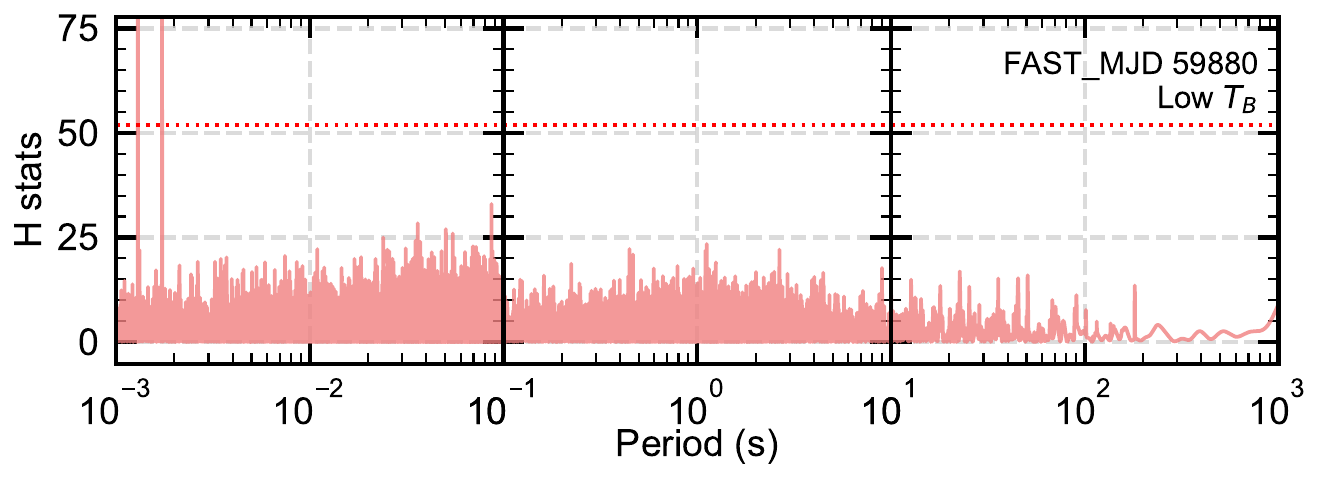}
   \includegraphics[width=0.32\textwidth]{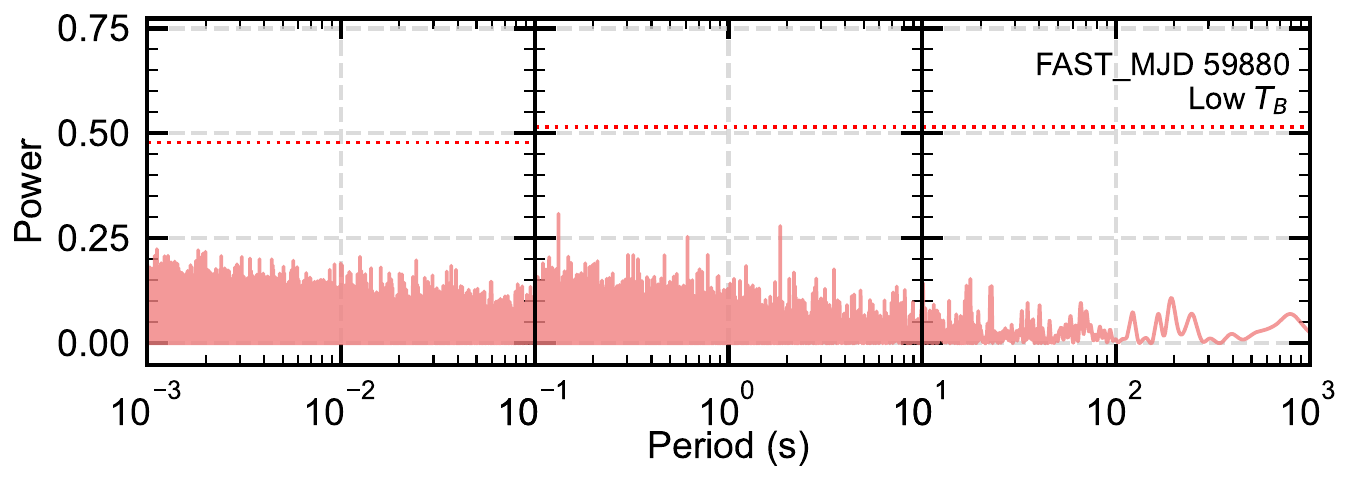}
   \includegraphics[width=0.32\textwidth]{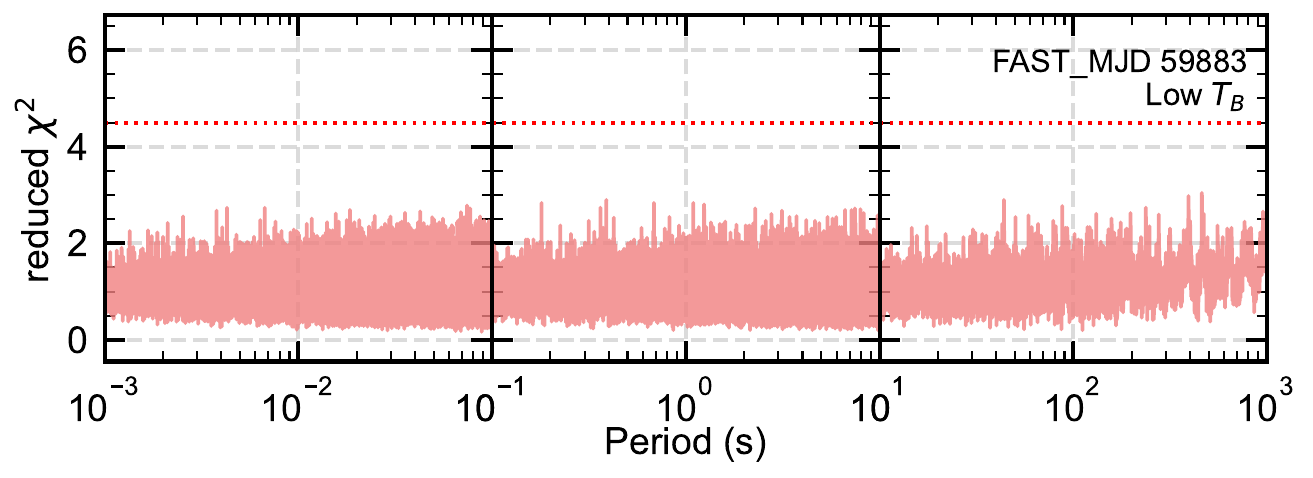}
   \includegraphics[width=0.32\textwidth]{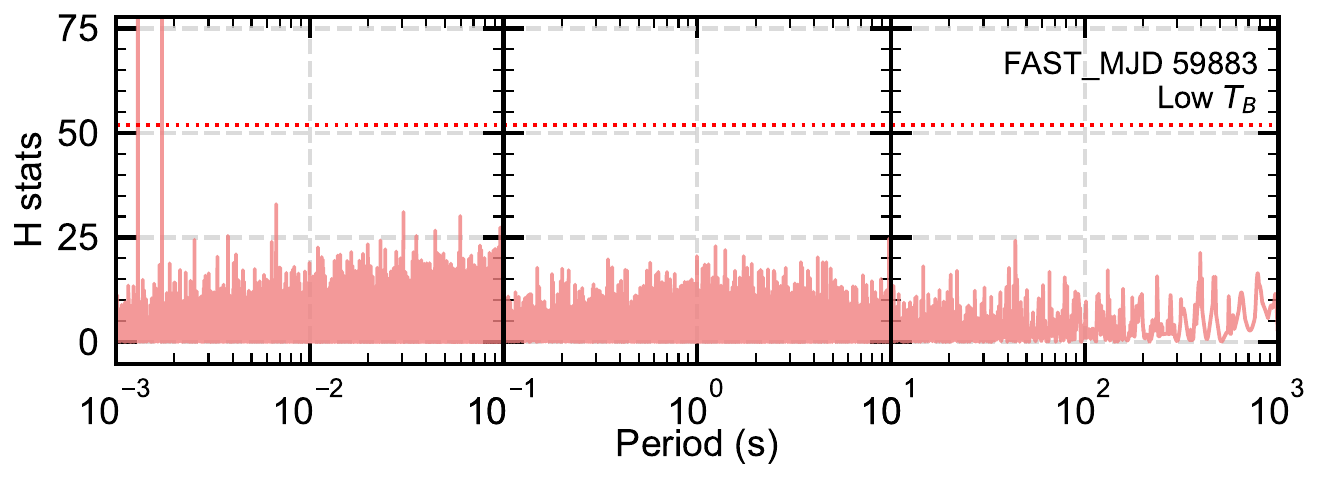}
   \includegraphics[width=0.32\textwidth]{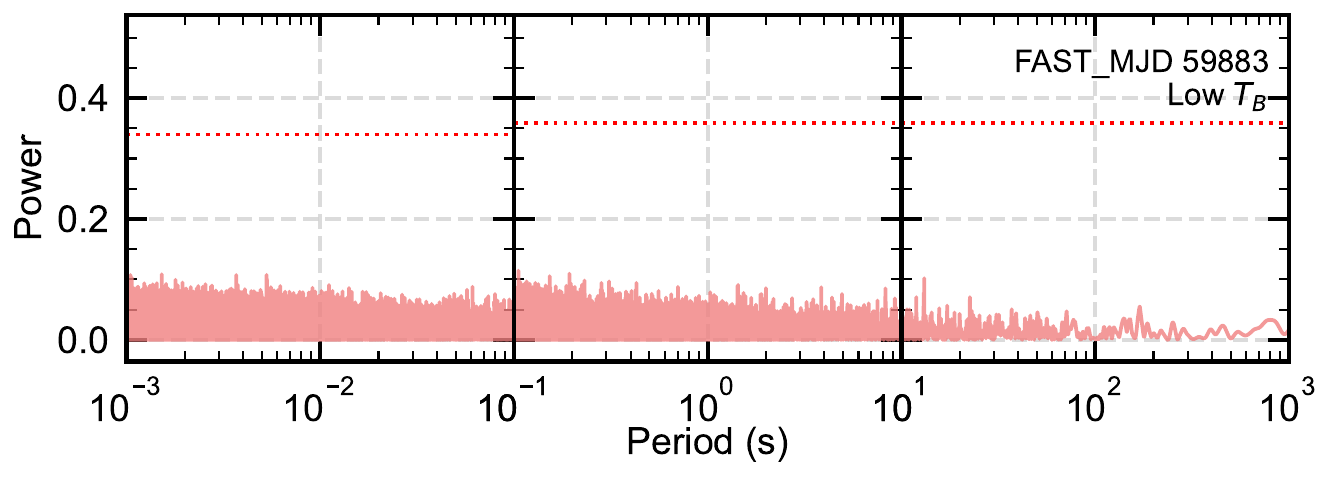}
   \caption{\footnotesize The period search results based on the FAST Dataset \#9 of
   FRB 20220912A, considering the brightness temperature of the bursts. The three columns
   from left to right correspond to the three search methods, i.e. the phase folding algorithm,
   the H-test, and the Lomb-Scargle periodogram, respectively. The rows from top to bottom
   correspond to high temperature or low temperature events on selected days: high $T_\mathrm{B}$
   bursts on MJD 59880 and MJD 59883; and low $T_\mathrm{B}$ bursts on MJD 59880 and MJD 59883.
   In the cases of phase folding and the H-test methods, a horizontal dotted line is plotted
   to mark the p-value of $10^{-9}$. In the cases of Lomb-Scargle periodogram, a horizontal
   dotted line indicating a FAP level of $10^{-9}$ is plotted. No clear evidence of
   periodicity is found in these plots. Note that in the middle panels,
   the peak structures in the 1 ms--2 ms
   range are fake signals due to the limited timing accuracy. }
   \label{Fig18}
\end{figure*}

For the GBT Dataset \#8 of FRB 20220912A, the high
$T_\mathrm{B}$ subset consists of 64 bursts on the only
observation day MJD 59876, of which 59 bursts have a waiting time
> 0.1 s. The low $T_\mathrm{B}$ subset of the GBT Dataset \#8
also consists of 64 bursts on the same day, of which 63 bursts
have a waiting time > 0.1 s. The results are shown in Figure
~\ref{Fig17}. For the FAST Dataset \#9 of FRB 20220912A, the high
$T_\mathrm{B}$ subset consists of bursts on two days: 105 bursts
on MJD 59880, of which 93 bursts have a waiting time > 0.1 s; and
137 bursts on MJD 59883, of which 129 bursts have a waiting time
> 0.1 s. The low $T_\mathrm{B}$ subset of the FAST Dataset \#9
also consists of bursts on two days: 90 bursts on MJD 59880, of
which 80 bursts have a waiting time > 0.1 s; and 140 bursts on MJD
59883, of which 131 bursts have a waiting time > 0.1 s. The
results are shown in Figure ~\ref{Fig18}. In the 1--2
ms range and the 1--20 ms range, several extremely high peaks
could be observed, which are all harmonics of 0.864 ms ($10^{-8}$
d). They are fake signals due to limited timing accuracy of
observations. In each plot, we have also examined by hand the
phase-folded histograms at the positions that correspond to the
relatively high peaks in the $\chi^2$/H-statistic/Lomb-Scargle
periodograms, but found that no marked features appear in the
histograms. Additionally, a peak occurring on one day usually does
not appear on other days, which further excludes its possibility
of being a true period. From Figures ~\ref{Fig17} and
~\ref{Fig18}, we conclude that no periodicity exceeding the
confidence level is found in the whole range of 0.001--1000 s
either in high $T_\mathrm{B}$ or in low $T_\mathrm{B}$ bursts from
FRB 20220912A.

\section{Conclusions and discussion}
\label{sect:discussion}

Whether short timescale periodicity exists or not in the
activities of repeating FRBs is an intriguing issue. In this
study, we perform a thorough search for periodicity in the range
of 0.001--1000 s on the four most active repeaters, i.e. FRBs
20121102A, 20200120E, 20201124A, and 20220912A. Observational data
obtained by various telescopes including FAST, Arecibo, uGMRT,
Effelsberg and GBT are collected and used in the analysis. Three
period searching methods are adopted in the process, which include
the phase folding algorithm, the H-test, and the Lomb-Scargle
periodogram. For the three most active repeating sources, FRBs
20121102A , FRB 20201124A, and 20220912A, more in-depth period
searches are conducted. We have tried to analyze the data by using
the pulse width, peak flux or the fluence as a weight for each
dataset. Periodicity analysis is also conducted separately for
different subsets featured by the brightness temperature. In all
the attempts, no evidence pointing to any clear short timescale
periodicity is found.

The recurrence of FRBs from a particular source strongly indicates
that the bursts are non-catastrophic events. The association of
FRB 20200428 with SGR J1935+2154 firmly establishes the connection
between magnetars and at least some FRBs
\citep{li2021HXMT,mereghetti2020INTEGRAL}. As a result, it is
naturally expected that periodicity connected with the spinning of
magnetars should exist in the activities of repeating FRBs. The
properties of FRBs, such as the pulse width, peak flux, fluence,
and the brightness temperature, may depend on the detailed trigger
mechanism and the position of radiation. However, our study shows
that there are no periodical signals in the detected bursts even
when the effects of pulse width, peak flux, fluence, and the
brightness temperature are considered. Such a null result could
provide useful constraints on the theoretical models of repeating
FRBs. It may indicate that the radiation is emitted in a region
that is relatively far away from the magnetar, at least outside
the light cylinder, where the radiation zone does not co-rotate
with the magnetar. However, recent studies using
scintillation to constrain the size of the emission region of FRB
20221022A \citep{nimmo2024Magnetospheric} suggest that, at least
for this source, the emission region is small. It further implies
that the bursts are produced inside the magnetosphere.
Additionally, the S-shaped swing in the polarization position
angle observed in FRB 20221022A also supports a magnetospheric
origin \citep{mckinven2024pulsarlikea}. Another possibility is
that the burst activities are not restricted to a particular
region of the magnetar. Instead, they could be triggered at
different positions on the surface or in the magnetosphere of the
compact star. It could also be possible that some external
mechanisms such as the collision of asteroids trigger these
repeating bursts \citep{geng2015fast}.

However, considering the fact that the observed FRBs from SGR
J1935+2154 may not be representative of all repeating FRB sources
due to their tendency to have broader bandwidths and lack of
evidence for downward frequency drifting compared to typical
repeating FRBs \citep{giri2023Comprehensive}, the magnetar origin
of FRBs still remains uncertain. Note that
\citet{katz2022absence}'s periodicity study on FRB 20121102A
suggests that if repeating FRBs indeed possess periodicity
connected with the rotation of magnetars, the periodogram should
reveal the periods, even when the random phase deviations exist.
In this regard, the absence of periodicity connected with the
rotation of magnetars challenges the magnetar origin model for
repeating FRBs. Consequently, alternative explanations for
repeating FRBs, such as black hole accretion disk model
\citep{katz2023Sources}, have been proposed.

It should be noted that for the four active repeating FRBs
studied in this work, the existence of periodicity below 100 ms,
especially below 20 ms, still cannot be completely excluded yet.
First, due to the limited timing accuracy of many observations,
the periodicity analysis is subject to serious interference when
searching for periods below 20 ms. Significantly improved timing
accuracy is necessary in the future to overcome this problem.
Second, constraining millisecond-level periodicity requires a
relatively high burst rate (exceeding $\sim 100$ hr$^{-1}$). A
high burst rate means a large number of bursts occur in a short
time span, making the influence of the period derivative
negligible. In fact, \citet{katz2022absence} conducted a
periodicity analysis on FRB 20121102A by considering dense
bursts detected over a short time span
\citep{gajjar2018Highestfrequency}. They did not find any clear
periodicity in the data. Note that some recent datasets also
include many high burst rate events
\citep{li2021bimodal,jahns2022FRB,zhang2022FASTa,zhang2023FAST},
which may further shed light on the extremely short periodicity.
Third, as mentioned earlier, the intrinsic pulse widthes of FRBs
are typically $\sim 10$ ms, and many FRBs have multi-peak
structures. These factors further make it difficult to accurately
determine the arrival time of FRBs. Finally, our simulations
indicate that to constrain millisecond-level periodicities, the
random ToA fluctuations should be less than $\sim 100$ $\mu$s.
Therefore, to better constrain the periodicity below 100 ms, more
precise timing accuracy, more standardized and accurate ToA
determination, and more observations of active sources are needed.
To conclude, constraining millisecond-level periodicity in FRB
activities is still extremely challenging for current
observations, which, however, deserves great efforts in the near
future.


\begin{acknowledgments}
 We thank the anonymous referee for helpful comments and
 suggestions that led to an overall improvement of this study.
 This work is supported
 by the National Key R\&D Program of China (2021YFA0718500),
 by the National SKA Program of China (No. 2020SKA0120300),
 by the National Natural Science Foundation of China (Grant Nos. 12041306, 12233002, 11988101, U2031118),
 and by the Major Science and Technology Program of Xinjiang Uygur Autonomous Region (No. 2022A03013-1).
 YFH also acknowledges the support from the Xinjiang Tianchi Program. We thank Yongkun Zhang for helpful discussion.
\end{acknowledgments}

%





\bibliography{ms}{}
\bibliographystyle{aasjournal}

\begin{deluxetable*}{ccccccc}[h!]
\centering
\renewcommand\arraystretch{1.6}
\tabletypesize{\scriptsize}\tabcolsep=12pt\tablecaption{Summary of
the observations of the four most active repeating FRBs. The data
listed here do not represent all the observations of these FRBs;
rather, they are the clustered bursts that are suitable for
periodicity analysis conducted in this study.}\label{Tab1}
\tablehead{
  \colhead{\multirow{3}{*}{FRB name}} &
  \colhead{\multirow{3}{*}{Dataset number}} &
  \colhead{\multirow{3}{*}{Telescope}} &
  \colhead{\multirow{1}{*}{Number of}} &
  \colhead{\multirow{1}{*}{Observation}} &
  \colhead{\multirow{3}{*}{Active time span$^{\star}$}} &
  \colhead{\multirow{3}{*}{Reference}} \\
  & & & \multirow{1}{*}{detected bursts} & \multirow{1}{*}{frequency}     &                                & \\
  & & &                                  & \multirow{1}{*}{{\tiny(GHz)}}  & \multirow{1}{*}{{\tiny(MJD)}}  &
}
\startdata
\multirow{3}{*}{20121102A}
& \multirow{1}{*}{\#1} & \multirow{1}{*}{Arecibo} & \multirow{1}{*}{478} & \multirow{1}{*}{0.98--1.78} & \multirow{1}{*}{57364--57666} & \multirow{1}{*}{\citet{hewitt2022Arecibo}} \\
& \multirow{1}{*}{\#2} & \multirow{1}{*}{Arecibo} & \multirow{1}{*}{849} & \multirow{1}{*}{1.15--1.73} & \multirow{1}{*}{58409--58450} & \multirow{1}{*}{\citet{jahns2022FRB}}  \\
& \multirow{1}{*}{\#3} & \multirow{1}{*}{FAST} & \multirow{1}{*}{1653} & \multirow{1}{*}{1.05--1.45} & \multirow{1}{*}{58724--58776} & \multirow{1}{*}{\citet{li2021bimodal}} \\
\hline
\multirow{3}{*}{20201124A}
& \multirow{1}{*}{\#4} & \multirow{1}{*}{FAST} & \multirow{1}{*}{1863} & \multirow{1}{*}{1.0--1.5} & \multirow{1}{*}{59307--59360} & \multirow{1}{*}{\citet{xu2022fast}} \\
& \multirow{1}{*}{\#5} & \multirow{1}{*}{uGMRT} & \multirow{1}{*}{48} & \multirow{1}{*}{0.55--0.75} & \multirow{1}{*}{59309--59309} & \multirow{1}{*}{\citet{marthi2021Burst}} \\
& \multirow{1}{*}{\#6} & \multirow{1}{*}{FAST} & \multirow{1}{*}{881} & \multirow{1}{*}{1.05--1.45} & \multirow{1}{*}{59482--59485} & \multirow{1}{*}{\citet{zhang2022FASTa}} \\
\hline
\multirow{1}{*}{20200120E}
& \multirow{1}{*}{\#7} & \multirow{1}{*}{Effelsberg} & \multirow{1}{*}{60} & \multirow{1}{*}{1.255--1.505} & \multirow{1}{*}{59593--59655} & \multirow{1}{*}{\citet{nimmo2023burst}} \\
\hline
\multirow{2}{*}{20220912A}
& \multirow{1}{*}{\#8} & \multirow{1}{*}{GBT} & \multirow{1}{*}{128} & \multirow{1}{*}{1.1--1.9} & \multirow{1}{*}{59876--59876} & \multirow{1}{*}{\citet{feng2023extreme}} \\
& \multirow{1}{*}{\#9} & \multirow{1}{*}{FAST} & \multirow{1}{*}{1075} & \multirow{1}{*}{1.0--1.5} & \multirow{1}{*}{59880--59935} & \multirow{1}{*}{\citet{zhang2023FAST}} \\
\enddata
\tablecomments{${^\star}$ The time span here refers to the
duration between the detection of the first and the last burst in
the dataset, not the total period of the telescope's monitoring.}
\end{deluxetable*}

\end{document}